\documentclass[12pt]{article}
\usepackage{epsfig}
\usepackage{amsfonts,amssymb,amsbsy,cite,enumerate}
\renewcommand\baselinestretch{1.2}
\DeclareMathAlphabet{\mbf}{OT1}{cmr}{bx}{it}
\newcommand{\rf}[1]{(\ref{#1})}

\def\bn{\begin{equation}}
\def\ed{\end{equation}}

\def\be{{\bf e}}
\def\a{\alpha}
\def\b{\beta}
\def\g{\gamma} 
\def\la{\lambda}
\def\l{\lambda}
\def\o{\omega}
\def\s{\sigma}

  \def\ma{\mu_\a} \def\mb{\mu_\b}
\def\q{q^{-1}}
\def\ot{\otimes} 
\def\A{{\mathbf A}}
\def\Ao{\overline{\mathbf A}}

\def\Q{{\mathbf Q}}
\def\Qt{\overline{{\mathbf Q}}}

\def\Tt{\overline{{\mathbf T}}}
\def\T{{\mathbf T}}
\def\Tb{{\mathbb T}}
\def\Tob{{\mathbb {\overline T}}}
\def\To{\overline{\mathbf T}}
\def\hf{{1\over2}}
\def\ab{{\mathbf a}}

\def\ds{\displaystyle}

\def\Go{\overline{\mathbf G}}
\def\G{{\mathbf G}}
\def\Ho{\overline{\mathbf H}}
\def\H{{\mathbf H}}
\def\To{\overline{\mathbf T}}
\def\Q{{\mathbf Q}}
\def\Qt{\overline{{\mathbf Q}}}

\newcommand{\ket}[1]{| #1 \rangle}

\def\Ev{{\mathcal Ev}}
\def\EEv{\overline{{\mathcal Ev}}}

\def\hf{{1\over2}}
\def\ab{{\mathbf a}}

\def\hbc{{\mathcal H}}
\def\rrho{{\overline \rho}}
\def\AAA{{\overline{\mathbf A}}}

\def\vev#1{\langle #1 \rangle}

\newtheorem{conjecture}{Conjecture}
\newtheorem{proposition}{Proposition}
\renewcommand{\theequation}{{\thesection}.{\arabic{equation}}}
\newcounter{app}
\newcounter{sapp}[app]
\def\theapp{\Alph{app}}
\newcommand{\app}[1]{
\refstepcounter{app}{\vspace{7mm}
\noindent\Large\bf Appendix
\theapp.
 \ #1 \par \vspace{5mm}}
\setcounter{equation}{0}
\def\theequation{\Alph{app}.\arabic{equation}}}

\def\t{\theta}

\renewcommand{\thefootnote}{\fnsymbol{footnote}}

\begin{document}
\renewcommand{\baselinestretch}{1.2}
\normalsize

\begin{titlepage}
\vskip 0.5cm
\begin{flushright}
{{hep-th/0105177}}
\end{flushright}
\vskip 2.4cm
\begin{center}
{\LARGE{\bf Integrable structure of ${\mathcal W}_3$ Conformal Field Theory, 
Quantum Boussinesq Theory and  Boundary Affine Toda Theory.}}
\end{center}
\vskip 0.6cm
\begin{center}{Vladimir V. Bazhanov${}^{1,2,}$%
\footnote[2]{e-mail: {\tt vladimir.bazhanov@rsphysse.anu.edu.au}},
Anthony N. Hibberd${}^{1,}$%
\footnote[3]{e-mail: {\tt anthony.hibberd@rsphysse.anu.edu.au}}\\
and Sergey M. Khoroshkin${}^{3,4}$%
\footnote[4]{e-mail: {\tt khor@heron.itep.ru}}}
\end{center}
\vskip 1.5cm
\centerline{${}^{1}$ Department of Theoretical Physics,\, } 
\centerline{Research School of Physical Sciences and Engineering,\, }
\centerline{Institute of Advanced Studies,\, }
\centerline{Australian National University,\, } 
\centerline{Canberra, ACT, 0200, Australia\, }
\vskip .5cm
\centerline{${}^{2}$Centre for Mathematics and its
Applications,\, }  
\centerline{Institute of Advanced Studies,\, }
\centerline{Australian National University,\, } 
\centerline{Canberra, ACT, 0200, Australia\, }
\vskip .5cm
\centerline{${}^{3}$Institute of Theoretical and
Experimental Physics\,} 
\centerline{Moscow, 117259, Russia\, }
\vskip .5cm
\centerline{${}^{4}$Max-Planck Institut fur Mathemetik,\,} 
\centerline{Vivatsgasse 7, 53111, Bonn, Germany\,}

\flushbottom
\begin{flushleft}
May, 2001
\end{flushleft}

\end{titlepage}

\renewcommand{\thepage}{\roman{page}}
\section*{Abstract}
\noindent
In this paper we study the Yang-Baxter integrable structure 
of  Conformal Field Theories
with extended conformal symmetry generated by the ${\cal W}_3$
algebra. We explicitly construct various ${\bf T}$ and ${\bf Q}$-operators
which act in the irreducible highest weight modules of the ${\cal W}_3$
algebra. These operators can be viewed as continuous field theory analogues of
the commuting transfer matrices and $Q$-matrices of the integrable lattice
systems associated with the quantum algebra $U_q(\widehat{sl}(3))$. 
We formulate several conjectures detailing certain analytic characteristics 
of the ${\bf Q}$-operators and propose exact asymptotic expansions of
the  
${\bf T}$ and ${\bf Q}$-operators at large values of the spectral parameter.
We show, in particular, that the asymptotic expansion of 
the ${\bf T}$-operators 
generates an infinite set of local integrals of motion of the ${\cal W}_3$ CFT
which in the classical limit reproduces an infinite set of conserved 
Hamiltonians associated with the classical Boussinesq equation.
We further study the vacuum eigenvalues of the ${\bf Q}$-operators 
(corresponding to the highest weight vector of the  ${\cal W}_3$
module) and show that they are simply related to the expectation values 
of the boundary exponential fields in the non-equilibrium boundary
affine Toda field theory with zero bulk mass.

\newpage

\renewcommand{\thefootnote}{\arabic{footnote}}
\setcounter{footnote}{0}

\renewcommand{\arraystretch}{1.5}
\newcounter{storeeqn}
\newenvironment{numlet}{
\addtocounter{equation}{1}
\setcounter{storeeqn}{\value{equation}}
\setcounter{equation}{0}
\renewcommand{\theequation}{\thesection.\arabic{storeeqn}\alph{equation}}}{
\setcounter{equation}{\value{storeeqn}}
\renewcommand{\theequation}{\thesection.\arabic{equation}}}

\tableofcontents

\newpage

\setcounter{equation}{0}
\section{Introduction}
\setcounter{page}{1}
\renewcommand{\thepage}{\arabic{page}}
\renewcommand{\baselinestretch}{1.2}
\normalsize
The Yang-Baxter equation and its 
associated integrability structures play an important 
role in the understanding of Integrable Quantum Field Theory (IQFT)
\cite{ZZ79}.
Recently, the authors of \cite{BLZ96,BLZ97a,BLZ99a,BLZ97b} 
developed a continuous 
version of the
Quantum Inverse Problem Method (QIPM) \cite{STF79} or Baxter's commuting
transfer matrix method \cite{Bax82} to study IQFT, especially the finite
volume IQFT. This new method inherits all the analytic power of
the theory of integrable lattice models but does not require intermediate 
lattice regularization and allows from the very beginning to work 
in terms of the continuous IQFT. 
The method has been most successfully applied
\cite{BLZ96,BLZ97a,BLZ99a} 
to the 
$c<1$ Conformal Field Theory (CFT) which is a simplest
nontrivial example of IQFT. The fundamental problem addressed in
these papers is  
the simultaneous diagonalization of an infinite set of mutually commuting
local integrals of motion (IM) in CFT.

In this paper we extend  the results of
\cite{BLZ96,BLZ97a,BLZ99a} to CFT's with  
extended conformal symmetry generated by the simplest 
$W$-algebra ${\cal W}_3$ \cite{Zam85}. 
Such a CFT involves additional local spin-3 fields  $W(u)$ and
$\overline{W}(\overline{u})$ besides the usual left and right 
components $T(u)$ and
$\overline{T}(\overline{u})$ of the stress-energy tensor. 
In what follows we will concentrate on the left chiral component ${\cal
W}_3$   of the full symmetry algebra ${\cal W}_3\times\overline{{\cal W }}_3$.
The algebra ${\cal W}_3$ 
is generated by the fields $T(u)$ and $W(u)$ along with 
all composite fields built from products of $T(u)$ and $W(u)$ and
their derivatives. We will interpret the variable $u$ as a 
 complex coordinate on a 2D cylinder of circumference $2\pi$ and impose
the  periodic boundary conditions
\begin{equation}
T(u+2\pi)=T(u),\qquad W(u+2\pi)=W(u).\label{BC-int}
\end{equation}

The algebra ${\cal W}_3$ possesses an infinite-dimensional  Abelian subalgebra 
\cite{LF90,KM89} generated by
the local integrals of motion (IM), ${\bf I}_k\in {\cal W}_3$ 
\begin{equation}
{\bf I}_k=\int_{0}^{2\pi}{du\over{2\pi}}\ T_{k+1}(u), \qquad  k\in Z_+,\qquad
k\not=0 \pmod 3 \label{locintintro}
\end{equation}
where the densities $T_{k}(u)$ are appropriately regularized 
polynomials in $T(u)$ and $W(u)$ and their derivatives. 
The first  two densities read 
\begin{equation}
T_2(u)=T(u),\qquad T_3(u)=W(u),
\end{equation}
while the higher densities are determined by the requirement of 
commutativity
\begin{equation}
[{\bf I}_k,{\bf I}_l]=0,
\end{equation}
and by the ``spin'' assignment \cite{LF90} (see Section~\ref{QBT} for
the details).

We explicitly construct operator valued functions ${\mathbb T}(t)$ and
${\overline {\mathbb T}}(t)$ where $t$ is a complex parameter. 
These operators act invariantly in each 
irreducible highest weight module 
of the ${\cal W}_3$ algebra 
\bn
{\mathbb T}(t),
{\overline {\mathbb T}}(t):\qquad 
{\cal V}_{\Delta_2,\Delta_3}\to{\cal V}_{\Delta_2,\Delta_3}\label{Tact}
\ed
where  $\Delta_2,\Delta_3$ are the highest weights
and they commute
among themselves for any values of $t$, i.e.,
\begin{equation}
[{\mathbb T}(t),{\mathbb T}(t')]=[{\mathbb T}(t),{\overline {\mathbb
T}}(t')]=[{\overline {\mathbb T}}(t),{\overline {\mathbb T}}(t')]=0. 
\end{equation}
These operators are defined in terms of certain monodromy
matrices associated with the three-dimensional 
representation of the quantum
algebra $U_q(\widehat{sl}(3))$ where  
\begin{equation}
q=e^{i\pi g},
\end{equation}
and $g$ is related to the central  charge $c$ of the ${\cal W}_3$ algebra
as
\begin{equation}
c=50-24(g+g^{-1}),
\end{equation}
Here we mostly restrict our considerations to the ``quasiclassical
domain'' \cite{BLZ96} 
\begin{equation}
-\infty<c<-2,
\end{equation}
however, the results are valid in a wider region $-\infty<c<2$,
corresponding to $0<g<1$.

Obviously, the operators ${\mathbb T}(t)$ and
${\overline {\mathbb T}}(t)$ are ${\cal W}_3$ analogues of Baxter's commuting
transfer matrices from lattice theory. They enjoy remarkable 
analyticity properties, namely, they are entire functions of $t$ with an
essential singularity at $t=\infty$ and their asymptotic expansion
near this point is described in terms of the local IM
\rf{locintintro}. For instance, 
\begin{equation}
\log {\mathbb T}(t)\simeq  m\ t^{\frac{1}{3(1-g)}}+
\sum_{k=1}^{\infty}C'_{k}\ t^{-\frac{k}{3(1-g)}}\ {\bf I}_{k}, 
\end{equation}
where $m$ and $C'_k$ 
are some explicitly known coefficients which depend on the central charge 
$c$ only and $C'_k\equiv0$ for $k=0\pmod 3$. A similar expansion holds 
for ${\overline {\mathbb T}}(t)$. 
Therefore the  operators ${\mathbb T}(t)$ and
${\overline {\mathbb T}}(t)$ can be thought of as generating functions of
the local IM since their eigenvalues contain all the information about 
the eigenvalues of the local IM. 

Another powerful method known in the lattice theory is based on the
so-called ${\bf Q}$-operators. This method was introduced by Baxter
 in his original solution of the eight-vertex model of statistical
mechanics \cite{Bax72}. Here we construct appropriate 
 ${\cal W}_3$ versions of Baxter's ${\bf
Q}$-operators (actually, there are six such operators 
${\bf Q}_i(t)$ and ${\overline {\bf Q}}_i(t)$, $i=1,2,3$). They are defined 
again as traces of certain monodromy matrices this time associated
with infinite-dimensional representations of the so-called
$q$-oscillator algebra. 
The ${\bf Q}$-operators satisfy the following relations with 
the operators ${\mathbb T}(t)$ and ${\overline {\mathbb T}}(t)$ discussed above
\begin{equation}\begin{array}{rcl}
\Q_i(tq^3)-\Tb(tq^{\hf})\,\Q_i(tq)+\Tob(tq^{-\hf})\,\Q_i(tq^{-1})
-\Q_i(tq^{-3})&=&0,\\&&\\
\Qt_i(tq^{3})-\Tob(tq^{\hf})\,\Qt_i(tq)+\Tb(tq^{-\hf})\,
\Qt_i(tq^{-1})-\Qt_i(tq^{-3})&=&0,
\end{array}\label{TQrel}
\end{equation}
where $i=1,2,3$. 
These relations are obviously the $U_q(\widehat{sl}(3))$-analogues
\cite{KR82,KLWZ97} of Baxter's famous  T-Q relation.  The relations \rf{TQrel}
are just the first representatives of a rich family of the functional
relations obeyed by the ${\bf Q}$-operators in the
$U_q(\widehat{sl}(3))$ related integrable systems. In particular,
in Section~4 
we define more general ${\bf T}$-operators ${\bf T}_\mu(t)$ 
and $\overline{{\bf T}}_\mu(t)$ corresponding to finite-dimensional 
irreducible representations $\pi_\mu$ of $U_q({gl}(3))$. Here
$\mu=(\mu_1,\mu_2,\mu_3)$ is the highest weight of the representation 
$\pi_\mu$ such
that $\mu_1-\mu_2$ and $\mu_2-\mu_3$ are non-negative
integers\footnote{The existence of the two different 
sets of operators  ${\bf T}_\mu(t)$  and $\overline{{\bf T}}_\mu(t)$
is related to the fact that there are two non-equivalent ``evaluation
maps'' from $U_q(\widehat{sl}(3))$ to $U_q({gl}(3))$.}. All these ${\bf
T}$-operators can be  expressed through $\Q_i(t)$ and $\Qt_i(t)$ 
by the means 
of only two elegant relations 
\bn
\begin{array}{rcl}
{\mathbf T}_{(\mu_1,\mu_2,\mu_3)}(t)&=&z_{0}^{-1}
\det\left\Vert\,\Q_{i}(t\, q^{+2{\mu'}_j})\right\Vert_{i,j=1,2,3}
\\&&\\
-\Tt_{(\mu_1,\mu_2,\mu_3)}(t)&=&z_{0}^{-1}
\det \left\Vert\,{\Qt}_{i}( t\,q^{-2{\mu'}_j})\right\Vert_{i,j=1,2,3}
\end{array}\label{Tweyl}
\ed
where ${\mu'}_1=\mu_1+1$,\ \  ${\mu'}_2=\mu_2$, \ \ ${\mu'}_3=\mu_3-1$, and 
where $z_0$ is a simple operator acting diagonally in each ${\cal W}_3$ 
module which can be interpreted as  the ``quantum Wronskian'' \cite{BLZ97a}
of the third order difference equations \rf{TQrel}
\bn
z_0=\det\left\Vert{\Q}_{i}(t q^{-2j})\right\Vert_{i,j=1,2,3} =
-\det \left\Vert{\Qt}_{i}(t q^{2j})\right\Vert_{i,j=1,2,3}
\label{qwron}
\ed
We prove the  relations \rf{Tweyl} algebraically by using decomposition 
properties of direct products of certain ``$q$-oscillator''
representations of the Borel subalgebra of
$U_q(\widehat{sl}(3))$\footnote{It should noted that some
particular cases of \rf{Tweyl} were previously obtained \cite{PS99,KLWZ97} 
from the
properties of the corresponding eigenvalues determined by the
so-called ``nested 
Bethe Ansatz''  solutions \cite{Sut75}.}.
These relations  
provide a very simply tool for derivation of various ``fusion relations''
for the ${\bf T}$-operators \cite{BR90,KNS94}. In particular, the
so-called RSOS-reductions \cite{BR90} of the ${\bf T}$-operators
simply follow from \rf{Tweyl} when two
columns of the three-by-three matrices therein become proportional.

The eigenvalues of the ${\bf Q}$-operators  can be studied by various methods.
In particular, 
they satisfy the Destri-de-Vega non-linear integral 
equations \cite{DdV95,KBP91,DT00a}. We solve these equations exactly 
in a particular case when $\Delta_2,\Delta_3\to\infty$, where 
$\Delta_2,\Delta_3$ are the highest weight in \rf{Tact}. On the basis 
of this result in Section~6 
we conjecture exact asymptotic asymptotic expansions 
of the eigenvalues of the ${\bf Q}$-operators at large values of $t$. 
These expansions 
exhibit remarkable duality properties of the ${\bf Q}$-operators
with respect to the
substitution $g\to g^{-1}$ similar to the case of $c<1$ CFT 
\cite{BLZ97a}. We support our conjectures by consideration of certain 
exactly solvable cases and numerical calculation of the zeroes of 
eigenvalues of the ${\bf Q}$-operators. 

As shown in \cite{BLZ96} the ${\bf T}$ and ${\bf Q}$-operators  
define integrable boundary conditions \cite{GZ94} for CFT in the geometry of
the half-infinite cylinder. In particular, the vacuum eigenvalues
of these operators can be interpreted \cite{BLZ97a} as finite-temperature 
partition functions of certain integrable QFT models with a non-trivial 
boundary interaction. These models are of interest in many branches of 
physics. For example, the integrable boundary sine-Gordon model
\cite{GZ94} (with zero bulk mass) finds important application in
dissipative quantum mechanics \cite{CL81,Sch83,CT90,FZ90};
as explained in \cite{KF92,MYKGF93} it also describes the universal
current through the  
point contact in the quantum Hall system\cite{FLS95a,FLS95b,BLZ97a,BLZnon}.
Here we consider a simplest generalisation of this model, namely, the
boundary $A^{(1)}_2$ affine Toda theory with zero bulk mass.
Its Hamiltonian has the form
\begin{equation}	
{\bf H}={1\over {4\pi g}}\int_{-\infty}^{0}dx\, 
\big(\, 
{{\boldsymbol \Pi}}^2 + 
{(\partial_x{\boldsymbol \Phi})}^2\,  \big)
-{\kappa\over 2g}\sum_{j=1}^3\ y_j\ e^{i({\mbf e}_j,{\boldsymbol \Phi}(0))} 
\ ,\label{BAT-int}
\end{equation}
where $g$, $\kappa$ are parameters and 
${\boldsymbol \Phi(x)}=(\Phi_1(x),\Phi_2(x))$,
${\boldsymbol{\Pi}}(x)=(\Pi_1(x),\Pi_2(x))$
are field operators obeying canonical 
commutation relations 
\bn
\big[\Pi_a(x)\, , \Phi_b(x')\big] = -2\pi i g\  \delta (x-x')\delta_{ab}
,\qquad a,b=1,2\label{bosecomm-int}.
\ed
and ${\mbf e}_j$, $j=1,2,3$ are the simple roots of the 
algebra $A_2^{(1)}$.
The ``amplitudes''   $y_1,y_2,y_3$ in \rf{BAT-int}
are not c-numbers, but quantum operators describing some additional
boundary degrees of freedom. They  satisfy the commutation relations
\begin{equation}
y_1y_2=qy_2y_1,\qquad y_2y_3=qy_3y_2, \qquad
y_3y_1=qy_1\,y_3,\qquad  
y_1y_2y_3=1,\label{ucomm-int}
\end{equation}
where $q=e^{i\pi g}$. It should be noted that the integrability of the 
boundary affine Toda systems has been studied previously both on classical
\cite{CDRS94} and quantum levels\cite{PZ95}. In \cite{CDRS94} it
was shown  
that (with the exception of the boundary sine-Gordon model) the 
integrability in the boundary affine Toda theory 
imposes severe restrictions on the amplitudes of the boundary interaction.
Namely, the boundary interaction should vanish when the bulk mass is zero. 
This conclusion was then confirmed \cite{PZ95} 
with the perturbation theory calculations in the quantum case.
We would like to stress here that the model \rf{BAT-int} does not 
fall under this ``no-go theorem'' since \rf{BAT-int} involves additional
boundary interaction, which was not considered in \cite{CDRS94,PZ95}. 
We claim here that the theory \rf{BAT-int} is integrable, but postpone 
detailed considerations of this question to a separate publication.
Actually, the  
sufficient condition for the integrability of \rf{BAT-int} 
is that operators $y_j$ obey the Serre
relations of quantum affine algebra  $U_q(A_2^{(1)})$ 
\begin{equation}
y_i^2\,y_j-(q+q^{-1})\,y_i\,y_j\,y_i+y_j\,y_i^2=0, \qquad i\not=j.
\end{equation}
and not necessarily restricted to the case \rf{ucomm-int}.
A Similar statement holds for generalisations of \rf{BAT-int} to other affine
algebras. In Section~\ref{BATFT} we consider a non-equilibrium variant of
\rf{BAT-int} 
with a particular time-dependent boundary interaction. We show that
expectation values of boundary exponential fields over a stationary
non-equilibrium state in this theory are simply related to the vacuum
eigenvalues of the ${\bf Q}$-operators of the ${\cal W}_3$ CFT. This
calculation generalises the corresponding result of \cite{BLZnon}
for the boundary sine-Gordon model.

\setcounter{equation}{0}
\section{Quantum Boussinesq theory} \label{QBT}
\subsection{Extended ${\mathcal W}_3$ conformal symmetry and local
integrals of motion}
In this paper we consider CFT with  
extended conformal symmetry generated by the simplest 
$W$-algebra ${\cal W}_3$ (sometimes called $WA_2$) \cite{Zam85, FL88}.
Such a CFT involves additional local spin-3 fields  $W(u)$ and
$\overline{W}(\overline{u})$ besides the usual left and right 
components $T(u)$ and
$\overline{T}(\overline{u})$ of the stress-energy tensor. 
The left chiral component ${\cal
W}_3$   of the full symmetry algebra ${\cal W}_3\times\overline{{\cal W }}_3$
is generated by the fields $T(u)$ and $W(u)$ along with 
all composite fields built from products of $T(u)$ and $W(u)$ and
their derivatives. We will interpret the variable $u$ as a 
 complex coordinate on a 2D cylinder of circumference $2\pi$ and impose
the  periodic boundary conditions \rf{BC-int}.
The Fourier mode expansions of $T(u)$ and $W(u)$
 \bn
T(u)=-{c\over 24}+\sum_{-\infty}^{+\infty}L_{-n}\ e^{inz},\qquad
W(u)=\sum_{-\infty}^{+\infty}W_{-n}\ e^{inz},\label{fourier}
\ed
are expressed in terms of the operators $L_n$ and $W_n$ satisfying the
commutation relations \cite{Zam85}
\begin{eqnarray}
[L_n, L_m]&=&(n-m)\ L_{n+m}+{c\over 12}\,(n^3-n)\ \delta_{n+m,0}, \nonumber\\{}
[L_n,W_m]&=&(2n-m)\ W_{n+m},\label{Walg}\\{}
[W_n,W_m]&=&\frac{1}{3}(n-m)\Lambda_{n+m}
+\frac{(n-m)}{3b^2}\left[\frac{1}{15}(n+m+2)(n+m+3)\right.\\
&&\left.-\frac{1}{6}(n+2)(m+2)\right]L_{n+m}
+\frac{c}{1080\,b^2}(n^2-4)(n^2-1)n\delta_{n+m,0},\nonumber
\end{eqnarray}
where $c$ is the central charge, $$b^2=16/(22+5c),$$ and 
the operators $\Lambda_n$ are expressed through $L_n$ as
\bn
\Lambda_n=\sum_{k=-\infty}^{+\infty}: L_k L_{n-k} :+{1\over5}\,x_n
L_n,\label{TTdef}  
\ed
with
$$
x_{2l}=(1+l)(1-l),\qquad x_{2l+1}=(2+l)(1-l).
$$
The symbol $:\ :$ in \rf{TTdef} 
denotes the normal ordering such that operators $L_n$ with bigger $n$
are placed to the right. 

The chiral space of state ${\cal H}_{chiral}$ is built up from a
suitable collection of highest weight modules
\bn
{\cal H}_{chiral}=\mathop{\oplus}_a{\cal V}_{(\Delta^{(a)}_2,\Delta^{(a)}_3)},
\label{Hchiral}
\ed
where the parameters $\Delta_2$ and $\Delta_3$ 
are the highest weights, and the 
associated highest weight vectors $|\Delta_2,\Delta_3\rangle\in
{\cal V}_{(\Delta_2,\Delta_3)}$ satisfy the
equations
\bn
L_n{|\Delta_2,\Delta_3\rangle=0},
\qquad W_n|\Delta_2,\Delta_3\rangle=0,\qquad n>0;
\ed
\bn
L_0|\Delta_2,\Delta_3\rangle=\Delta_2|\Delta_2,\Delta_3\rangle,\qquad W_0
|\Delta_2,\Delta_3\rangle=
\Delta_3|\Delta_2,\Delta_3\rangle.
\ed

The algebra ${\cal W}_3$ possesses an infinite Abelian subalgebra 
\cite{LF90,KM89} generated by
the local integrals of motion (IM), ${\bf I}_k\in {\cal W}_3$ 
\bn
{\bf I}_k=\int_{0}^{2\pi}{du\over{2\pi}}\ T_{k+1}(u), \qquad  k\in Z_+,\qquad
k\not=0 \pmod 3 \label{locint}
\ed
where the densities $T_{k}(u)$ are appropriately regularized 
polynomials in $T(u)$ and $W(u)$ and their derivatives. The integer  $k+1$ in 
\rf{locint} denotes   the spin of the local field $T_{k+1}$ 
\begin{equation}
\oint_{\cal C}{dw\over{2\pi i}}\ 
(w-u){\cal T}\big(T(w)T_{k+1}(u)\big)
=(k+1)\ T_{k+1}(u), \label{spin}
\end{equation}
where ${\cal T}$ stands for the chronological ordering 
\begin{equation}{\cal T}\big(A(w)B(u)\big)=\cases{
A(w)B(u)\ ,\ \    & if\ \ \  $  \Im m\  u > \Im m\  w$\ ;\cr
B(u)A(w)
\  ,\ \  & if\ \ \  $  \Im m\  w > \Im m\  u$\ .\cr}
\end{equation}
The set of possible values of $k$ in \rf{locint} is determined 
by the allowed values of spins of the densities $T_{k+1}(u)$  \cite{LF90}
\begin{equation}
{\rm spin}[T_{k+1}(u)]=k+1, \qquad k\in {\mathbb Z}_+,\qquad k\not=3\pmod 3
\label{spinas}
\end{equation}

The first few densities
$T_{k}(u)$ can be written as \cite{KM89}\footnote{A 
misprint in the expression for $T_6(u)$ in \cite{KM89} is corrected here.}
\bn
T_2(u)=T(u),\qquad T_3(u)=W(u),\qquad T_5(u)=:T(u)W(u):, 
\ed
\bn
T_6(u)=:T(u)^3:+9:W(u)^2:+\frac{c-10}{32}:T'(u)^2:,\label{dens}
\ed
where the prime denotes the derivative and $:\ :$ denotes the regularized 
product of the fields, such that the product $:A(u)B(u):$ is 
defined as the regular term (of order
$O((u-v)^0$) in the operator product expansion of
$A(u)B(v)$.\footnote{
We assume that these regularized
product are calculated in the cylinder geometry,
as they are not invariant under conformal
transformations.}
The densities \rf{dens} are
defined up to total derivatives which, obviously, have no effect in
\rf{locint}. A general expression for the  densities \rf{dens} is not 
known, however, once the first two densities in \rf{dens} are fixed
all other higher densities can be uniquely determined (up to a
normalisation) by the requirement of 
commutativity
\bn
[{\bf I}_k,{\bf I}_l]=0,\
\ed
and the spin assignment \rf{spinas}.
Alternatively one can formulate  the last requirement as the ``degree assignment'', namely, if one assigns degrees $2$ and
$3$ to the 
fields $T(u)$ and $W(u)$, respectively, and the degree 1 to each
derivative then $T_{k+1}(u)$ is a homogeneous polynomial of the degree
$k+1$. We normalise the densities $T_{k+1}(u)$ by fixing 
coefficients to the terms containing highest power of $T(u)$, 
\begin{equation} 
T_{2n}(u)=(T(u))^n+\ldots,\qquad
T_{2n+1}(u)=(T(u))^{n-1}\,
W(u)+\ldots,\qquad n=1,2,3,\ldots\label{Inorm}
\end{equation}

The integrals \rf{locint} define operators ${\bf I}_k$: ${\cal
V}_{\Delta_2,\Delta_3}\to{\cal V}_{\Delta_2,\Delta_3}$ 
which can be expressed in terms of
the generators $L_n$ and $W_n$, for example
\pagebreak
\begin{eqnarray}
{\bf I}_1&=&L_0-{c\over24}\ ,\nonumber\\{\bf I}_2&=&W_0\ ,\nonumber\\
{\bf
I}_4&=&\sum_{n=1}^{+\infty}(L_{-n}W_{n}+W_{-n}L_{n})+W_{0}L_{0}-{{c+6}\over
24}W_{0}\ ,\label{firstIM}\\
{\bf I}_5&=&\sum_{n_1+n_2+n_3=0}:L_{n_1}L_{n_2}L_{n_3}:
+\sum_{n=1}^{\infty}\left(\frac{14+c}{16}n^2-2-\frac{c}{4}\right)L_{-n}L_{n}+ 
\frac{3}{2}\sum_{r=1}^{\infty}L_{1-2r}L_{2r-1}\nonumber\\
&&+18\sum_{n=1}^{\infty}W_{-n}W_{n}
+9W_0^2-\frac{c+8}{8}L_0^2+\frac{(c+15)(c+2)}{192}L_0
-\frac{c(c+23)(7c+30)}{96768}.  \nonumber\\
&&\nonumber
\end{eqnarray}
\renewcommand{\arraystretch}{1.5}
Here the symbol $:\ :$ denotes the same normal ordering as in \rf{TTdef}
(where $L_n$ with bigger $n$ are placed to the right).
Note that although these operators are not polynomials in $L_n$ and
$W_n$ their action in ${\cal V}_{\Delta_2,\Delta_3}$ is well defined.

\subsection{Relation to classical Boussinesq theory}
In this paper we will study the problem of simultaneous
diagonalization of the infinite set of the operators ${\bf I}_k$ in 
${\mathcal V}_{(\Delta_2,\Delta_3)}$.
This problem can be regarded as the quantum version of the related theory 
of the classical Boussinesq
 equation \cite{Bou72}, which is a well known example of a completely
integrable Hamiltonian system possessing an infinite set of local IM
\cite{Zak73}. 
As is known \cite {LF90} 
the ${\cal W}$-algebras can regarded as the quantization 
of the Gelfand-Dickey Poisson bracket algebras \cite{GD78}. 
The classical limit of \rf{Walg} corresponds to  $c\to-\infty$. In this
limit the substitution
\begin{equation}
T(u)\to\frac{c}{24}\,U(u),\qquad W(u)\rightarrow
i\Big(-\frac{c}{24}\Big)^{3\over2}V(u), 
\qquad [\ ,\ ]\to \frac{48\pi}{ic}\{\ 
,\ \},\label{climit}
\end{equation}
where 
\begin{equation}
U(u+2\pi)=U(u),\qquad V(u+2\pi)=V(u), \label{UVper}
\end{equation}
 reduces the algebra
\rf{Walg} to the following Poisson bracket algebra \cite{Mat88}
\begin{eqnarray} 
\{U(u),U(v)\}&=&-\big({U(u)+U(v)}\big)\ \delta'(u-v)
+2\ \delta'''(u-v), \nonumber \\
\{U(u),V(v)\}&=&-\big({V(u)+2V(v)}\big)\ \delta'(u-v),\label{PB}\\
\{V(u),V(v)\}&=&-\Big[{\frac{1}{4}({U''(u)+U''(v)})
+\frac{1}{3}({U^2(u)+U^2(v)})}\Big]\  
\delta'(u-v)
\nonumber\\
&+&\frac{5}{12}({U(u)+U(v)})\delta'''(u-v)-\frac{1}{6}\delta^{(5)}(u-v),
\nonumber
\end{eqnarray}
which is a particular example of the Gelfand-Dickey bracket algebra.
The substitution \rf{climit} brings the local IM  \rf{locint}  
to their classical counterparts 
\bn
{\bf I}_k\to
i^{-k-1}\Big(-\frac{c}{24}\Big)^{k+1\over2}I^{(cl)}_k,\qquad
c\to-\infty \label{qclass}
\ed
which form a commutative Poisson bracket algebra
$\{{I^{(cl)}_k,I^{(cl)}_l}\}=0$ .
For example, the first few classical IM $I^{(cl)}_k$ read 
\begin{eqnarray}
 I^{(cl)}_1&=&\int_0^{2\pi}\frac{du}{2\pi}\  U(u)\ ,\\
 I^{(cl)}_2&=&\int_0^{2\pi}\frac{du}{2\pi}\  V(u),\\
 I^{(cl)}_4&=&\int_0^{2\pi}\frac{du}{2\pi}\ U(u) V(u),\\
 I^{(cl)}_4&=&\int_0^{2\pi}\frac{du}{2\pi}\
\Big(U(u)^3+9V(u)^2+\frac{3}{4}U'(u)^2\Big).
\label{classint}
\end{eqnarray}
The Poisson bracket algebra \rf{PB} is known to describe the {\it second}
Hamiltonian structure  of the Boussinesq equation \cite{McK78,Mat88} provided
one of the infinite set of the classical IM $I^{(cl)}_k$ 
is taken as a Hamiltonian.

The evolution equations\footnote{
The original Boussinesq equation of \cite{Bou72}
$$ 
\partial_t \,U=-2\partial_u \,V,\qquad \partial_t\,V=\frac{1}{3}\,\partial_u
\,\Big[\frac{1}{2}\partial_u^2\, U-U^2\Big],\nonumber
$$
corresponds to $k=2$ in \rf{hamflow}} 
\begin{equation}  
\partial_{t_k} U = \{I^{(cl)}_k,U\},\qquad \partial_{t_k} V =
\{I^{(cl)}_k,V\},\qquad t_1=u, \qquad k\not =0 \pmod 3,\label{hamflow}
\end{equation}
describe isospectral deformations of the third-order differential
operator 
\begin{equation} 
L=\partial_u^3-\,U(u)\,\partial_u+V(u)-\frac{1}{2} 
U'(u)-\lambda^3,\label{Lthird}
\end{equation}
where $\lambda$ is the spectral parameter. 
In particular, if one defines the monodromy matrix ${\bf M}(\lambda)$, 
which is an element of the group $SL(3)$, as 
\begin{equation} 
(\psi_1(u+2\pi),\psi_2(u+2\pi),\psi_3(u+2\pi))=
(\psi_1(u),\psi_2(u),\psi_3(u)){\bf M}(\lambda), \label{Mdef}
\end{equation}
where $\psi_1(u),\psi_2(u)$ and $\psi_3(u)$ are three linearly
independent solutions of the equation $L\psi=0$, 
then the eigenvalues of ${\bf M}(\lambda)$ are in involution among
themselves  and 
with all local IM \rf{classint} (with respect to the Poisson bracket
algebra \rf{PB}). The trace of the monodromy matrix 
\begin{equation} 
{\bf T}^{(cl)}(\lambda)={\rm Tr}\  {\bf M}(\lambda),\label{T-one}
\end{equation}
can be regarded as a generating function 
for the local IM \rf{classint} as it expands in the asymptotic series 
\cite{McK78}
\begin{equation} 
\frac{1}{2\pi}\log {\bf T}^{(cl)}(\lambda)=
\lambda+{\sum_{k=1}^{\infty}}c_k\ I_k^{(cl)} \lambda^{-k}, \qquad
\lambda\to+\infty \label{cexpan}
\end{equation}
where $c_k$ ($c_k=0$ for $k=0 \pmod 3$)  are some numerical
coefficients. The 
simplest  way to calculate these coefficients 
is to consider the case when both $U(u)$ and $V(u)$
are constant (i.e., independent of $u$) and take real values. 
In this case \rf{cexpan}
reduces to the $\lambda\to+\infty$ expansion of the positive root of
the cubic equation $x^3+2Ux+V-\lambda^3=0$ which can be easily
calculated. 
Similarly to \rf{Inorm}  we normalise 
the classical IM $I_k^{(cl)}$ by the term containing
the highest power of $U(u)$ 
\renewcommand{\arraystretch}{2.5}
\begin{equation}
I^{(cl)}_{2n-1}=\ds\int_0^{2\pi}{du\over
2\pi}(U(u))^n+\ldots,\qquad 
I^{(cl)}_{2n}=\ds\int_0^{2\pi}{du\over
2\pi}(U(u))^{n-1}V(u)+\ldots,\qquad n=1,2,3,\ldots
\end{equation}
Note that this normalisation is consistent with 
\rf{Inorm}, \rf{climit} and \rf{qclass}. It follows then that
\begin{equation}	\
c_{2n-1}=\ds
{4\pi\cos\big(\frac{(n-2)\pi}{3}\big)
\Gamma\big(\frac{2n-1}{3}\big)\over 3\,
n!\Gamma\big(\frac{2-n}{3}\big)},\qquad
c_{2n+2}=\ds{\frac {4\pi
\,\cos\big(\frac{(n+1)\pi}{3}\big)\Gamma\big(\frac{2(n+1)}{3}\big)} 
{n!\,\Gamma\big(-\frac{n+1}{3}\big)}},
\qquad 
\end{equation}	
where $n=0,1,2,\ldots$. 
The required calculations are given in Appendix~C; the above 
expressions are simple corollaries of \rf{lamclas}.
\renewcommand{\arraystretch}{1.5}

For our purposes it will be more convenient to work with a first order 
differential operator instead of \rf{Lthird}. To define it we need
some additional notation. 
The algebra $gl(3)$ is generated by the elements $H_i$, $i=1,2,3$
and $E_{ij}$, $(i,j)=\,(1,2),\, (2,1),\, (2,3),\, (3,2)$, for which we use
also the notation
 $$E_\a=E_{12},\quad E_\b=E_{23},\quad F_\a=E_{21},\quad F_\b=E_{32},$$
and
$$H_\a=H_1-H_2,\qquad H_\b=H_2-H_3,\qquad H_{\a+\b}=H_1-H_3.$$
They satisfy the relations
\begin{equation}	
[H_i,H_j]=0,\qquad [H_i,E_{kl}]=(\delta_{ik}-\delta_{il})E_{kl},
\qquad [E_{\a_i},F_{\a_j}]=\ds
\delta_{\a_i,\a_j}\,{H_{\a_i}},\label{sl3}
\ed
\begin{equation}	
\begin{array}{rcl}
E_{\a_i}^2\,E_{\a_j}-2\,E_{\a_i}\,E_{\a_j}\,E_{\a_i}+
E_{\a_j}\,E_{\a_i}^2&=&0,\\
F_{\a_i}^2\,F_{\a_j}-2\,F_{\a_i}\,F_{\a_j}\,F_{\a_i}+
F_{\a_j}\,F_{\a_i}^2&=&0,
\end{array}\qquad \a_i\not=\a_j.\label{cserre}
\end{equation}	
where in the last three relations  greek indices $\a_i$ and $\a_j$
take two values $\a$ or $\b$.

Introduce two-dimensional vectors 
\begin{equation}
\begin{array}{rclrclrcl}
\be_1&=&(\frac{1}{2},-\frac{\sqrt{3}}{2}),
\qquad&\be_2&=&(-1,\phantom{-}0),\qquad
\qquad&\be_3&=&(\phantom{-}\frac{1}{2},\frac{\sqrt{3}}{2}),\\
{\mbf w}_1&=&(1,\phantom{-}\frac{1}{\sqrt{3}}),\qquad 
&{\mbf w}_2&=&(\phantom{-}0,-\frac{2}{\sqrt{3}}),\qquad
&{\mbf w}_3&=&(-1,\frac{1}{\sqrt{3}}),\\
\end{array}\label{edef}
\end{equation}
such that 
\begin{equation}	
{\mbf e}_1+{\mbf e}_2+{\mbf e}_3=0,\qquad {\mbf w}_1+{\mbf w}_2+{\mbf w}_3=0,
\end{equation}
and 
\begin{equation}
{\mbf e}_i=\frac{1}{2}({\mbf w}_j-{\mbf w}_k),\qquad
{\mbf w}_i=-\frac{2}{3}({\mbf e}_j-{\mbf e}_k),
\end{equation}
where $(i,j,k)$ is a cyclic permutation of $(1,2,3)$.	
Moreover,
\begin{equation}
{\mbf e}_i{\mbf e}_j=\ds\frac{1}{2}a_{ij},\qquad
{\mbf w}_i{\mbf w}_j=\ds\frac{2}{3}a_{ij},\qquad
{\mbf e}_i{\mbf w}_j=\epsilon_{ij},\qquad i,j=1,2,3,
\end{equation}	
where ${\mbf a}{\mbf b}$ denotes the (Euclidean) scalar
product, $a_{ij}$, $i,j=1,2,3$, is the Cartan matrix of the algebra $A_2^{(1)}$
\begin{equation}
a_{ii}=2,\qquad a_{ij}=-1,\qquad i\not=j,\qquad i,j=1,2,3,\label{cartanm}
\end{equation}	
and 
\begin{equation}
\epsilon_{11}=\epsilon_{22}=\epsilon_{33}=0,\qquad
\epsilon_{12}=\epsilon_{23}=\epsilon_{31}=1,\qquad 
\epsilon_{21}=\epsilon_{32}=\epsilon_{13}=-1.\label{epsdef}
\end{equation}	
It is also useful to define
\begin{equation}
E_{31}=[F_\b,F_\a],\qquad 
{\mbf H}={\mbf w}_1 H_1+{\mbf w}_2 H_2+{\mbf w}_3 H_3.\label{defH}
\end{equation}

For any  $\mu=(\mu_1,\mu_2,\mu_3)$, such that $\mu_\a=\mu_1-\mu_2$ and
$\mu_\b=\mu_2-\mu_3$ are non-negative integers,  denote by 
$\pi_\mu$ the finite dimensional irreducible representation of $gl(3)$ 
with the highest 
weight $\mu$ and highest weight vector $\ket{\,0}$ defined as
\begin{equation}
E_\a\ket{\,0}=E_\b\ket{\,0}=0, \qquad H_i\ket{\,0}=\mu_i\ket{\,0},
\qquad i=1,2,3.\label{hwvector}
\end{equation}
The dimension of such a representation is 
${\rm dim}[\pi_\mu]=\frac{1}{2}(\mu_\a+1)(\mu_\b+1)(\mu_\a+\mu_\b+2)$.

Consider the following differential operator 
\begin{equation}
{\mathcal L}=\partial_u-\boldsymbol{\phi}'(u)\,{\mbf
H}-\lambda(E_{23}+E_{31} +E_{12}), 
\end{equation}
where the components of the vector 
$\boldsymbol{\phi}(u)=(\phi_1(u),\phi_2(u))$ are the canonical
variables with Poisson bracket
\begin{equation}
\{\phi_i(u),\phi_j(v)\}=\frac{1}{2}\delta_{ij}\epsilon(u-v),
\qquad\epsilon(u)=\frac{u}{2\pi}+\frac{1}{\pi} 
\sum_{n=1}^{\infty}\frac{\sin (nu)}{n},\qquad
0\le u,v<2\pi;\label{phiPB}
\end{equation}
which are related to $U(u)$ and $V(u)$ by the generalised 
Miura transform \cite{DS84}
\begin{eqnarray}
U(u)&=&\phi_1'(u)^2+\phi_2'(u)^2+2\phi_1''(u),\nonumber
\\
\sqrt{3} V(u)&=&\frac{2}{3}\phi_2'(u)^3-2\phi_1'(u)^2\phi_2'(u)
-3\phi_1'(u)\phi_2''(u)-\phi_1''(u)\phi_2'(u)-\phi_2'''(u).\label{cmiura}
\end{eqnarray}
For the periodic potentials $U(u)$ and $V(u)$, \rf{UVper} the field
$\boldsymbol{\phi}(u)$ is, in general,  quasiperiodic 
\bn
\boldsymbol{\phi}(u+2\pi)=\boldsymbol{\phi}(u)+2\pi i {\mbf p}.
\ed
The monodromy 
matrix associated with the equation $\pi_\mu[{\mathcal L}]\Psi(u)=0$
(where $\pi_\mu$ denotes a representation  of $gl(3)$) is given by
\bn
{\bf M}_\mu(\lambda)=\pi_\mu\left[e^{2\pi i {\mbf p}{\mbf H}}{\mathcal
P}\exp\Big(\lambda \int_0^{2\pi} (V_1 E_{23}+V_2(u)E_{31}+V_3(u)E_{12})
du\Big)\right], 
\ed
where the symbol ${\mathcal P}$ denotes the ``path ordered''
exponential and 
\bn
V_i(u)=e^{-2\,{\mbf e}_i \,{\boldsymbol \phi}(u)}\ ,\qquad i=1,2,3.
\ed
The associated $L$-operator
\bn 
{\bf L}^{(cl)}_\mu(\lambda)=\pi_\mu[{e^{-\pi i {{\mbf p} {\mbf H}}}}]{\bf
M}_\mu(\lambda), 
\label{Lclass}\ed
satisfies the ``$r$-matrix'' Poisson bracket algebra \cite{Skl79,FT87}
\bn
\lbrace {\bf L}^{(cl)}_\mu (\lambda)\matrix{{}\cr\otimes\cr{} {^{,}}}
{\bf L}^{(cl)}_{\mu'}(\lambda')\rbrace =
[{\bf r}_{\mu \mu'}({\lambda/\lambda'}),{\bf L}^{(cl)}_\mu (\lambda)\otimes 
{\bf L}^{(cl)}_{\mu'}(\lambda')]\ ,\label{rPB}
\ed
The quantity 
${\bf r}_{\mu  \mu'}(\lambda)=\pi_\mu\otimes\pi_{\mu'}[\ {\bf  r}\ ]$
is the ``classical $r$-matrix'' given by 
\begin{equation}
{\bf r}(\lambda)=\frac{\lambda^{\frac{3}{2}}+\lambda^{-\frac{3}{2}}}
{\lambda^{\frac{3}{2}}-\lambda^{-\frac{3}{2}}}\sum_{i=1}^{3}H_i\otimes H_i
+\frac{2}{\lambda^{\frac{3}{2}}-\lambda^{-\frac{3}{2}}}
\sum_{(ij)}\left(\lambda^{-\frac{1}{2}}
E_{ij}\otimes {E}_{ji}+
\lambda^{\frac{1}{2}}E_{ji}\otimes {E}_{ij}\right),
\end{equation}
where the second sum is taken over the values $(ij)=(12),(23),(31)$.

It follows from \rf{rPB} that the quantities 
\bn
{\bf T}^{(cl)}_\mu(\lambda)={\rm Tr}_{\pi_\mu}\, 
{\bf M}_\mu(\lambda),\label{Tclass}
\ed
are in involution with respect to the Poisson brackets
\rf{PB},\rf{phiPB}
\bn
\{{{\bf T}^{(cl)}_\mu(\lambda),
{\bf T}^{(cl)}_{\mu'}(\lambda')}\}=0.
\ed
Note, in particular, that ${\bf T}^{(cl)}_{(1,0,0)}(\lambda)$ 
coincides with \rf{T-one}. 

\setcounter{equation}{0}
\section{Quantum Yang-Baxter equation} 

\subsection{Universal $R$-matrix}
The quantum Kac-Moody algebra
${\mathcal A}=U_q(\widehat{sl}(3))$ is generated by the elements
$x_i$, $y_i$ and $h_i$, \ $i=1,2,3$, \ and the grading element $d$,
 satisfying the commutation relations
\begin{equation}
 \left[h_i,h_j\right]=0\, ,\qquad
\left[h_i, x_j\right]= -a_{ij} x_j\, ,\qquad
\left[h_i, y_j\right]= a_{ij} y_j\, ,\label{Acomm}
\end{equation}
\begin{equation}
 [d,y_i]=\delta_{i,3} y_i\, ,\qquad [d,x_i]=-\delta_{i,3}x_i\, ,
\qquad
\left[y_i, x_j\right]=
\delta_{i j}\ {q^{h_i}-q^{-h_i}\over q-q^{-1}}\ ,
\end{equation}
and the cubic Serre relations
\begin{equation}\begin{array}{rcl}
x_i^2\,x_j-[2]_q\,x_i\,x_j\,x_i+x_j\,x_i^2&=&0,\cr
y_i^2\,y_j-[2]_q\,y_i\,y_j\,y_i+y_j\,y_i^2&=&0,
\end{array} \qquad i\not=j.
\label{serre}
\end{equation}
Here the indices $i,j$ take three values $i,j=1,2,3$; \ the Cartan
matrix  $a_{ij}$ is the same as \rf{cartanm}
and the ``$q$-numbers'' $[n]_q$ are defined as
\begin{equation}
[n]_q={q^n-q^{-n}\over q-q^{-1}}.
\end{equation}
The element $k=h_1+h_2+h_3$ is the central element of the algebra
${\mathcal A}$. In the following we will always work with the
representations of the subalgebra of ${\mathcal A}$ with suppressed
 grading element $d$ on which the central element $k$ acts by zero:
\begin{equation}
h_1+h_2+h_3=0.
\end{equation}
The algebra ${\cal A}=U_q(\widehat{sl}(3))$ is a Hopf algebra with the
co-multiplication
$$
\delta :  \qquad  {\cal A} \longrightarrow {\cal A}\otimes {\cal A},
$$
defined as
\begin{equation}
\begin{array}{rcl}
\delta(x_i)&=&x_i\otimes 1+q^{-h_i}\otimes x_i\ ,\cr
\delta(y_i)&=&y_i\otimes q^{h_i}+1\otimes y_i\ ,\cr
\delta(h_i)&=&h_i\otimes 1+1\otimes h_i\ ,\cr
\delta(d)&=&d\otimes 1+1\otimes d\ ,
\end{array}\label{3.6}
\end{equation}
where $i=1,2,3$.
As usual, we  introduce the ``twisted'' co-multiplication
\begin{equation}
\delta'=\sigma\circ\delta\, , \qquad \sigma\circ(a\otimes b)=b\otimes a
\qquad  (\, \forall a,b\in {\cal A}\, )\ .
\end{equation}
Define also two Borel subalgebras ${\cal B}_-\subset {\cal A}$
and  ${\cal B}_+\subset {\cal A}$ generated by $h_{i},\,
x_i$, \ $i=1,2,3$,   and
$h_{i},\, y_i$, \ $i=1,2,3$, \   respectively. Let
$\widehat{{\cal B}}_\pm $ be their extensions by $d$,
that is, the subalgebras of ${\cal A}$, generated by
${\cal B}_\pm$ and $d$ .

There exists a unique element \cite{Dri87,KT92}
\begin{equation}
{\cal R}\in \widehat{{\cal B}}_+\otimes \widehat{{\cal B}}_-\ ,
\end{equation}
 satisfying the following relations
\begin{eqnarray}
\delta'(a)\ {\cal R}&=&{\cal R}\ \delta(a)
\qquad (\forall\ a\in {\cal A})\, ,\label{comult}\\
(\delta\otimes 1)\, {\cal R}&=&{\cal R}^{13}\, {\cal R}^{23}\,
,\label{Rdef1}\\
(1\otimes \delta)\, {\cal R}&=&{\cal R}^{13}\, {\cal R}^{12}\,
,\label{Rdef2}
\end{eqnarray}
where ${\cal R}^{12},\, {\cal R}^{13},\, {\cal R}^{23}\in
{\cal A}\otimes{\cal A}\otimes {\cal A}$ and
${\cal R}^{12}={\cal R}\otimes 1$, ${\cal R}^{23}=1\otimes {\cal R}$,
${\cal R}^{13}=(\sigma\otimes 1)\, {\cal R}^{23}$.
The element ${\cal R}$ is called the universal $R$-matrix.
It satisfies the Yang-Baxter equation
\begin{equation}
{\cal R}^{12}{\cal R}^{13}{\cal R}^{23}={\cal R}^{23}{\cal
R}^{13}{\cal R}^{12}\ ,\label{uybe}
\end{equation}
which is a simple corollary of the definitions \rf{comult}-\rf{Rdef2}.
The universal
$R$-matrix is understood as a formal series in generators in
$\widehat{{\cal
B}}_+ \otimes \widehat{{\cal B}}_-$. Its dependence on the
Cartan elements
can be isolated as a simple factor
\begin{equation}
{\mathcal R}=q^{\mathsf t}\, {\overline{\mathcal R}},\label{Rred}
\end{equation}
following
where ${\mathsf t}$ is an invariant tensor in the tensor square of the
extended
Cartan subalgebra. It can be chosen to be \cite{TK92, Gould}
${\mathsf t}=h_1\otimes h^1 + h_2\otimes h^2 +k\otimes d+ d\otimes k$,
$$h^1=\frac{2}{3}h_1+\frac{1}{3}h_{2}, \quad
h^2=\frac{2}{3}h_2+\frac{1}{3}h_1,$$ and  $q^{{\mathsf t}}$ satisfies
the
 following defining properties:
\begin{eqnarray}
\begin{array}{lllllll}
q^{\mathsf t}(y_i\otimes 1)&=&(y_i\otimes q^{h_i})q^{\mathsf t},&\qquad
&q^{\mathsf t}(x_i\otimes 1)&=&(x_i\otimes q^{-h_i})q^{\mathsf t},\\
q^{\mathsf t}(1\otimes y_i)&=&(q^{h_i}\otimes y_i)q^{\mathsf t},&\qquad
&q^{\mathsf t}(1\otimes x_i)&=&(q^{-h_i}\otimes x_i)q^{\mathsf t}.
\end{array}
\label{1}
\end{eqnarray}
The factor
${\overline{\mathcal R}}$ is the ``reduced''  universal $R$-matrix
${\overline{\mathcal R}}\subset {\cal B}_+ \otimes {\cal B}_-$
which is a formal power series in $y_i\otimes 1$ and $1\otimes x_i$ only

(with no dependence on the Cartan elements)
\begin{equation}
{\overline{\mathcal R}}
=1+(q-q^{-1})(y_1\otimes x_1+y_2\otimes x_2+y_3\otimes
x_3)+\ldots\label{Rser}
\end{equation}
Below we will use the following important property of the reduced
universal $R$-matrix.
\begin{proposition}Suppose elements $I$, $A_i$ and $B_i$,
$i=1,2,3$, satisfy the relations
\begin{equation}
[I,x_i]={A_i-B_i\over q-q^{-1}},\label{cond1}
\end{equation}
\begin{equation}
A_i \, x_j=q^{a_{ij}} \, x_j\, A_i,\qquad
B_i \, x_j=q^{-a_{ij}} \, x_j\, B_i,\qquad\label{cond2}
\end{equation}
where $i,j=1,2,3$. Then
\begin{equation}
[1\otimes I,{\overline{\mathcal R}}]=\big(\sum_{i=1}^3 y_i\otimes
A_i\big)
\ {\overline{\mathcal R}}-{\overline{\mathcal R}} \
\big(\sum_{i=1}^3 y_i\otimes B_i\big).\label{Rprop}
\end{equation} \label{prop1}
\end{proposition}
The proof of this statement is given in Appendix A.

\subsection{Free field representation and the vertex operators}
The quantum version of the Miura transformation \rf{cmiura} 
is the free field representation \cite{FZ87} of the algebra
${\mathcal W}_3$, 
\begin{eqnarray}
-gT(u)&=&:\phi'_1(u)^2:+:\phi'_2(u)^2:+2(1-g)\phi''_1(u)+\frac{g}{12},\nonumber
\\
-\frac{(3g)^{\frac{3}{2}}i}{2}W(u)&=&:\phi'_2(u)^3:
-3:\phi'_1(u)^2\phi'_2(u):+\frac{3}{2}(g-1):\phi''_1(u)\phi'_2(u):
\\ 
&+&\frac{9}{2}(g-1):\phi'_1(u)\phi''_2(u):
-\frac{3}{2}(g-1)^2\phi'''_2(u),\nonumber 
\end{eqnarray}
in terms of a two-component free chiral Bose field 
 ${\boldsymbol \varphi}(u)=(\varphi_1(u),\varphi_2(u))$,  
\bn
{\boldsymbol \varphi}(u)
=i{\bf X}+i{\mbf P}\,u+\sum_{n\neq 0}\frac{{\bf b}_{-n}}{n}e^{inu}. 
\ed
The mode operators ${\mbf X}=(X_1,X_2)$, ${\mbf P}=(P_1,P_2)$ and
${\bf b}_{n}=(b_{1,n},b_{2,n})$ ($n=\pm1,\pm2,\ldots$) 
satisfy the commutation relations of the
Heisenberg algebra
\bn
[X_{i},P_{j}]=\frac{ig}{2}\delta_{i,j}, \qquad [b_{i,n}, b_{j,m}]
=\frac{ng}{2}\delta_{n+m, 0}\ \delta_{i,j},\qquad i,j=1,2\label{heis}
\ed
where the parameter $g$ is related to the central charge 
$c$ as 
\bn
g={50-c-\sqrt{(2-c)(98-c)}\over48},\qquad c=50-24(g+g^{-1}).\label{gdef}
\ed
Let ${\mathcal{F}}_{{\mbf p}}$ be the highest weight module
over the algebra \rf{heis}, with highest weight
vector $\ket{\,{\mbf p}}$,\  ${\mbf p}=(p_1,p_2)$ defined as 
\bn
P_{i}\ket{\,{\mbf p}}=p_{i}\ket{\,{\mbf p}}, 
\qquad a_{i,n}\ket{\,{\mbf p}}=0, \qquad n>0, \qquad  i=1,2.
\ed
For generic $c$ and ${\mbf p}$, 
the Fock space ${\mathcal{F}}_{{\mbf p}}$ 
is isomorphic to the space ${\mathcal V}_{\Delta_2,\Delta_3}$ with the highest weights
\bn
\Delta_2={p_{1}^2+p_{2}^2\over g}+{c-2\over 24},\qquad
\Delta_3={2p_2\,(p_2^2-3p_{1}^2)\over(3 g)^{3/2}}.\label{cweights}
\ed

The space 
\bn
{\hat{{\mathcal F}}}_{{\mbf p}}=\mathop{\oplus}_{n_1,n_2=-\infty}^{\infty}
{\mathcal{F}}_{{\mbf p}+g(n_1{\mbf e}_1+n_2{\mbf e}_2)},\label{qspace}
\ed
supports the action of the vertex operators
\bn
V_i(u)=q^{\hf}
:e^{-2\,{\mbf e}_i \,{\boldsymbol \varphi}(u)}:\ ,\qquad i=1,2,3,\label{Vop}
\ed
defined as
\bn
:e^{-2\,{\mbf e}_i \/ {\boldsymbol \varphi}(u)}:\equiv
\exp\big(-2\sum_{n=1}^\infty{{\mbf e}_i\, {\mbf b}_{-n}\over n}e^{inu}\big)
\exp\big(-2i{\mbf e}_i({\mbf X}+{\mbf P}u)\big)
\exp\big(+2\sum_{n=1}^\infty{{\mbf e}_i\, {\bf b}_n\over n}e^{-inu}\big),
\ed
where the root vectors ${\mbf e}_i$ are given in \rf{edef} and 
\begin{equation}	
q=e^{i\pi g}\ .
\end{equation}	
Using \rf{heis} it is easy to check that 
\begin{equation} \begin{array}{ccl}
V_i(u_1)V_j(u_2)&=&q^{a_{ij}}V_j(u_2)V_i(u_1),\qquad u_1>u_2,\\
{[}{\mathbf P}\/\,,\,V_j(u){]}&=&-g\,\be_j\,V_j(u),
\end{array}\label{VVcomm}\end{equation}	
where the $a_{ij}$ are given by \rf{cartanm} and
\bn
V_j(u+2\pi)=q^{-2}e^{-4i\pi {\mbf P}\be_j}V_j(u).
\label{Vper}
\ed
It will sometimes be convenient  to use the operators
\begin{equation}
z_j=e^{2\pi i{\mbf w}_j{\mbf P}},\qquad z_1\,z_2\,z_3=1,\qquad
j=1,2,3,\label{zdef} 
\end{equation}
where the vectors ${\mbf w}_j$ are defined in \rf{edef}.
For example, with these operators the relations \rf{VVcomm},\rf{Vper}
take the form
\begin{equation}	
z_iV_j(u)=q^{2\epsilon_{ij}}V_j(u)z_i,
\end{equation}	
where $\epsilon_{ij}$ is defined in \rf{epsdef} and 
\begin{equation}	
V_i(u+2\pi)=q^{-2}\, z_j^{-1}\, z_k\, V_i(u),
\end{equation}
where $(i,j,k)$ is a cyclic permutation of $(1,2,3)$.

\subsection{The universal $L$-operator}\label{section:Lop}
Consider the following operator \cite{FL92,FL93}
\begin{equation}
 \mbox{$\mathcal{L}$}=e^{i\pi\/{\mbf P}{\mbf h}}\mbox{$\mathcal{P}$}
 \exp\left(\int_{0}^{2\pi}K(u)\,du\right),
 \label{Loper}
\ed
where 
\begin{equation}
{\mbf h}=\frac{2}{3}({\mbf e}_1 \, h_1+{\mbf e}_2 \, h_2+{\mbf e}_3 \,
h_3),
\end{equation}
with the vectors ${\mbf e}_1$, ${\mbf e}_2$ and  ${\mbf e}_3$ defined 
in \rf{edef} and 
\begin{equation}	
K(u)=
V_{1}(u)y_{1}
 +V_{2}(u)y_{2}+V_{3}(u)y_{3}.\label{Kdef}
\end{equation}
Here $h_i$, $y_i$, $i=1,2,3$, $h_1+h_2+h_3=0$, are the generators of the Borel
subalgebra ${\cal B}_+\subset U_q(\widehat{sl}(3))$ defined above.
The
ordered exponential in \rf{Loper} (the symbol ${\mathcal P}$ denotes
the path ordering) is defined as a series 
\bn
 \mbox{$\mathcal{L}$}=e^{i\pi\/{\mbf P}{\mbf h}}\sum_{n=0}^\infty
\int_{2\pi\ge u_1\ge u_2\ge\ldots\ge
u_n\ge0}K(u_1)K(u_2)\cdots K(u_n)du_1du_2\ldots du_n.\label{Lseries}
\ed
The integrals in \rf{Lseries} converge when
\bn
-\infty<c<-2,\label{quasiclas}
\ed
(which corresponds to the range $0<g<2/3$). For the values of $c$ outside
this {\it quasiclassical domain}, \rf{quasiclas}, the integrals in \rf{Lseries}
can be defined  via analytic continuation in $c$ \cite{BLZ97a}.
The operator \rf{Loper} is an element of the algebra ${\mathcal B}_+$ 
whose coefficients are operators acting in the quantum space \rf{qspace}.
Following the arguments of \cite{BLZ99a} it is not difficult to show that 
this $L$-operator satisfies the Yang-Baxter equation
\bn
{\cal R}\,\,(\,{\cal L}\otimes1\,) \,(\,1\otimes{\cal L}\,)=
(\,1\otimes{\cal L}\,)\,(\,{\cal L}\otimes1\,) \,\, {\cal R}\ \label{qybe}.
\ed
where $\mathcal R$ is the universal $R$-matrix for the algebra 
$U_q(\widehat{sl}(3))$ defined above. 
The only property of the universal $R$-matrix 
required for \rf{qybe} to hold is relation \rf{comult}.

An important remark is in order. As the reader might expect, 
the $L$-operator
\rf{Loper} can be obtained  as a 
specialization  of the universal $R$-matrix and the
equation \rf{qybe} is, in fact, a special case of the   Yang-Baxter
equation \rf{uybe}. The exact correspondence (first established in
\cite{BLZ99a}  
for the case of $U_q(\widehat{sl}(2))$) is formulated as follows.
Each side of the equation \rf{uybe} is an element 
of the direct product ${\mathcal B}_+\otimes {\mathcal A}\otimes
{\mathcal B}_-$. Therefore to reduce \rf{uybe} to \rf{qybe} it is
enough to find an appropriate realization of the Borel subalgebra 
${\mathcal B}_-$ (appearing in the third factor of the product) in the
quantum space \rf{qspace}. 
Let us identify the generators $x_i\in {\mathcal B}_-$,
$i=1,2,3$, of this Borel subalgebra with the integrals of the vertex
operators 
\begin{equation}	
x_i={1\over q-q^{-1}}\int_{0}^{2\pi} V_i(u)d u,\qquad i=1,2,3.\label{Vint}
\end{equation}
One can check that these generators satisfy \cite{BMP96} the Serre relations
\rf{serre}. 
Substituting \rf{Vint} into the reduced universal
$R$-matrix, \rf{Rser}, one gets an element of ${\cal B}_+$ whose
co-ordinates are operators acting in the quantum space \rf{qspace}. As
expected, it exactly coincides with the ${\cal P}$-ordered exponent
from  \rf{Loper}.

\begin{proposition}
The specialization of the 
``reduced" universal $R$-matrix ${\overline{\cal R}}\in
{\cal B}_+\otimes {\cal B}_-$\   for the case when 
$x_i\in {\cal B}_-$, $i=1,2,3$, \  are realized as in  \rf{Vint} 
coincide with the ${\cal P}$-exponent 
\begin{equation}
\overline{{\mathcal R}}\,\Big|_{x_i=\rf{Vint}}=
{\cal P}\exp\Big( \int_0^{2\pi} {\cal K}(u)\, du \Big)\equiv
{\overline{\cal L}},\label{conj}
\end{equation}	
where ${\cal K}(u)$ is defined in \rf{Kdef}.\label{prop2}
\end{proposition}

This is precisely the conjecture stated in  
ref. \cite{BLZ99a}. It was formulated there
for $U_q(\widehat{sl}(2))$, but the statement is rather general and
(with obvious modifications) holds for any quantized Kac-Moody
algebra. A proof of the conjecture of \cite{BLZ99a} based on 
combinatorial properties of the
 representations of quantum groups has been obtained by 
E.Frenkel and F.A.Smirnov \cite{FS98}.
Here we present an alternative  (and, in fact, rather elementary) 
proof of \rf{conj} which is based on the defining properties of the universal 
$R$-matrix.

The reduced universal $R$-matrix \rf{Rser} is a formal 
power series in the generators $\overline{y}_i=y_j\otimes 1$ and 
$\overline{x}_i=1\otimes x_i$, $i=1,2,3$, such that 
$\overline{x}_i\overline{y}_j=\overline{y}_j\overline{x}_i$,
therefore we can write it as ${\overline{\cal R}}= {\overline{\cal
R}}(\overline{y}_i,\overline{x}_i)$. 
Then, substituting \rf{Rred} into \rf{Rdef2} one
obtains
\begin{equation}
{\overline{\mathcal R}}
(\overline{y}_j,{x}'_i+{x}''_i)=
{\overline{\mathcal R}}
(\overline{y}_j,{x}'_i)\,
{\overline{\mathcal R}}
(\overline{y}_j,{x}''_i)
\label{comx}
\end{equation}
where $\overline{y}_i=y_i\otimes 1\otimes 1$,
${x}'_i=1\otimes q^{-h_i}\otimes x_i$, 
${x}''_i=1\otimes x_i\otimes 1$.
Both sets ${x}'_i$ and ${x}''_i$ satisfy the Serre relations
\rf{serre} of ${\cal
B}_-$, commute with $\overline{y}_j$ 
\bn
{x}'_i\overline{y}_j=\overline{y}_j{x}'_i,\qquad
{x}''_i\overline{y}_j=\overline{y}_j{x}''_i \label{xycomm}
\ed
and also obey the relations
\begin{equation}
{x}'_i {x}''_j=q^{a_{ij}}{x}''_j{x}'_i
\label{primerel}
\end{equation}
Conversely, we can regard \rf{comx} as an identity for
${\overline{\cal
R}}(\overline{y}_i,\overline{x}_i)$ 
 which should be valid once all the arguments 
involved there are subjected to the above relations (namely,
\rf{serre}, \rf{xycomm} and \rf{primerel}).

Consider integrals more general than \rf{Vint} which also satisfy the
Serre relations \rf{serre}
\begin{equation}	
\overline{x}_i={1\over q-q^{-1}}\int_{u_2}^{u_1} V_i(u)d u,
\qquad i=1,2,3,\label{Vint1}
\end{equation}
where $2\pi \ge u_1>u_2\ge0$. Denote by ${\overline{\mathcal
L}}(u_1,u_2)$ the result of substituting these generators
$\overline{x}_i\in {\cal B}_-$
into the reduced universal $R$-matrix 
\begin{equation}	
{\overline{\mathcal L}}(u_1,u_2)=\overline{{\mathcal
R}}(\overline{y_i},\overline{x}_i)\,\Big|_{\overline{x}_i=\rf{Vint1}} 
\end{equation}	
Note that from \rf{Rser} it follows that for $(u_1-u_2)\to0$ 
\bn
{\overline{\mathcal L}}(u_1,u_2)=1+\int_{u_2}^{u_1}\, K(u)\, du +O\big(
{(u_1-u_2)^2}\big)
\label{smallu}\ed
where $K(u)$ is defined in \rf{Kdef}.

Now, let $2\pi>u_1\ge u_2\ge u_3\ge0$, then  the identity \rf{comx}
implies that
\begin{equation}
{\overline{\mathcal L}}(u_1,u_3)={\overline{\mathcal L}}(u_1,u_2)
{\overline{\mathcal L}}(u_2,u_3), 
\label{Lmult}
\end{equation}
since two sets of quantities 
$$	
{x}'_i={1\over q-q^{-1}}\int_{u_2}^{u_1} V_i(u)d u,
$$
$$	
{x}''_i={1\over q-q^{-1}}\int_{u_3}^{u_2} V_i(u)d u,
$$
satisfy all the required relations \rf{serre}, \rf{xycomm} and
\rf{primerel}. In particular, the relation \rf{primerel} readily follows
from \rf{VVcomm}. With an account of \rf{smallu} the functional
equation \rf{Lmult} has a unique solution
\begin{equation}
{\overline{\mathcal L}}(u_1,u_2)=\mbox{$\mathcal{P}$}
 \exp\left(\int_{u_2}^{u_1}K(u)\,du\right).
\end{equation}	
Finally, setting $u_2=0$ and $u_1=2\pi$ one arrives at the statement of
Proposition~\ref{prop2}.

\setcounter{equation}{0}
\section{Quantum ${\bf T}$- and ${\bf Q}$-operators}

\subsection{Symmetry relations} \label{symmetry}
Consider the symmetry properties of the $L$-operator \rf{Loper}.
Any linear orthogonal operator ${\bf U}$ in two-dimensional Euclidean space 
induces an automorphism $U$ of the two-component Heisenberg algebra
\rf{heis} by the rule $U[{\boldsymbol\varphi}(u)]={\mathbf U}
{\boldsymbol \varphi}(u)$. In particular, the Weyl reflections
associated with the vectors ${\mbf w}_i$, $i=1,2,3$, generate the 
symmetric group $S_3$ permuting the vertex operators
\begin{equation}
\sigma_{ij}[V_i(u)]=V_j(u),\qquad \sigma_{ij}[V_j(u)]=V_i(u),\qquad
\sigma_{ij}[V_k(u)]=V_k(u),
\end{equation} 
\begin{equation}	
\sigma_{ij}[z_i]=z_j^{-1}, \qquad  \sigma_{ij}[z_j]=z_i^{-1},\qquad
\sigma_{ij}[z_k]=z_k^{-1}.
\end{equation}	
where $(i,j,k)$ is any permutation of $(1,2,3)$.
Define also the cyclic automorphism $\tau=\sigma_{12}\sigma_{23}$ 
corresponding to the rotation of the Dynkin diagram of $A_2^{(1)}$
\begin{equation}	
\tau [V_{i}(u)]=V_{i+1  \pmod 3}(u).
\end{equation}	

The weights \rf{cweights} can be written in a symmetric form
\bn
\Delta_2=-{r_1r_2+r_1r_3+r_2r_3\over g}+{c-2\over 24},\qquad
\Delta_3=-{r_1r_2r_3\over\,g^{3/2}}\ ,\label{cweights1}
\ed
where 
\begin{equation}
r_i={\mbf w}_i {\mbf p}, \qquad i=1,2,3,
\end{equation}
with
${\mbf w}_i$ given in \rf{edef}. From the symmetry properties
\begin{equation}
\tau[r_i]=r_{i+1\pmod 3}, \qquad \sigma_{ij}[r_i]=-r_j, \qquad
\sigma_{ij}[r_j]=-r_i,\qquad 
\sigma_{ij}[r_k]=-r_k,
\end{equation}
it is obvious that the cyclic automorphism $\tau$ transforms the 
Fock module ${\cal F}_{\mbf p}$ to an isomorphic module
\begin{equation}
\tau[{\cal F}_{\mbf p}]\sim
{\cal F}_{\mbf p},
\end{equation}
while the Weyl reflections $\sigma_{ij}$ negate ${\mbf p}$ and the
weight $\Delta_3$
\begin{equation}
\sigma_{ij}[{\cal F}_{\mbf p}]\sim
{\cal F}_{-{\mbf p}}\sim{\cal V}_{\Delta_2,-\Delta_3}.
\end{equation}

Further, the  algebra of outer automorphisms of ${\mathcal B}_+\subset 
U_q(\widehat{sl}(3))$ also contains the
symmetric group $S_3$: $\sigma[y_i]=y_{\sigma(i)}$,
$\sigma[h_i]=h_{\sigma(i)}$. Again by
$\tau$ we denote   the automorphism corresponding to rotation of the Dynkin
 diagram of $A_2^{(1)}$
\begin{equation}	
\tau [y_{i}]=y_{i+1\pmod 3},\qquad \tau [h_{i}]=h_{i+1\pmod 3}\ ,\label{tauaut}
\end{equation}
and by $\sigma_{ij}$ the transpositions
\begin{equation}\begin{array}{rclrclrcl}	 
\sigma_{ij}[y_i]&=&y_j,\qquad&\sigma_{ij}[y_j]&=&y_i,\qquad
&\sigma_{ij}[y_k]&=&y_k,\\
\sigma_{ij}[h_i]&=&h_j,\qquad
&\sigma_{ij}[y_h]&=&h_i,\qquad&\sigma_{ij}[h_k]&=&h_k.
\end{array}\label{sigmaaut}
\end{equation}	
The $L$-operator \rf{Loper} is invariant with respect to the diagonal
action of the automorphisms 
\bn
(\sigma\otimes\sigma)[{\mathcal
L}]={\mathcal L}, \qquad \sigma\in S_3.\label{Laction}
\ed

\subsection{The ${\bf T}$-operators}
We will need several specializations of the universal $L$-operator 
\rf{Loper} corresponding to various matrix representations of the
Borel subalgebra ${\mathcal B}_+\subset U_q(\widehat{sl}(3))$. 
First consider the so-called {\it evaluation 
maps} from ${\mathcal B}_+$ to the finite dimensional 
algebra $U_q({gl}(3))$. This latter algebra is generated by the elements
$H_i$, $i=1,2,3$
and $E_{ij}$, $(i,j)=\,(1,2),\, (2,1),\, (2,3),\, (3,2)$, for which we use
also the notations
 $$E_\a=E_{12},\quad E_\b=E_{23},\quad F_\a=E_{21},\quad F_\b=E_{32},$$
and
$$H_\a=H_1-H_2,\qquad H_\b=H_2-H_3,\qquad H_{\a+\b}=H_1-H_3.$$
They satisfy the relations
$$
[H_i,H_j]=0,\qquad [H_i,E_{kl}]=(\delta_{ik}-\delta_{il})E_{kl},
$$
\bn
[E_{\a_i},F_{\a_j}]=\ds
\delta_{\a_i,\a_j}\,{q^{H_{\a_i}}-q^{-H_{\a_i}}\over
q-q^{-1}},\label{Uqsl3}
\ed
$$
\begin{array}{rcl}
E_{\a_i}^2\,E_{\a_j}-[2]_q\,E_{\a_i}\,E_{\a_j}\,E_{\a_i}+
E_{\a_j}\,E_{\a_i}^2&=&0,\\
F_{\a_i}^2\,F_{\a_j}-[2]_q\,F_{\a_i}\,F_{\a_j}\,F_{\a_i}+
F_{\a_j}\,F_{\a_i}^2&=&0,
\end{array}\qquad \a_i\not=\a_j,
$$
where $[n]_q=(q^n-q^{-n})/(q-q^{-1})$
and the greek indices $\a_i$ and $\a_j$ in the last three relations   
take two values $\a$ or $\b$.
The algebra $U_q({gl}(3))$ has an outer automorphism
$\sigma_{13}$, induced by the automorphism of its Dynkin
diagram:
\begin{equation}
\begin{array}{c}
 \sigma_{13}[E_\a]=E_\b,\qquad \sigma_{13}[E_\b]=E_\a,\qquad
\sigma_{13}[F_\a]=F_\b,\qquad
\sigma_{13}[F_\b]=F_\a,
\\
\sigma_{13}[H_1]=-H_3,\qquad
\sigma_{13}[H_2]=-H_2,\qquad
\sigma_{13}[H_3]=-H_1.
\end{array}
\label{sigma13}
\ed
Let us  introduce  the following root vectors:
$$F_{\a+\b}=q^{\frac{1}{2}} F_\b F_\a-q^{-\frac{1}{2}}F_\a F_\b,\qquad
\overline{F}_{\a+\b}=q^{-\frac{1}{2}}F_\b
F_\a-q^{\frac{1}{2}} F_\a F_\b,\qquad 
$$
and
$$E_{31}=F_{\a+\b}q^{H_1+H_3},\qquad
\overline{E}_{31}=\overline{F}_{\a+\b}q^{-H_1-H_3}.\qquad
$$

There are two non-equivalent evaluation maps ${\mathcal B}_+\to
U_q({gl}(3))$. The first one is 
\bn
\Ev_t(y_1)=E_{23},\qquad \Ev_t(y_2)=tE_{31},\qquad \Ev_t(y_3)=E_{12},
\ed
\bn
\Ev_t(h_1)=H_2-H_3,\qquad \Ev_t(h_2)=H_{3}-H_{1},\qquad \Ev_t(h_3)=H_{1}-H_{2}.
\label{eval1}
\ed
where $t$ is a spectral parameter.
The second evaluation map $\EEv_t:{\mathcal B}_+\to U_q({gl}(3))[t]$ has the 
form
\bn
\EEv_t=\sigma_{13}\cdot\Ev_{-t}\cdot\sigma_{13},
\ed
where the automorphisms $\sigma_{13}$ entering this formula are
defined in \rf{sigmaaut}, \rf{sigma13}. Explicitly one has
\bn
\EEv_t(y_1)=E_{23},\qquad \EEv_t(y_2)=t\overline{E}_{31},\qquad
\EEv_t(y_3)=E_{12}, \ed
\bn \EEv_t(h_1)=H_2-H_3,\qquad
\EEv_t(h_2)=H_{3}-H_{1},\qquad \EEv_t(h_3)=H_{1}-H_{2}.
\label{eval2}
\ed
One can easily check that both these maps are algebra homomorphisms 
as all the defining relations
\rf{serre}, \rf{Acomm} of the algebra ${\mathcal B}_+$ become
simple corollaries of \rf{Uqsl3}. 

For any $\mu=(\mu_1,\mu_2, \mu_3)$,  such that $\mu_\a=\mu_1-\mu_2$
and $\mu_\b=\mu_2-\mu_3$ are 
nonnegative integers denote by $\pi_\mu$ the finite dimensional representation
of $U_q(gl(3))$ with the highest weight $\mu$ and highest weight
vector $\ket{\,0}$  defined as
\begin{equation}	
E_{12}\ket{\,0}=E_{23}\ket{\,0}=0, \qquad
H_i\ket{\,0}=\mu_i\ket{\,0},\qquad i=1,2,3.
\end{equation}
Further let $\pi_\mu(t)$ be the 
representation of ${\mathcal B}_+$
given by the composition of $\pi_\mu$ with the evaluation
homomorphism $\Ev_t$. Evidently, the operator-valued matrices 
\bn
{\bf L}_\mu(t)=\pi_\mu(t)[{\mathcal L}], \label{Lquant}
\ed
(whose matrix elements are operators acting in \rf{qspace}) satisfy
suitable specializations of the quantum Yang-Baxter equation \rf{qybe}. 
These operators \rf{Lquant} are 
quantum analogues of the classical $L$-operators \rf{Lclass}  whereas the 
operators 
\bn
{\bf T}_\mu(t)={\rm Tr}_{\pi_\mu(t)}\,[{ e^{i\pi{\mbf P\,h}} {\cal L}}],
\label{Tquant}
\ed
are the quantum analogues of ${\bf T}^{(cl)}(\l)$ in \rf{Tclass}
(the spectral 
parameter $t$ is related to $\l$ in \rf{Lclass} and \rf{Tclass} as $t=\l^3$).
From the standard arguments based on the Yang-Baxter equation 
it follows that
\bn
[{\bf T}_\mu(t),{\bf T}_{\mu'}(t')]=0.\label{Tcomm}
\ed
Let us show that they also commute with all local IM \rf{locint},
\bn
[ {\bf T}_\mu(t),{\bf I}_k]=0.\label{TIcomm}
\ed
The  characteristic property of the local IM is that their commutators 
with vertex operators \rf{Vop} reduce to total derivatives \cite{FF96}
\begin{equation}
[\,{\bf I}_k\,, V_i(u)\,]=\partial_u \Big\{ :O_k^{(i)}(u) V_i(u):
\Big\}\equiv \partial_u \Theta^{(i)}_k(u), \label{totder}
\end{equation}	
where $O^{(i)}_k(u)$ are some polynomials with respect to the 
field $\partial_u{\boldsymbol \phi}(u)$ and its derivatives.
Obviously, the operators $\Theta^{(i)}_k(u)$ in \rf{totder} 
obey the same periodicity properties as the vertex operators $V_i(u)$, 
\rf{Vper}, 
\bn
\Theta^{(j)}_k(u+2\pi)=q^{-2}e^{-4i\pi {\mbf P}\be_j}\Theta^{(j)}_k(u),
\label{thetaper}
\ed
and satisfy the commutation relations
\begin{equation} \begin{array}{rcl}
\Theta^{(i)}_k(u_1)V_j(u_2)&=&q^{a_{ij}}V_j(u_2)\Theta^{(i)}_k(u_1),
\qquad u_1>u_2,\\
{[}{\mbf P}\/\,,\,\Theta^{(j)}_k(u){]}&=&-g\,\be_j\,\Theta^{(j)}_k(u)\ ,
\end{array}\label{thetacomm}\end{equation}	
where $a_{ij}$ is defined in \rf{cartanm}.
It is easy to check that the triple
\begin{equation}	
I={\bf I}_k,\qquad A_i=\Theta_k^{(i)}(2\pi),\qquad
B_i=\Theta_k^{(i)}(0),
\end{equation}
satisfies the conditions \rf{cond1},\rf{cond2} of Proposition~\ref{prop1}
with the generators $x_i\in {\cal B}_+$ realized as in \rf{Vint}
Therefore taking into account \rf{conj} and applying 
\rf{Rprop} one obtains 
\begin{equation}
[{\bf I}_k,\overline{{\mathcal
L}}]=\Theta_k(2\pi)\,\overline{{\mathcal L}}-
\overline{{\mathcal L}}\,\Theta_k(0),\label{ILcomm}
\end{equation}	
where 
\begin{equation}	
\Theta_k(u)=\sum_{i=1}^3\, y_i \Theta_k^{(i)}(u).
\end{equation}	
Now substitute  \rf{Tquant} into the RHS of \rf{TIcomm}. Using the relation 
\rf{ILcomm}, the commutation relations \rf{Acomm}, 
\rf{thetacomm} and the cyclic property of the trace it is not
difficult to show that the commutator \rf{TIcomm} actually vanishes. 

The most important ${\bf T}$-operators correspond to  the
three-dimensional (fundamental) representations of $\pi_{\omega_\a}$,
and $\pi_{\omega_\b}$ of $U_q({gl}(3))$ 
with the highest weights $\omega_\a=(1,0,0)$
and $\omega_\b=(1,1,0)$ (The corresponding representation matrices are given
explicitly in Appendix~\ref{repmatrix}). 
We will introduce special notation for
these ${\bf T}$-operators (adjusting the normalisation of the spectral 
parameter for later convenience)
\bn
\Tb(t)\equiv\T_{(1,0,0)}(tq^{-{3\over2}}),
\qquad
\Tob(t)\equiv\T_{(1,1,0)}(tq^{-\hf}).
\ed
Computing the trace in \rf{Tquant} in this case one obtains
\bn
\Tb(t)=\sum_{n=0}^{\infty} \ t^n\, {\bf G}_n\ ,
\qquad
\Tob(t)=\sum_{n=0}^{\infty}\  t^n\, \overline{{\bf G}}_n.\label{Tfund}
\ed
Here 
\bn
\G_0=z_1+z_2+z_3,\qquad \Go_0=z_1^{-1}+z_2^{-1}+z_3^{-1}\ ,
\ed
with $z_1,z_2$ and $z_3$ defined in \rf{zdef}
and for $n\ge1$
\begin{equation}\begin{array}{rcl}
\G_n&=&z_1\  J^{(n)}_{312}+z_2\ J^{(n)}_{123}
+z_3\  J^{(n)}_{231},\\
(-1)^n\Go_n&=&z_1^{-1} J^{(n)}_{213}+z_2^{-1}
J^{(n)}_{321}+
z_3^{-1}  J^{(n)}_{132},
\end{array}\label{Gdef}\end{equation}
\begin{eqnarray}
J^{(n)}_{ijk}
&=&\int_{2\pi\ge u_1\ge u_2\ldots\ge u_{3n}}^{}
\Big(V_i(u_1)V_j(u_2)V_k(u_3)
\ V_i(u_4)V_j(u_5)V_k(u_6)\cdots\label{Jdef}\\
&&\cdots V_i(u_{3n-2})V_j(u_{3n-1})V_k(u_{3n})\Big) \ du_1\,du_2\ldots 
du_{3n}, \nonumber 
\end{eqnarray}
where $(i,j,k)$ is a permutation of $(1,2,3)$.

It follows from \rf{Tcomm} and \rf{TIcomm} that the operators 
$\G_n$ and $\Go_n$ can be regarded as the ``nonlocal'' IM which
commute among themselves and with all the local IM \rf{locint}
\bn
[\G_k,\G_l]=[\G_k,\Go_l]=[\Go_k,\Go_l]=[\G_k,{\bf I}_l]=[\Go_k,{\bf
I}_l]=0.
\ed

Definitions similar to those in  \rf{Tquant} but with
the map \rf{eval2} instead of \rf{eval1}
lead to a new set of ${\bf T}$-operators 
\bn
\overline{\bf T}_\mu(t)=
{\rm Tr}_{\overline{\pi}_\mu(t)}\,[{ e^{i\pi{\mbf P\,h}} {\mathcal
L}}],\label{Tquant2} 
\ed
where $ \overline{\pi}_\mu(t)$ denotes the representation of
${\mathcal B}_+$ obtained by the composition of 
the representation $\pi_\mu$ of $U_q({gl}(3))$
with the evaluation map $\rf{eval2}$. 
The operators $\overline{\bf
T}_\mu(t)$ with different values of $t$ commute among
themselves and with all 
${\bf T}_\mu(t)$,
\bn
[\overline{\bf T}_\mu(t),\overline{\bf T}_{\mu'}(t')]=
[\overline{\bf T}_\mu(t),{\bf T}_{\mu'}(t')]=0.
\ed
In general for the same $\pi_\mu$ the formulae
\rf{Tquant} and \rf{Tquant2} define different operators. However, when 
$\pi_\mu$ is restricted to symmetric powers of a fundamental representation 
$\pi_{\omega_\a}$ and $\pi_{\omega_\b}$ then the two sets of ${\bf
T}$-operators coincide (up to a rescaling of spectral
parameter). For future references it will  be convenient to define 
\bn
\Tb_m(t)\equiv\T_{(m,0,0)}(tq^{-m-{\hf}})=\Tt_{(m,0,0)}(tq^{m+{1\over2}}), 
\label{T}
\ed
and
\bn
\Tob_m(t)\equiv\T_{(m,m,0)}(tq^{-m+\hf})=\Tt_{(m,m,0)}(tq^{m-\hf}),
\label{Tbar} 
\ed
where $m=0,1,2,\dots,\infty$. In particular,
$\Tb_0(t)=\Tob_0(t)\equiv1$ while $\Tb_1(t)\equiv\Tb(t)$, 
$\Tob_1(t)\equiv\Tob(t)$ 
are exactly the same as in \rf{Tfund}.

\subsection{The ${\bf Q}$-operators}

The operators \rf{Tquant} and \rf{Tquant2}  are obviously the CFT versions of
Baxter's 
transfer-matrices of the lattice models related to the
$U_q(\widehat{sl}(3))$ algebra. It is well known that the 
theory of solvable lattice models involves other important objects, namely
Baxter's $Q$-matrix. Similarly to the standard CFT case \cite{BLZ97a}  
appropriate ${\mathcal W}_3$ CFT versions of ${\bf Q}$-operators can 
be obtained as certain 
specializations of the universal $L$-operator \rf{Loper}.

The generators  $y_i$, $h_i$, $i=1,2,3$, $h_1+h_2+h_3=0$, of the 
Borel subalgebra $ {\mathcal B}_+\subset U_q(\widehat{sl}(3))$ satisfy 
the commutation relations \rf{Acomm} 
\begin{equation}
[h_i,h_j]=0,\qquad [h_i,y_i]=a_{ij},\label{Hcomm}
\end{equation}	
and the Serre relations 
\rf{serre}. The later can be written as 
\begin{equation}
y_i y_{ij}=q^{-1}y_{i,j}y_i,\qquad i,j=1,2,3,\qquad i\not=j,\label{Bserre}
\end{equation}
where $y_{ij}$ stands for the $q$-commutator of $y_i$ and $y_j$:
\begin{equation}
y_{ij}=y_iy_j-qy_jy_i.\label{qcomm}
\end{equation}
Let $\ab_i^+,\ab^-_i,\hbc_i$, $i=1,2$  be the generators
of two independent q-oscillator algebras ${\mathsf H}_q$,
\bn
[{\mathcal H}_i,\ab^\pm_i]=\pm\ab_i^\pm,\qquad
q\ab_i^+\ab_i^--q^{-1}\ab_i^-\ab_i^+=
{1\over q-q^{-1}}.\label{qos}
\ed
Define the following homomorphism  
$\rho_t:$\ ${\mathcal B}_+\to {\mathsf H}_q\otimes
{\mathsf H}_q$
\begin{equation}
{\begin{array}{rclrclrcl}	
\rho_t(y_1)&=&q^{\frac{1}{2}}\,t(q-q^{-1})\,q^{-\hbc_2}\ab_1^-\ab_2^+,\quad&
\rho_t(y_2)&=&\ab_2^-,\quad&
\rho_t(y_3)&=&\ab_1^+ q^{\hbc_2},
\end{array}\atop
\begin{array}{rclrclrcl}	
\rho_t(h_1)&=&-\hbc_1+\hbc_2,\quad&
\rho_t(h_2)&=&-\hbc_1-2\hbc_2,\quad&
\rho_t(h_3)&=&2\hbc_1+\hbc_2,
\end{array}}
\label{rhomap}
\end{equation}	
where $t$ is a spectral parameter.
Note that this map subjects the generators $y_i$, $i=1,2,3$ to 
some additional relations which are not implied by \rf{Hcomm} and the
Serre relations \rf{Bserre}.
\begin{equation}
\rho_t\left(qy_3 y_{21}-q^{-1}y_{21}y_3\right)=
\rho_t\left(q
y_{13}y_2-q^{-1}y_2y_{13}\right)=-\frac{q^{\frac{1}{2}}\,t}
{q-q^{-1}},
\qquad
\rho_t\left( y_{23}\right)=0,\label{extrarel}
\end{equation}
where $y_{ij}$ is defined in \rf{qcomm}. Define also the maps
$\rho_i(t)$ and
$\rrho_i(t):\ {\mathcal B}_+\to {\mathsf H}_q\otimes
{\mathsf H}_q$, $i=1,2,3$ obtained as  compositions of  
\rf{rhomap} with the
automorphisms \rf{tauaut},\rf{sigmaaut} of  the algebra ${\mathcal B}_+$
\bn
\rho_i(t)=\rho_t\cdot \tau^{-i+1},\qquad
\rrho_i(t)=\rho_{-t}\cdot \tau^{-i+1}
\cdot \sigma_{jk}\qquad i=1,2,3,
\label{rhoi}
\ed
where $\rho_t$ is given by \rf{rhomap} and $\rho_{-t}$ is
obtained from \rf{rhomap} by the replacement $t\to-t$.
Let $\pi_i$ and $\overline{\pi}_i$ be  any representations of
${\mathsf H}_q\otimes {\mathsf  H}_q$ such that
the traces
\bn
Z_i={\rm Tr}_{\pi_i\rho_i(t)}[ e^{2i\pi{\mbf p\,h}}]
\qquad {\rm and}\qquad
\overline{Z}_i={\rm Tr}_{{\overline{\pi}}_i\rrho_i(t)}[ e^{2i\pi{\mbf p\,h}}], 
\ed
converge and do not vanish for 
\begin{equation}
\Im m\,{\mbf e}_1 {\mbf p}<0, \qquad \Im m\,{\mbf e}_3 {\mbf p}<0, \label{Pdom}\
\end{equation}
where ${\mbf p}=(p_1,p_2)\in{\mathbb C}^2$.

Define the operators 
\bn
\A_i(t)=Z_i^{-1}\ {\rm Tr}_{\pi_i\rho^+_i(t)} [ e^{i\pi{\mbf P\,h}}{\cal L}]
\qquad
{\rm and}\qquad
\AAA_i(t)={\overline Z}_i^{-1}\
{\rm Tr}_{\overline{\pi}_i\rrho^+_i(t)} [ e^{i\pi{\mbf P\,h}}{\cal L}],
\label{Adef}
\ed
where ${\mathcal L}$ is given in \rf{Loper}. Since we are interested 
in the action of  these operators in ${\cal F}_{\mbf p}$ the operator 
${\mbf P}$ in \rf{Adef} can be substituted by its eigenvalue
${\mbf p}=(p_1,p_2)$. The 
definitions \rf{Adef} obviously apply to the domain \rf{Pdom}.
However they  can be extended  for all complex ${\mbf 
p}\in{\mathbb C}^2$ (except
for some set of singular points for real values $p_1$ and
$p_2$) by an analytic continuation in ${\mbf p}$.
 
Obviously, the operators \rf{Adef} can be written as a power series. For
example,
\begin{equation}
\A_i(t)=1+\sum_{n=1}^\infty \sum_{\{\sigma_i=1,2,3\}}
a^{(i)}_{3n}(\s_1,\s_2,\ldots,\s_{3n}){\mathbb J}_{3n}(\s_1,\s_2,\ldots,\s_{3n}),
\label{Aser}
\end{equation}
where 
\begin{equation}
{\mathbb J}_{3n}(\s_1,\s_2,\ldots,\s_{3n})=
\int_{2\pi\ge u_1\ge \cdots \ge u_{3n}} V_{\s_1}(u_1)V_{\s_2}(u_2)
\cdots V_{\s_{3n}}(u_{3n}),\label{JJint}
\end{equation}
and 
\begin{equation}
a^{(i)}_{3n}(\s_1,\s_2,\ldots,\s_{3n})=
Z_i^{-1}{\rm Tr}_{\pi_i \rho_i(t)} \Big(e^{2\pi i {\mbf P h}}
y_{\s_1}y_{\s_2}\cdots y_{\s_{3n}}\Big).\label{Acoeff}
\end{equation}
Note that these coefficients vanish unless the ``total charge'' of the 
product of vertex operators in \rf{JJint} 
\begin{equation}
{\mbf e}_{\s_1}+{\mbf e}_{\s_2}+\cdots +{\mbf e}_{\s_{3n}}=0,
\end{equation}
is zero. The spectral parameter $t$ is absorbed in the coefficients 
$a^{(i)}_{3n}\sim O(t^n)$ so that \rf{Aser} is a power series in $t$.

A remarkable feature of the definitions \rf{Adef} is that the
coefficients \rf{Acoeff} are  
completely determined by the commutation relations
\rf{Hcomm}, \rf{Bserre}, \rf{extrarel} and the
cyclic property of the trace, so the specific choice of the
representations $\pi_i$ and $\overline{\pi}_i$ is not significant 
as long as the above convergence property is maintained. The operator 
coefficients 
in the power series \rf{Aser} can be expressed in terms of basic
nonlocal IM \rf{Gdef}. For example, calculating the first nontrivial 
coefficient in \rf{Aser} one gets
\begin{equation}\renewcommand{\arraystretch}{3.0}
\begin{array}{rcl}
\ds\A_i(t)&=&\ds 1+{q^{\frac{5}{2}}(q z_i \G_1- \Go_1)\over
(q^2-1)(z_j-q^2z_i)(z_k-q^2z_i)}\,t +O(t^2),\\
\ds\Ao_i(t)&=&\ds 1-{q^{\frac{5}{2}}(z_i \G_1- q\Go_1)\over
(q^2-1)(q^2z_j-z_i)(q^2z_k-z_i)}\,t +O(t^2),
\end{array}\label{AGrel}\end{equation}
where $z_i$, $i=1,2,3$, defined in \rf{zdef}. For further references
introduce an alternative set  of nonlocal IM, \ $\H^{(n)}_i$,
$\Ho_i^{(n)}$, \ $i=1,2,3$, $n=1,2\ldots \infty$ defined as coefficients in 
the expansions
\bn
\log \A_j(t)=-\sum_{n=1}^{\infty} s^n\,\H^{(j)}_n ,\qquad
\log \Ao_j(t)=-\sum_{n=1}^{\infty}s^n\, \Ho^{(j)}_n,\label{Hdef}
\ed
where
\begin{equation}	
s=i\,(g^{-1}\Gamma(1-g))^3\, t. \label{sdef}
\end{equation}

The operators $\A_i(t)$ and $\Ao_i(t)$ simplify considerably when
acting in the space ${\cal F}_{\mbf p}$ with values of $2{\mbf p}$
belonging to the weight lattice of the algebra $sl_3$
\begin{equation}	
2{\mbf p}=n_1{\mbf w}_1+n_2{\mbf w}_2,\qquad n_1,n_2\in {\mathbb Z}.\label{pspec}
\end{equation}	
In this case the coefficients \rf{Acoeff} can be written as
\begin{equation}	
a_{3n}(\s_1,\s_2,\ldots,\s_{3n})={\rm Tr}_u
\Big(u_{\s_1}u_{\s_2}\cdots u_{\s_{3n}}\Big),\label{ucoeff}
\end{equation}
where $u_1$, $u_2$ are generators of the Weyl algebra 
\begin{equation}	
u_1u_2=qu_2u_1,\label{ucomm}
\end{equation}	
with the trace defined as \cite{Fad95}
\begin{equation}	
{\rm Tr}_u(u_1^nu_2^m)=\delta_{n,0}\delta_{m,0},\qquad n,m\in {\mathbb Z} 
\label{faddtrace}
\end{equation}
and 
\begin{equation}	
u_3=\frac{q^{1\over2}\,t}{(q-q^{-1})^3}\,u_2^{-1}u_1^{-1}\ .\label{u3}
\end{equation}	
Obviously, the expression \rf{ucoeff} is 
invariant with respect to the cyclic 
automorphism $\tau$ in \rf{tauaut} therefore all three operators 
$\A_i(t)$, $i=1,2,3$  coincide. With the help of \rf{Aser},
\rf{ucoeff} they can be written as
\begin{equation}	
\A_i(t)\Big|_{2{\mbf p}=\rf{pspec}}
={\rm
Tr}_u\Big[{\cal P}\exp\int_0^{2\pi}\Big(u_1V_1(v)+u_2V_2(v)+u_3V_3(v)
\Big)dv\Big],\qquad i=1,2,3.
\end{equation}
Exactly the same expression but with $u_1$ and $u_2$ interchanged
(which is equivalent to the replacement $q\to q^{-1}$ in \rf{ucomm})
holds for the operators $\Ao_i(t)$.

The ${\bf Q}$-operators 
(the CFT analogues of Baxter's $Q$-matrix for $U_q(\widehat{sl}(3)$ 
related lattice 
models) are defined as 
\bn
\Q_i(t)=t^{{{\mbf w}_i{\mbf P}\over g}}\ \A_i(t),\qquad
\Qt_i(t)=t^{-{{\mbf w}_i{\mbf P}\over g}}\ \Ao_i(t),\qquad
\ed
where ${\mbf w}_i$, $i=1,2,3$, are defined in \rf{edef}. 

Evidently, the operators $\Q_i(t)$ and  $\Qt_i(t)$ with different spectral 
parameters commute among themselves and with all the operators
$\T_\mu(t)$ and $\To_\mu(t)$ constructed above. They also commute with 
all the local IM defined in \rf{locint} 
\begin{equation}
[ \Q_i(t),{\bf I}_k]=0,\qquad [ \Qt_i(t),{\bf I}_k]=0.\label{QIcomm}
\end{equation}
Indeed, the proof of the similar statement \rf{TIcomm} for the ${\bf
T}$-operators given above does not depend on a
particular choice of the representation of ${\cal B}_+$ and relies only on 
the commutation relations \rf{Hcomm}, \rf{Bserre} and the cyclic
property of the trace. Therefore this proof applies to \rf{QIcomm} as well. 

As is known \cite{SY88,EY89,FF96} the local IM ${\bf I}_k$ 
do not change under the transformation $g\to
g^{-1}$, provided one makes simultaneous substitution  ${\boldsymbol
\phi}(u)\to g^{-1}{\boldsymbol \phi}(u)$
\begin{equation}
{\bf I}_k\{g,{\boldsymbol \phi}(u)\}=
{\bf I}_k\{g^{-1},g^{-1}{\boldsymbol \phi}(u)\}.
\end{equation}
Obviously, the nonlocal IM $\G_n$ and $\Go_n$ (as well as ${\bf
H}^{(n)}_i$ and ${\overline {\bf H}_i^{(n)}}$) do change and there
exists an infinite set of {\it dual nonlocal IM} \cite{BLZ97a} 
obtained from \rf{locint} by the above 
substitution. For instance,
\begin{equation}	
\widetilde{\G}_n=z_1^{1\over g}\  \widetilde{J}^{(n)}_{312}+z_2^{1\over g}\
\widetilde{J}^{(n)}_{123}
+z_3^{1\over g}\  \widetilde{J}^{(n)}_{231}
\label{Gdual},\end{equation}
\begin{eqnarray}
\widetilde{J}^{(n)}_{ijk}
&=&\int_{2\pi\ge u_1\ge u_2\ldots\ge u_{3n}}^{}
\Big(\widetilde{V}_i(u_1)\widetilde{V}_j(u_2)\widetilde{V}_k(u_3)
\ \widetilde{V}_i(u_4)\widetilde{V}_j(u_5)\widetilde{V}_k(u_6)\cdots
\label{Jdual}\\
&&\cdots
\widetilde{V}_i(u_{3n-2})\widetilde{V}_j(u_{3n-1})\widetilde{V}_k(u_{3n})
\Big) \ du_1\,du_2\ldots 
du_{3n}, \nonumber 
\end{eqnarray}
where $(i,j,k)$ is a permutation of $(1,2,3)$ and 
\begin{equation}
\widetilde{V}_i(u)=\widetilde{q}^{1\over2}e^{-2{\mbf e}_i{\boldsymbol
\phi}(u)/g},\qquad \widetilde{q}=e^{i\pi/g}.
\end{equation}
Similarly, define
\begin{equation}	
\widetilde{\Go}_n\{g,{\boldsymbol \phi}(u)\}= \Go_n\{g^{-1},g^{-1}{\boldsymbol
\phi}(u)\},
\end{equation}	
and 
\begin{equation}	
\widetilde{\bf H}^{(n)}_i\{g,{\boldsymbol \phi}(u)\}= 
{\bf H}^{(i)}_i\{g^{-1},g^{-1}{\boldsymbol \phi}(u)\},\qquad
\widetilde{\overline{\bf H}}^{(n)}_i\{g,{\boldsymbol \phi}(u)\}= 
\overline{\bf H}^{(n)}_i\{g^{-1},g^{-1}{\boldsymbol
\phi}(u)\}. \label{Hdual}
\end{equation}	 
For $g<3/2$ the above definitions require analytic continuation
in $g$ as described in \cite{BLZ97a}.

\subsection{Vacuum eigenvalues}
The Fock space \rf{qspace} naturally splits into 
level subspaces ${\cal F}_{{\mbf p}}^{(\ell)}$, $\ell=0,1,2,\ldots$, defined as 
\begin{equation}
  {\cal F}_{{\mbf p}}=\sum_{\ell=0}^\infty {\cal F}_{{\mbf
p}}^{(\ell)},\qquad L_0{\cal F}_{{\mbf p}}^{(\ell)}=(\Delta_2+\ell)
{\cal F}_{{\mbf p}}^{(\ell)}.
\end{equation}
All the IM (both local and nonlocal) act invariantly 
in each level subspace ${\cal F}_{{\mbf p}}^{(\ell)}$. In particular
the vacuum state is an eigenstate of all these operators,
\begin{equation}
{\bf I}_k \,\ket{{\mbf p}}=I^{(vac)}_k \,\ket{{\mbf p}},\qquad
\G_n\,\ket{{\mbf p}}=G^{(vac)}_n \,\ket{{\mbf p}},\qquad
\Go_n\,\ket{{\mbf p}}={\overline G}^{(vac)}_n \,\ket{{\mbf p}}.
\end{equation}
Below we list a few vacuum eigenvalues $I_k^{(vac)}$ which readily
follow from \rf{firstIM}
\begin{eqnarray}
I^{(vac)}_1&=&\Delta_2-{c\over24}\ ,\nonumber\\I^{(vac)}_2&=&\Delta_3\
,\nonumber\\
I^{(vac)}_4
&=&(\Delta_2-{{c+6}\over
24})\,\Delta_3\ ,\label{Ivac}\\
I^{(vac)}_5&=&\Delta_2^3
+\frac{4}{3}\Delta_3^2-\frac{c+8}{8}\Delta_2^2+\frac{(c+15)(c+2)}{192}\Delta_2
-\frac{c(c+23)(7c+30)}{96768}.  \nonumber\\
&&\nonumber
\end{eqnarray}

Using \rf{Gdef} one can express the vacuum eigenvalues of the nonlocal 
IM as Coulomb-type integrals in a circle. In general they cannot be
evaluated in a closed form. Even the simplest eigenvalues $G^{(vac)}_1$ and
$\overline{G}^{(vac)}_1$
reduce to the generalised hypergeometric integral
\begin{eqnarray}
J(x,y)&=&\int_0^1 {d\o_1\over\o_1}\int_0^{1}{d\o_2\over\o_2}
{\o_1^{x+g}\o_2^{y+g}\over(1-\o_1)^g(1-\o_2)^g(1-\o_1\o_2)^g}\nonumber\\
&=&{\Gamma(g+x)\Gamma(g+y)\Gamma^2(1-g)\over\Gamma(1+x)\Gamma(1+y)}\ 
{}_3F_2\left({g,\ g+x,\ g+y\atop 1+x,\ 1+y}\ ; 1\right).\label{hyper}
\end{eqnarray}
For example, 
\begin{equation}	
G^{(vac)}_1=z_2\int_0^{2\pi}du_3\int_{u_3}^{u_3+2\pi}du_1
\int_{u_3}^{u_1}du_2  {\ds e^{-2i{\mbf p}({\mbf e}_1u_1+{\mbf e}_2u_2+
{\mbf e}_3u_3)}\over\ds
\Big[(2\sin{u_1-u_2\over2})(2\sin{u_1-u_3\over2})
(2\sin{u_2-u_3\over2})\Big]^{g}  },\label{g1vac}
\end{equation}
can be expressed through \rf{hyper} in the following alternative 
forms
\begin{eqnarray}	
G_1^{(vac)}&=&-2\pi q^{-\frac{3}{2}} \Big[q(q^2z_1-z_2)J_{32}-
(q^2z_2-z_3)J_{12}\Big], \nonumber\\
&=&-2\pi q^{-\frac{3}{2}} \Big[q(q^2z_2-z_3)J_{13}-
(q^2z_3-z_1)J_{23}\Big], \label{g1vac1}\\
&=&-2\pi q^{-\frac{3}{2}} \Big[q(q^2z_3-z_1)J_{21}-
(q^2z_1-z_2)J_{31}\Big], \nonumber
\end{eqnarray}
where
\begin{equation}
J_{ij}=J(2{\mbf e}_i{\mbf p},-2{\mbf e}_j{\mbf p}),
\end{equation}
with $J(x,y)$ defined in \rf{hyper} and    $z_1$, $z_2$ and $z_3$
denote the vacuum eigenvalues of the corresponding operators in \rf{zdef}.
The vacuum eigenvalue of $\overline{G}^{(vac)}_1$ is obtained 
from ${G}^{(vac)}_1$ 
 by the negation of the overall sign in \rf{g1vac}, 
\rf{g1vac1} and the replacement
\begin{equation}
z_1\to z_3^{-1},\qquad z_2\to z_2^{-1},\qquad z_3\to z_1^{-1},
\qquad {\mbf e}_1\leftrightarrow {\mbf e}_3,\qquad {\mbf e}_2\to {\mbf e}_2,
\end{equation}
induced by the automorphism $\sigma_{13}$,
\begin{eqnarray}
\overline{G}_1^{(vac)}&=&2\pi q^{-\frac{3}{2}} \Big[q(q^2z_3^{-1}-z_2^{-1})
J_{12}-
(q^2z_2^{-1}-z_1^{-1})J_{32}\Big], \nonumber\\
&=&2\pi q^{-\frac{3}{2}} \Big[q(q^2z_2^{-1}-z_1^{-1})
J_{31}-
(q^2z_1^{-1}-z_3^{-1})J_{21}\Big], \label{g1bvac1}\\
&=&2\pi q^{-\frac{3}{2}} \Big[q(q^2z_1^{-1}-z_3^{-1})
J_{23}-
(q^2z_3^{-1}-z_2^{-1})J_{13}\Big]. 
\end{eqnarray}

Using \rf{AGrel}, \rf{g1vac1} and \rf{g1bvac1} one can obtain the 
vacuum eigenvalues of the nonlocal IM $\H^{(1)}_i$ and 
$\Ho_i^{(1)}$ defined in \rf{Hdef},
\begin{equation}
H^{(i)(vac)}_1=-{\pi\,g^3\,J_{kj}\over \Gamma^3(1-g)\,\sin(\pi g)},\qquad
\overline{H}^{(i)(vac)}_1={\pi\,g^3\,J_{jk}\over \Gamma^3(1-g)\,\sin(\pi g)},\label{H1vac}
\end{equation}
where $(i,j,k)$ is a cyclic permutation of $(1,2,3)$.

It will be useful to consider a special choice of the vacuum parameter ${\mbf
p}=(p,0)$. In this case, we have
\begin{equation}	 	
({\mbf e}_1{\mbf p})={p\over2},\qquad({\mbf e}_2{\mbf p})=-p,\qquad
({\mbf e}_3{\mbf p})={p\over2},\label{pspece}
\end{equation}
and
\bn
z_1=z_3^{-1}=e^{2\pi ip}, \qquad z_2=1. \label{pspecz}
\ed
Both $G_1^{(vac)}$ and $H_1^{(1)(vac)}$ simplify considerably in this
case,
\bn
G_1^{(vac)}(p,0)={4\pi^3\Gamma^2(1-g)\Gamma(1-\frac{3g}{2})\over
\Gamma(1-\frac{g}{2})\Gamma(1-g+p)\Gamma(1-g-p) 
\Gamma(1-g/2+p)\Gamma(1-g/2-p)},  
\label{G1vacspec}
\ed
\bn
H^{(1)(vac)}_1(p,0)=-\frac{\pi}{2\sin (\pi
g)}\frac{g^3\Gamma\left(1-\frac{3g}{2}\right)}{\Gamma\left(1-\frac{g}{2}\right)\Gamma\left(1-g\right)}\frac{\Gamma(g+p)\Gamma(g/2+p)}{\Gamma(1-g+p)\Gamma(1-g/2+p)},
\label{A22H1} 
\ed
where we have used eq. (4.4(5)) of \cite{Bateman} to simplify the
hypergeometric functions $J_{ij}$. 

\setcounter{equation}{0}
\section{Functional Relations}
The $\T$- and $\Q$-operators  introduced  in the previous
Section enjoy a variety of important functional relations (FR). 
In fact, many such functional relations 
have been previously 
obtained in the context of the solvable lattice models associated
with $U_q(\widehat{sl}(3))$ algebra (see, e.g. \cite{KR82, BR90, KNS94,
KLWZ97, PS99}). 
It turns out that the 
explicit construction of the $\Q$-operators given in the previous Section,
(which defines them as 
special quantum transfer matrices associated with the $q$-oscillator algebra),
can be used to considerably simplify the derivation of
these FR. In fact,
 we found it rather remarkable that all known FR relevant here
(as well as some new ones) can be obtained  as elementary corollaries 
of just two basic relations 
which express the operators ${\bf T}_\mu(t)$,\ \rf{Tquant}, and 
$\To_\mu(t)$,\ \rf{Tquant2},  
corresponding to finite dimensional representations $\pi_\mu$ of $U_q(gl(3))$
(i.e. those with integral non-negative  weights $\mu_\a=\mu_1-\mu_2\ge0$ and 
$\mu_\b=\mu_2-\mu_3\ge0$) in terms of the operators $\Q_i(t)$ and $\Qt_i(t)$ 
\begin{eqnarray}
{\mathbf T}_{(\mu_1,\mu_2,\mu_3)}(t)&=&z_{0}^{-1}
\det\left\Vert\,\Q_{i}(t\, q^{+2{\mu'}_j})\right\Vert_{i,j=1,2,3}
\label{aweyl}\\&&\nonumber\\
-\Tt_{(\mu_1,\mu_2,\mu_3)}(t)&=&z_{0}^{-1}
\det \left\Vert\,{\Qt}_{i}( t\,q^{-2{\mu'}_j})\right\Vert_{i,j=1,2,3}
\label{bweyl}
\end{eqnarray}
where ${\mu'}_1=\mu_1+1$,\ \  ${\mu'}_2=\mu_2$, \ \ ${\mu'}_3=\mu_3-1$, and 
\bn
z_0=(z_1-z_2)(z_1-z_3)(z_2-z_3),
\ed with $z_1$, $z_2$ and $z_3$ defined in \rf{zdef}. The operators
$\Q_i(t)$, $\Qt_i(t)$ are not functionally independent. They satisfy
the relations 
\begin{eqnarray}
c_i\, \Qt_i(t)&=&\Q_j(tq)\Q_k(tq^{-1})-\Q_k(tq)\Q_j(tq^{-1}), \label{QQ}
\\
-c_i
\,\Q_i(t)&=&\Qt_j(tq)\Qt_k(tq^{-1})-\Qt_k(tq)\Qt_j(tq^{-1}),\label{QQa} 
\end{eqnarray}
where $(i,j,k)={\rm cycle}(1,2,3)$, and 
\bn
c_i=z_i^{1\over2}(z_j-z_k).
\ed

The proof of these relations is given in Appendix B  (see
also the discussion at the end of this Section). Below we present 
some simple corollaries of \rf{aweyl},
\rf{bweyl}, \rf{QQ}, and \rf{QQa}.

The operators ${\mathbb T}(t)$ and ${\overline {\mathbb T}}(t)$ defined in
\rf{Tfund}  corresponding to
the three-dimensional fundamental representations of $U_q(gl(3))$
satisfy the following relations with the ${\bf Q}$-operators
\begin{eqnarray}
\Q_i(tq^3)-\Tb(tq^{\hf})\,\Q_i(tq)+\Tob(tq^{-\hf})\,\Q_i(tq^{-1})
-\Q_i(tq^{-3})&=&0,\label{TQa}\\
\Qt_i(tq^{3})-\Tob(tq^{\hf})\,\Qt_i(tq)+\Tb(tq^{-\hf})\,
\Qt_i(tq^{-1})-\Qt_i(tq^{-3})&=&0,\label{TQb}
\end{eqnarray}
where $i=1,2,3$, 
These relations are obviously the $U_q(\widehat{sl}(3))$-analogues \cite{KR82}
of Baxter's famous  T-Q relation \cite{Bax72}. 
Given $\Tb(t)$ and $\Tob(t)$ each of the equations \rf{TQa} 
is a third order difference equation for ${\bf Q}$'s 
which should have three linearly independent solutions. Since 
$\Tb(t)$ and $\Tob(t)$ are single-valued functions of $t$,
(i.e. periodic functions of $\log t$) the operators $\Q_i(t)$ and
$\Qt_i(t)$ can be interpreted as the ``Bloch wave'' solutions of
\rf{TQa} which satisfy the ``quantum Wronskian'' condition
\bn
z_0=\det\left\Vert{\Q}_{i}(t q^{-2j})\right\Vert_{i,j=1,2,3} =
-\det \left\Vert{\Qt}_{i}(t q^{2j})\right\Vert_{i,j=1,2,3}
\label{wron}
\ed
which is just a particular case of \rf{aweyl}, \rf{bweyl} for  
$\mu_i=0$. 

The equations \rf{TQa} and \rf{TQb} imply more familiar expressions 
\bn
\Tb(t)={{\Q(t\,q^{5\over2})}\over
\Q(t\,q^{\hf})}+{\Q(t\,q^{-{3\over2}})\,\Qt\,(t\,q^{3\over2})\over 
\Q(t\,q^{\hf})\, \Qt\,(t\,q^{-\hf})}
+{\Qt\,(t\,q^{-{5\over2}})\over \Qt\,(t\,q^{-\hf})},\label{Tusuala}
\ed
\bn
\Tob(t)={\Qt(t\,q^{5\over2})\over
\Qt(t\,q^{\hf})}+{\Qt(t\,q^{-{3\over2}})\,\Q\,(t\,q^{3\over2})\over 
\Qt(t\,q^{\hf})\, \Q\,(t\,q^{-\hf})}
+{\Q\,(t\,q^{-{5\over2}})\over \Q\,(t\,q^{-\hf})},\label{Tusualb}
\ed
where $\{\Q(t),\Qt(t)\}$ is any of the six pairs
$\{\Q_i(t),\Qt_j(t)\}$ with $i\neq j$. More general expressions 
\cite{BR90} of this type 
for the $\T$-operators with arbitrary weight $\mu$ can be obtained,
for example, from \rf{Tusuala}, \rf{Tusualb} and the relation 
\rf{sweyl} given below.

Further, using \rf{QQ} and
\rf{QQa} in \rf{aweyl} and \rf{bweyl} one can easily express the operators 
\rf{T} and \rf{Tbar} as 
\begin{eqnarray}
\Tb_m(t)&=&z_0^{-1}\,\sum_{i=1}^3 c_i\, \Q_i(tq^{m+{3\over2}})
\Qt_i(tq^{-m-{3\over2}}),\\
\Tob_m(t)&=&z_0^{-1}\,\sum_{i=1}^3 c_i \,\Q_i(tq^{-m-{3\over2}})
\Qt_i(tq^{m+{3\over2}}),
\end{eqnarray}
Using these two results, the well known 
{\it fusion relations}, \cite{KNS94}, can be deduced,  
\bn
\Tb_m(tq)\Tb_m(tq^{-1})=\Tob_m(t)+\Tb_{m-1}(t)\Tb_{m+1}(t),\label{fusrel}
\ed
\bn
\Tob_m(tq)\Tob_m(tq^{-1})=\Tb_m(t)+\Tob_{m-1}(t)\Tob_{m+1}(t),
\ed

Note that the formulae \rf{aweyl} and \rf{bweyl} strongly
resemble the first Weyl formula for  
characters of $gl(3)$. Note also, 
that the corresponding analogue of the  second Weyl
formula for the transfer matrices (sometimes called the quantum
Jacobi-Trudi identity \cite{KNS94})
has been known for some time \cite{BR90}
\bn
\T_{(\mu_1,\mu_2,0)}(t)=
\det\left\Vert\,{\boldsymbol\tau}^{(\mu^T_i-i+j)}(t\,q^{2(j-1)})
\,\right\Vert_{1\le i,j\le \mu_1}\label{sweyl}
\ed
where
\bn
{\boldsymbol\tau}^{(0)}(t)=
{\boldsymbol\tau}^{(3)}(t)\equiv1,
\qquad {\boldsymbol\tau}^{(1)}(t)=\T_{(1,0,0)}(t),\qquad
{\boldsymbol\tau}^{(2)}(t)=\T_{(1,1,0)}(t),
\ed
and ${\boldsymbol\tau}^{(k)}(t)\equiv0$,\  for $k<0$ and $k>3$. The
integers $\mu^T_i$,\  
$i=1,\ldots,\mu_1$, read
$$
\mu^T_i=2, \quad 1\le i\le \mu_2; \qquad
\mu^T_i=1, \quad \mu_2<i\le\mu_1. \qquad 
$$
Again, the formula \rf{sweyl} easily follows from \rf{aweyl}. Also, 
a similar expression for $\To_\mu(t)$ is obtained from \rf{sweyl}
by the replacement $q\to q^{-1}$.
 
The proof of the basic relations \rf{aweyl} and \rf{bweyl} given in
Appendix~\ref{Functional relations}  
is a straightforward (but rather more technically complicated) generalisation 
of the corresponding  $U_q(\widehat{sl}(2))$ results of \cite{BLZ99a}.
We consider more general $\T$-operators which correspond to the infinite
dimensional highest weight representations of $U_q(gl(3))$. These new 
 $\T$-operators are defined by the same formulae as \rf{Tquant}
\bn
{\bf T}^+_\mu(t)=
{\rm Tr}_{{\pi}^+_\mu}\,[{ e^{i\pi{\bf P\,h}} 
{\bf L}^+_\mu(t)}], \qquad {\bf L}^+_\mu(t)=
{\pi}^+_\mu(t)[{{\mathcal L}}],\label{Tplus}
\ed
except the trace is now taken over the infinite
dimensional representations $\pi^+_\mu$ of $U_q(gl(3))$ with arbitrary 
highest weights $\mu=(\mu_1,\mu_2,\mu_3)$ the highest weight vector 
$\ket{\,0}$ defined by \rf{hwvector}. Similar modifications to \rf{Tquant2} 
allow us to define the operators $\To^+_\mu(t)$ related to the second
evaluation map \rf{eval2}. Then, by considering the tensor
product of three $q$-oscillator algebras corresponding to the product
of three $\Q_i(t)$ operators, we show that
\bn
{\T}^+_{\mu_1,\mu_2,\mu_3}(t)=z_0^{-1}
\Q_1(t q^{2\mu_1+2})\Q_2(t q^{2\mu_2})\Q_3(t q^{2\mu_3-2}), \label{Tplusa}
\ed
and 
similarly 
\bn
{\Tt}^+_{\mu_1,\mu_2,\mu_3}(t)=-z_0^{-1}
\Qt_1(t q^{-2\mu_1-2})\Qt_2(t q^{-2\mu_2})\Qt_3(
q^{-2\mu_3+2}).\label{Tplusb} 
\ed
Then, we make use of the Bernstein-Gel'fand-Gel'fand (BGG) resolution of the
finite-dimensional modules. The BGG result allows one to express the finite 
dimensional highest weight modules in terms of a direct sum of
infinite dimensional highest weight modules. This implies that the $\T$-operators for a finite
dimensional module can be written in terms of \rf{Tplusa}, \rf{Tplusb} 
as 
\begin{eqnarray}
\T_\mu(t)=\sum_{\sigma\in S_3}(-1)^{l(\sigma)}\T^+_{\sigma(\mu+\rho)-\rho}(t),
\end{eqnarray}
where $\rho=(1,0,-1)$, the summation is over all permutations
$\sigma\in S_3$ 
of three elements, and 
$l(\sigma)$ is the parity of the permutation. There is an analogous result 
for the operators $\Tt_{\mu_1, \mu_2, \mu_3}(t)$. The determinant formulas
\rf{aweyl} and \rf{bweyl} follow immediately.

\setcounter{equation}{0}
\section{Conjectures: Exact Asymptotic Expansions} \label{Asymptotic
Expansions} 

\subsection{Assumptions}
We will make the following assumptions

{\it i) Analyticity.} The operators ${\bf A}_i(t)$ and $\Ao_i(t)$ (and 
hence all the ${\bf T}$-operators  introduced above)
are entire functions of the complex variable $t$ in the sense that all 
their matrix elements and eigenvalues are entire functions of this variable.
Their leading asymptotics at large $t$ are given by
\begin{equation}
\log \A_i(t)\sim M \,(+it)^{1\over 3(1-g)},
\qquad
\log \Ao_i(t)\sim M \,(-it)^{1\over 3(1-g)},\qquad |t|\to\infty\label{Aas}
\end{equation}
where $M$ is some c-number 
constant depending on the central charge $c$ only.

{\it ii) Location of zeroes.} For real ${\mbf p}$ and large $t$ the zeros 
of the eigenvalues of $A_i(t)$ and ${\overline A}_i(t)$ accumulate
along the positive imaginary and negative imaginary axes of $t$
respectively. The asymptotics \rf{Aas} are valid everywhere except
in the vicinity of the lines of zeroes (i.e. when $|\arg(it)|<\pi$\  for 
${\bf A}_i(t)$ and when $|\arg(-it)|<\pi$ \ for $\Ao_i(t)$).

As is known the zeroes of the eigenvalues of ${\bf Q}$-operators 
satisfy the Bethe Ansatz equations. In our case there are six
sets of these equations involving different pairs of eigenvalues
$\{A_i(t),\overline{A}_j(t)\}$ with $i\not=j$. To simplify the
notation let us choose a particular pair of eigenvalues, say  
$\{A_1(t),\overline{A}_3(t)\}$, and denote by $\{t_k\}|_{k=1}^\infty$
and $\{\overline{t}_k\}|_{k=1}^\infty$ their sets of zeroes  
\begin{equation}	
A_1(t_k)=0,\qquad \overline{A}_3(\overline{t}_k)=0,
\end{equation}
ordered as $|t_1|\le|t_2|\le\ldots$ and similarly $\overline{t}_k$.
The equations \rf{Tusuala}, \rf{Tusualb} (together with the standard arguments based on the
analyticity of ${\bf T}$-operators) imply the following Bethe-Ansatz
equations
\begin{equation}\begin{array}{rcl}	
\ds{1\over 2\pi i}\log\left[
\frac{A_{1}(t_{k}q^{2}){}^{\phantom{-}}}{A_{1}(t_kq^{-2})}
\frac{\overline{A}_{3}(t_kq^{-1})}{\overline{A}_{3}(t_kq){}^{\phantom{-}}}
\right]&=&-2{\mbf e}_3{\mbf p}-n^{(1)}_k+{1\over2}, 
\\
\ds{1\over 2\pi i}\log\left[
\frac{\overline{A}_{3}(\overline{t}_kq^{2}){}^{\phantom{-}}}
{\overline{A}_{3}(\overline{t}_kq^{-2})} 
\frac{A_{1}(\overline{t}_kq^{-1})}{A_{1}(\overline{t}_kq){}^{\phantom{-}}}
\right]&=&-2{\mbf e}_1{\mbf p}-n^{(3)}_k+{1\over2}, 
\end{array}
\label{BAE}
\end{equation}	
where $n^{(1)}_k,n^{(3)}_k\in{\mathbb Z}$. 
Different eigenvalues of  the operators ${\bf A}_1(t)$ and $\Ao_3(t)$ 
correspond to different sets of integer phases 
$\{n^{(1)}_k,n^{(3)}_k\}_{k=1}^\infty$.

The following assumption concerns the structure of the vacuum
eigenvalues  ${A}^{(vac)}_1(t)$ and $\overline{A}^{(vac)}_3(t)$ of
operators ${\bf A}_1(t)$ and $\Ao_3(t)$ acting in ${\cal F}_{\bf p}$.

{\sl iii). For real values of \ $2({\mbf w}_1{\mbf p})<g$ \ and 
 \  $2({\mbf w}_3{\mbf p})>-g$ \ 
all zeroes of $A_1(t)$ (\ $\overline{A}_3(t)$\ ) located on the positive
(negative) imaginary axis of $t$ and the integer phases in \rf{BAE}
run over consecutive positive integers
\begin{equation}	
n^{(1)}_k=n^{(3)}_k=k,\qquad k=1,2,\ldots,\infty. \label{phases}
\end{equation}
}
\subsection{Results}
Following \cite{BLZ97a} consider the following operator valued function
\begin{equation}	
{\bf \Psi}_i(\nu)={\Gamma(1-g)^{-i\nu(1+\xi)}
\over \Gamma(i\nu\xi/3)\Gamma(-i\nu(\xi+1)/3)
\Gamma((-1+i\nu)/3)\Gamma((1+i\nu)/3)}
\int_{-i\infty}^0 {dt\over t}(it)^{-i\nu(1+\xi)/3}\, \A_i(t)\, ,
\label{psidef}
\end{equation}	
where the integration contour goes along negative imaginary axis and
\begin{equation}
\xi={g\over 1-g}.
\end{equation}
This integral converges for $3(g-1)<\Im m\, \nu <-1$, however the
definition of ${\bf \Psi}_j(\nu)$ can be extended to the whole complex
plane of $\nu$ by analytic continuation. For example, using the
product representation which holds for any eigenvalue $A_j(t)$ of $\A_j(t)$
\begin{equation}	
A_j(t)=\prod_{n=1}^\infty \Big(1-\frac{t}{t^{(j)}_n}\Big),\label{Aprod}
\end{equation}
one can write the corresponding eigenvalue of ${\bf \Psi}_j(\nu)$
as a Dirichlet series 
\begin{equation}
\Psi_j(\nu)={\Big(\Gamma(1-g)\Big)^{-i\nu(1+\xi)}\Gamma(i\nu(\xi+1)/3)
\over \Gamma(i\nu\xi/3)
\Gamma((-1+i\nu)/3)\Gamma((1+i\nu)/3)}
\sum_{n=1}^\infty (-it^{(j)}_n)^{-i\nu (1+\xi)/3}.\label{psiser}
\end{equation}	
For $0<g<2/3$ this series converges absolutely for $\Im m\, 
{\nu}<-1$ and can be analytically continued to the whole complex $\nu$-plane
by means of the standard technique \cite{Mum83}. We expect that
the function  
${\bf \Psi}(\nu)$ thus defined enjoys the following analytic
properties

\begin{conjecture} The function ${\bf \Psi}(\nu)$ is an entire function
of the variable  
$\nu$.
\end{conjecture}

Converting the integral transform in \rf{psidef} one obtains
\begin{eqnarray}
\log \A_i(t)&=&{1\over 2\pi i}\int_{C_\nu}{d\nu\over \nu}\,
\Gamma(i\nu\xi/3)\Gamma(1-i\nu(\xi+1)/3)
\Gamma((-1+i\nu)/3)\times\nonumber\\
&&\phantom{{1\over 2\pi i}\int_{C_\nu}}
\Gamma((1+i\nu)/3)\Big(\Gamma(1-g)\Big)^{i\nu(1+\xi)}
(it)^{i\nu(1+\xi)/3} \ {\bf \Psi}(\nu), \label{Aint}
\end{eqnarray}
where the integration contour $C_\nu$ goes along the line $\Im m\, 
\nu=-1-\varepsilon$ with arbitrary small positive $\varepsilon$.
 
The values of ${\bf \Psi}_j(\nu)$ at special  points 
on the imaginary axis of $\nu$ (where the Gamma-functions display
poles) are of particular interest. For
example, using \rf{psiser}, \rf{Aprod} and the definition of the 
nonlocal IM $\H^{(j)}_n$ in \rf{Hdef} it is easy to see that
\begin{equation}
{\bf \Psi}_j(-3in(1-g))={(-1)^n\,n!\, g^{-3n}\over
\Gamma(ng)\Gamma(-1/3+n(1-g))\Gamma(1/3+n(1-g))}\ \H_n^{(j)}.
\end{equation}
The other special values of  ${\bf \Psi}_j(\nu)$ of the imaginary axis
are determined by the
following 
\begin{conjecture} \label{points} The function ${\bf \Psi}(\nu)$ 
has the following
special values on the positive imaginary axis 
\begin{eqnarray}
{\bf \Psi}_j((2n-1)i)&=&
{1\over 6\sqrt{3}\pi}\Big({2^{\frac{2}{3}}\over 3}\Big)^{2n-1}
{\Gamma({\frac{1}{6}-\frac{n}{3}})
\over \Gamma({\frac{1}{2}-n})}\ g^n\ 
 {\bf I}_{2n-1},\qquad n=0,1,2,\ldots\label{con2-1}
\\
{\bf \Psi}_j(2ni)&=&{1\over 2\sqrt{3}\pi}\Big({2^{\frac{2}{3}}\over
3}\Big)^{2n} 
{\Gamma({\frac{1}{2}-\frac{n}{3}})
\over \Gamma({-\frac{1}{2}-n})}\ g^{n+\frac{1}{2}}\ 
 {\bf I}_{2n},\qquad n=1,2,\ldots\label{con2-2}
\\
{\bf \Psi}_j(3in(g^{-1}-1))&=&
{(-1)^n\,n!\,g^{\frac{3n}{g}+1}\ {\widetilde\H}_n^{(j)}
\over
\Gamma(ng^{-1})\Gamma(-\frac{1}{3}+n(1-g^{-1}))\Gamma(\frac{1}{3}+n(1-g^{-1}))}
\ , \qquad
n=1,2,\ldots  \nonumber\\ \label{con2-3}
\\
{\bf \Psi}_j(0)&=&{({\mbf w}_j{\mbf P})\over 2\pi\sqrt{3}}, \label{psi0}
\end{eqnarray}	
where $j=1,2,3$ and ${\bf I}_{-1}\equiv {\bf I}$ is the identity operator.
\end{conjecture}

Below it will be more convenient to work with the variable $s$ introduced in 
\rf{sdef} instead of the spectral parameter $t$ and exhibit all arguments
of $\A(s)$
\begin{equation}	
\A_j\{s,g,{\boldsymbol \phi}(u)\}
\equiv\exp\Big(-\sum_{n=1}^{\infty}\H^{(j)}_n s^n\Big).
\end{equation}

The above conjectures allow us to derive the asymptotic expansion 
of the operators $\A_j(t)$ for large $t$.

Calculating the integral
 \rf{Aint} as a formal sum over residues in the upper half-plane $\Im
m\, \nu \ge-1$  one has
\begin{equation}	
\A_j\{s,g,{\boldsymbol \phi}(u)\}
\simeq{\bf C}_j\{g,{\boldsymbol \phi}(u)\}\, 
s^{-{{\mbf w}_j{\mbf P}\over g}}
 \exp\Big\{{\sum_{k=-1}^\infty}
B_k \,s^{-k/3(1-g)}\,{\bf I}_k\Big\}\A_j\{s^{-{1\over g}},g^{-1},
g^{-1}{\boldsymbol \phi}(u)\}, \label{Qas}
\end{equation}
where 
\begin{equation}
\A_j\{s^{-{1\over g}},g^{-1},
g^{-1}{\boldsymbol \phi}(u)\} \equiv 
\exp\Big(-\sum_{n=1}^{\infty}{\widetilde\H}^{(j)}_n s^{-{n\over g}}\Big),
\end{equation}
and the coefficients $B_n$ read
\begin{eqnarray}
B_{2n-1}&=&{\ds(-1)^{n+1}\Gamma\Big({2n-1\over 3(1-g)}\Big)
\Gamma\Big({2n-1\over 3(1-g^{-1})}\Big)\over\ds
3\,(1-g)\, n! \,\Gamma\Big({2-n\over3}\Big)}\ \ds
g^{{n(1+g)-1\over g-1}},\qquad n=0,1,2,\ldots\qquad \label{Bodd}\\
B_{2n}&=&{\ds (-1)^{n+1} 
\Gamma\Big({2n\over 3(1-g)}\Big)
\Gamma\Big({2n\over 3(1-g^{-1})}\Big)\,\over\ds
(1-g)\, n!\, \Gamma\Big(-{n\over3}\Big)}\ \ds
g^{{n(1+g)\over g-1}
+{1\over 2}},\qquad n=1,2,\ldots \label{Beven}
\end{eqnarray}
An exact form for the operator ${\bf C}_j\{g,{\boldsymbol
\phi}(u)\}$ is unknown. However, an asymptotic vacuum eigenvalue in a
special case can be calculated (see \rf{Htilde}). 
Note, in particular, that the coefficient $M$ in the leading asymptotics 
\rf{Qas} is given by
\begin{equation}	
M=\frac{\Gamma\left(\frac{g}{3(1-g)}\right)
\Gamma\left(\frac{2-3g}{3(1-g)}\right)}
{\Gamma\left(\frac{2}{3}\right)}\left(\Gamma(1-g)\right)^{\frac{1}{1-g}}.
\label{Mvalue} 
\ed

Naturally, considerations similar to those above are 
valid for the operators $\Ao_j(t)$ as well. 
The corresponding formulae
are obtained from 
\rf{Aint} simply by a replacement of  $\A_j(t)$, ${\bf \Psi}_j(\nu)$, ${\bf
H}^{(j)}_n$ by their ``barred'' counterparts 
\begin{equation} 
\A_j(t)\to \Ao_j(t),\qquad {\bf \Psi}_j(\nu)\to\overline{\bf
\Psi}_j(\nu),\qquad {\bf H}^{(j)}_n\to \Ho^{(j)}_n, \label{changes1}
\end{equation}	
accompanied by  sign changes
\begin{equation}	
{\mbf w}_j\to-{\mbf w}_j, \qquad {\bf I}_{2n}\to-{\bf I}_{2n}, 
\label{changes2}
\end{equation}	
where $j=1,2,3$ and $n=1,2,\ldots,\infty$ (the sign of the ``odd'' local
IM ${\bf I}_{2n-1}$ remains the same). Also, in the integrand of
\rf{Aint}, the change $t\rightarrow -t$ needs to be made.

We can use these asymptotic results for the operators $\A_i(t)$,
$\Ao_i(t)$ to calculate an asymptotic expansion for ${\mathbb
T}(t)$. Using \rf{Tusuala}, we can write ${\mathbb T}(t)$ in the
following form
\begin{equation}
{\mathbb T}(t)={\bf \Lambda}_i(q^{\frac{3}{2}}t)
+{\bf \Lambda}^{-1}_i(q^{-\frac{1}{2}}t)
{\overline {\bf \Lambda}}_j(q^{\frac{1}{2}}t)
+{\overline {\bf \Lambda}}^{-1}_j(q^{-\frac{3}{2}}t),\qquad i\not=j,
\label{Tlam} 
\end{equation} 
where
\begin{equation}
{{\bf \Lambda}}_j(t)={\Q_j(qt)\over\Q_j(q^{-1}t)},\qquad
{\overline {\bf \Lambda}}_j={\Qt_j(qt)\over\Qt_j(q^{-1}t)}. \label{Tas1}
\end{equation}
Then, using \rf{Aint} we have
\begin{eqnarray}
\log{\bf \Lambda}_j&=&2\pi i ({\mbf w}_j{\mbf P})+
\int_{C_\nu}{d\nu\over \nu}\,
{\Gamma(1-i\nu(\xi+1)/3)
\Gamma((-1+i\nu)/3)\Gamma((1+i\nu)/3)
\over\Gamma(1-i\nu\xi/3)}\times\nonumber\\
&&\phantom{{1\over 2\pi i}\int_{C_\nu}}
\Big(\Gamma(1-g)\Big)^{i\nu(1+\xi)} 
(it)^{i\nu(1+\xi)/3} \ {\bf \Psi}_j(\nu), \label{lamint}
\end{eqnarray}
which gives via the sums of residues in the upper half-plane $\Im
m\,\nu\geq -1$
\begin{equation}	
\log{\bf \Lambda}_j(t)\simeq i\,m\,{\bf I}\,\big(it\big)^{1\over3(1-g)}
+\mathop{{\sum}'}_{k=1}^{\infty} C_k \,\big(it\big)^{{-k\over3(1-g)}}\
{\bf I}_k,\qquad t\to+\infty\label{lamas}
\end{equation}
where
\begin{equation}	
m={2\pi\Gamma({2\over3}-{\xi\over3})\over
\Gamma(1-{\xi\over3})\Gamma({2\over3})}\Big(\Gamma(1-g)\Big)^{1\over1-g},
\end{equation}
and
\begin{equation}
C_k=-2i\,g^{k(1+\xi)}\,\sin\Big({\pi\xi
k\over3}\Big)\,\Big(\Gamma(1-g)\Big)^{-{k\over1-g}}\,
B_k.
\end{equation}
A similar expression is obtained for $\log{\overline {\bf
\Lambda}}_i(t)$, using \rf{changes1},
\rf{changes2}. 
It is easy then to see that only the second term in \rf{Tas1} will
contribute in the asymptotic limit, as the other two will be
exponentially small. Thus we have 
\begin{eqnarray}
\log {\mathbb T}(t)&\simeq & mt^{\frac{1}{3(1-g)}}{\bf
I}-2\sum_{n=1}^{\infty}C_{2n}\cos\left(\frac{\pi
n}{3}\right)(t)^{-\frac{2n}{3(1-g)}}{\bf I}_{2n} \nonumber \\&+&
2i\sum_{n=1}^{\infty}C_{2n-1}
\sin\left(\frac{\pi(2n-1)}{6}\right)(t)^{-\frac{2n-1}{3(1-g)}}{\bf
I}_{2n-1}. \label{Tas2} 
\end{eqnarray}
The corresponding formula for $\overline{\mathbb T}(t)$ is obtained 
from the above one by the replacement \rf{changes2}.
This result can be regarded as the the quantum version of
\rf{cexpan}. In fact, if we take the limit $\xi\rightarrow 0$
(equivalently $q\rightarrow 1$), then, using the correspondence
\rf{qclass}, we recover exactly the classical expression
\rf{cexpan}.\footnote{Recall the correspondence of variables 
$t=\la^3$.}

As in \cite{BLZ97a} the main motivation to the conjectures given above 
came from the study of the large $p$ asymptotics of the vacuum
eigenvalues of the operators $\A_j(t)$ (see Appendix \ref{DDV}) as well as 
consideration of the classical limit $g\to0$ in the
$\mathbf{T}$-operator \rf{Tlam}. In this way we
obtain
\begin{equation}	
\Psi_j(\nu)|_{{\mbf p}\rightarrow\infty}={\ds
2^{-\frac{2i\nu}{3}}\Big(-\frac{g_2}{3}\Big)^{(1-i\nu)/2}
\over\ds
\sqrt{3\pi}\Gamma(1/2+i\nu/3)}
{}_2F_1(-\frac{1}{6}+
\frac{i\nu}{6},\frac{1}{6}+
\frac{i\nu}{6},\frac{1}{2}+
\frac{i\nu}{3},1+\frac{27 g_3^2}{4g_2^3})|_{j{\rm -th}}, \label{psiq}
\end{equation}
where the subscript $j{\rm -th}$ means that an appropriate branch of
the hypergeometric function is chosen such that \rf{psi0} holds, and
$g_2$, $g_3$ are defined by
\bn
g_2=x_1x_2+x_1x_3+x_2x_3, \qquad g_3=-x_1x_2x_3, \label{quantg}  
\ed
where
\bn
x_j=({\mbf w}_j{\mbf P}), \qquad j=1,2,3. \label{quantx}
\ed
The expression \rf{psiq} together with \rf{Aint} allows one to derive 
the asymptotic expansion \rf{Qas} in the limit ${\mbf p}\to\infty$
which provided a basis for the conjectures \rf{con2-1}, \rf{con2-2},
\rf{con2-3}, \rf{psi0}. 



The formulae \rf{Qas} allow us to obtain asymptotic expansions for the
zeroes of the eigenvalues of the operators ${\bf A}_j(t)$ and 
$\overline{{\bf A}}_j(t)$. Let us introduce new (rescaled) 
variables $E^{(j)}_k$,
$\overline{E}^{(j)}_k$, $k=1,2,\ldots$, 
 (whose meaning will become clear in Section~7.2) 
to denote the positions of these zeroes
\begin{equation}	
E^{(j)}_k=-i\rho\, t^{(j)}_k,\qquad
\overline{E}^{(j)}_k=i\rho \,\overline{t}^{(j)}_k,\qquad A_j(t^{(j)}_k)=
\overline{A}_j(\overline{t}^{(j)}_k)=0,\qquad k=1,2,\ldots,\infty
\end{equation}	
where
\begin{equation}
\rho=\Big[(3/g)^{1-g}\Gamma(1-g)\Big]^3. \label{rho}
\end{equation}	
Substituting \rf{Qas} into the Bethe Ansatz equations one obtains
\begin{equation}	
N^{(j)}_k\equiv n^{(j)}_k-{1\over 2} + {{\mbf w}_j{\mbf p}\over g}\simeq
\sum_{n=-1}^{\infty}\beta_n\ 
 I_n \ \bigg(E^{(j)}_k\bigg)^{-{n\over3(1-g)}}\phantom{(-1)^{n+1}}+
\sum_{n=1}^{\infty}
\gamma_n \ \widetilde{H}^{(j)}_n \ \Big(E^{(j)}_k
\Big)^{-{n\over g}}, \label{series1}
\end{equation}	
\begin{equation}	
\overline{N}^{(j)}_k\equiv \overline{n}^{(j)}_k-{1\over 2} -
 {{\mbf w}_j{\mbf p}\over g}\simeq
\sum_{n=-1}^{\infty}(-1)^{n+1}\beta_n\ 
 I_n \ \bigg(\overline{E}^{(j)}_k\bigg)^{-{n\over3(1-g)}}+
\sum_{n=1}^{\infty}
\gamma_n \ \widetilde{\overline{H}}^{(j)}_n \ \Big(\overline{E}^{(j)}_k
\Big)^{-{n\over g}}, \label{series2}
\end{equation}	
where
\begin{equation}	
\beta_n=-\pi^{-1}\ 3^n \ g^{gn/(1-g)}\  \sin({\pi n/ 3(1-g)})\ B_n,
\qquad
\gamma_k=\pi^{-1}\ 3^{3(1-g)n/g}\  g^{3n}\ \sin(\pi n/g),
\end{equation}	
the coefficients $B_n$ are given in \rf{Bodd}, \rf{Beven} and $I_k$, $\widetilde{H}^{(j)}_n$ and
$\widetilde{\overline{H}}^{(j)}_n$ are the corresponding eigenvalues
of the local IM \rf{locint} and the dual nonlocal IM, defined in \rf{Hdual}.
Inverting the series \rf{series1}, \rf{series2} one can express $E^{(j)}_k$ and $E^{(j)}_k$ as 
asymptotic series 
\begin{equation}	
E^{(j)}_k\simeq\Upsilon(N^{(j)}_k),\qquad
\overline{E}^{(j)}_k\simeq\overline{\Upsilon}(\overline{N}^{(j)}_k), \label{asseries}
\end{equation}	
where, for example, for $g<3/5$ one has
\begin{equation}	
\Upsilon(x)\simeq\alpha_0 \ x^{3-3g}\ \Big(1+\alpha_2x^{-2}+\alpha_3 x^{-3}+
O(x^{-4})+O(x^{2-3/g})\Big),\qquad x\to \infty \label{asseries1}
\end{equation}	
\begin{equation}	
\overline{\Upsilon}(x)\simeq\alpha_0 \ x^{3-3g}\ 
\Big(1+\alpha_2x^{-2}-\alpha_3 x^{-3}+
O(x^{-4})+O(x^{2-3/g})\Big),\qquad x\to\infty \label{asseries2}
\end{equation}

\bn
\alpha_0=(\beta_{-1})^{3(g-1)}=\left[{\ds3\Gamma({2\over3})
\Gamma({(\xi+1)/3})  
\over
g\Gamma({\xi\over3})}\right]^{3(1-g)},
\end{equation}	
\begin{equation}	
\alpha_2=\beta_1\beta_{-1}\ I_1= {3(1-g)I_1\over \pi(\cot
(\pi(\xi+1)/3)-\cot\pi/3)}, 
\ed
\begin{equation}	
\alpha_3= \beta_2\beta_{-1}^2\ I_2={\ds
3\,g^{\frac{3}{2}}\,(1-g)\,\Gamma\Big(-\frac{2}{3}\xi\Big)
\Gamma\Big(\xi/3\Big)^2  
\over\ds
2\Gamma\Big(\frac{2}{3}\Big)^3\,
\Gamma\Big(-\frac{2}{3}(\xi+1)\Big)\, \Gamma\Big((\xi+1)/3\Big)} \ I_2.
\end{equation}

The Bethe Ansatz equations \rf{BAE} can be solved numerically. We have
performed  such calculations for vacuum eigenvalues of $\A_1(t)$ and
$\Ao_3(t)$ 
for a range of different values of $g$ and ${\mbf p}$ and compared the
results with the asymptotic expansions \rf{asseries}. A typical
example of these  
calculations with 
\bn
g={1\over 5}, \qquad {\mbf p}=( \begin{array}{rl}
			\frac{1}{2}, & \frac{3}{10}\end{array}),
\label{gpvalue} 
\ed
is presented in Table~\ref{Table1} where the numerical values for $E$'s
are compared with the corresponding 
values given by six terms of the asymptotic series \rf{asseries} (the
value of the  
leading term in \rf{asseries} is also shown).

\begin{table}
\renewcommand{\baselinestretch}{1.0}
\normalsize

\centering
\small
\begin{tabular}{|r|r|r|r||r|r|r|}
\hline
$n\hfill$&$E^{(1)}_n$(BA)& $E^{(1)}_n$(AS)&$E^{(1)}_n$(LT)&
$\overline{E}^{(3)}_n$(BA)& $\overline{E}^{(3)}_n$(AS)&
$\overline{E}^{(3)}_n$(LT)\\ \hline
 $1$&$426.44998$&$429.22624$&$615.63$&$42.72214$&$25.28503$&$147.89$\\ \hline
 $2$&$871.85882$&$872.59835$&$ 1069.31$&$232.30688$&$229.38814$&$371.98$\\ 
\hline
 $3$&$ 1466.71208$&$ 1466.98579$&$ 1674.60$&$563.02921$&$562.24812$&$723.05$\\
\hline
 $4$&$ 2224.92354$&$ 2225.04588$&$ 2443.32$&$ 1039.72260$&$
1039.43412$&$ 1216.12$\\ \hline 
 $5$&$ 3157.57647$&$ 3157.63845$&$ 3386.10$&$ 1673.43618$&$
1673.30798$&$ 1864.09$\\ \hline 
 $6$&$ 4274.50955$&$ 4274.54393$&$ 4512.70$&$ 2474.97610$&$
2474.91149$&$ 2678.44$\\ \hline 
 $7$&$ 5584.74586$&$ 5584.76628$&$ 5832.14$&$ 3454.37424$&$
3454.33857$&$ 3669.55$\\ \hline 
 $8$&$ 7096.68686$&$ 7096.69967$&$ 7352.86$&$ 4620.94608$&$
4620.92497$&$ 4846.97$\\ \hline 
 $9$&$ 8818.22784$&$ 8818.23622$&$ 9082.79$&$ 5983.39245$&$
5983.37925$&$ 6219.54$\\ \hline  
 $10$&$10756.83708$&$10756.84278$&$11029.43$&$ 7549.88941$&$
7549.88079$&$ 7795.56$\\ \hline 
 $11$&$12919.61477$&$12919.61876$&$13199.92$&$ 9328.16145$&$
9328.15561$&$ 9582.83$\\ \hline 
 $12$&$15313.33879$&$15313.34165$&$15601.06$&$11325.54050$&$11325.53642$&
$11588.77$\\ \hline   
 $13$&$17944.50152$&$17944.50363$&$18239.38$&$13549.01388$&$13549.01096$&
$13820.41$\\ \hline
 $14$&$20819.34016$&$20819.34174$&$21121.13$&$16005.26365$&$16005.26150$&
$16284.47$\\ \hline
 $15$&$23943.86187$&$23943.86307$&$24252.34$&$18700.69926$&$18700.69765$&
$18987.41$\\ \hline
 $16$&$27323.86522$&$27323.86615$&$27638.82$&$21641.48502$&$21641.48379$&
$21935.42$\\ \hline
 $17$&$30964.95845$&$30964.95918$&$31286.21$&$24833.56334$&$24833.56239$&
$25134.47$\\ \hline
 $18$&$34872.57528$&$34872.57586$&$35199.94$&$28282.67468$&$28282.67393$&
$28590.32$\\ \hline
 $19$&$39051.98867$&$39051.98914$&$39385.30$&$31994.37469$&$31994.37410$&
$32308.54$\\ \hline
 $20$&$43508.32291$&$43508.32328$&$43847.43$&$35974.04923$&$35974.04875$&
$36294.55$\\ \hline
 $21$&$48246.56426$&$48246.56456$&$48591.33$&$40226.92739$&$40226.92700$&
$40553.57$\\ \hline
 $22$&$53271.57046$&$53271.57072$&$53621.85$&$44758.09307$&$44758.09276$&
$45090.72$\\ \hline
 $23$&$58588.07919$&$58588.07940$&$58943.76$&$49572.49523$&$49572.49497$&
$49910.94$\\ \hline
 $24$&$64200.71562$&$64200.71580$&$64561.67$&$54674.95691$&$54674.95669$&
$55019.08$\\ \hline
 $25$&$70113.99930$&$70113.99945$&$70480.11$&$60070.18348$&$60070.18330$&
$60419.85$\\ \hline
\end{tabular}
\caption{Comparison of the numerical solution of the Bethe Ansatz
Equations \rf{BAE}, \rf{phases} for $E^{(1)}_n$, $E^{(3)}_n$ with $g$,
${\mbf p}$ 
given by \rf{gpvalue}, to the value from asymptotic
series \rf{asseries}, $E^{(i)}_n$(AS), and the leading term of the
series $E^{(i)}_n$(LT). \label{Table1}} 
\end{table}
\renewcommand{\baselinestretch}{1.2}
\normalsize



\subsection{Special cases} \label{special}
The following special cases are related to a particular  choice of 
the vacuum parameter ${\mbf p}=(p,0)$ such that
\begin{equation}	
({\mbf w}_1{\mbf p})=p,\qquad({\mbf w}_2{\mbf p})=0,\qquad
({\mbf w}_3{\mbf p})=-p,\label{pspec2}
\end{equation}	
and the weights \rf{cweights} read
\begin{equation}	
\Delta_2=\frac{p^2}{g}+\frac{c-2}{24},\qquad \Delta_3=0.
\end{equation}	
 
The vacuum eigenvalues in this case obey the symmetry
\begin{equation}
A_1^{(vac)}(t)=\overline{A}_3^{(vac)}(-t),\qquad
A_2^{(vac)}(t)=\overline{A}_2^{(vac)}(-t),\qquad
A_3^{(vac)}(t)=\overline{A}_1^{(vac)}(-t),\label{Abarrel}
\end{equation}	
\begin{equation}	
{\mathbb T}^{(vac)}(t)={\overline{\mathbb T}}^{(vac)}(-t),
\end{equation}	
while  the expression \rf{H1vac} reduces to
\begin{equation}	
H_1^{(1)(vac)}=-\frac{\pi}{2\sin (\pi g)}\frac{g^3\Gamma\left(1-\frac{3g}{2}\right)}{\Gamma\left(1-\frac{g}{2}\right)\Gamma\left(1-g\right)}\frac{\Gamma(g+p)\Gamma(g/2+p)}{\Gamma(1-g+p)\Gamma(1-g/2+p)}.\label{H1vacc}
\end{equation}	
Further the asymptotic expansion of the
vacuum eigenvalues ${\mathbb T}^{(vac)}(t)$  given by \rf{Tas2}
now reads
\begin{equation}
{\mathbb T}^{(vac)}(t)=m \,t^{\frac{1}{3(1-g)}}+
2i\sum_{n=1}^\infty\cos((n-2)\pi/3)\,C_{2n-1}\,
I^{(vac)}_{2n-1} t^{\frac{1-2n}{3(1-g)}},\qquad
|\arg(t)|<\frac{\pi}{2}(1-g). \label{Tvacsp}  
\end{equation}

\subsubsection{The case $g=1/2$, $c=-10$}\label{sect632}
Consider the case when
\begin{equation}
c=-10,\qquad g=1/2,\qquad q=e^{i\pi g}=i.
\end{equation}	
Using \rf{pspec} and \rf{Abarrel} 
it  easy to see that the Bethe Ansatz equations \rf{BAE} in this case 
reduce 
to those for the $U_q(\widehat{sl}_2)$ related $\bf Q$-operators  
studied in \cite{BLZ97a}. This fact has been noticed in \cite{DT00a}
and recently discussed in \cite{SW00}. 
Therefore  the functions $A_1^{(vac)}(t)$ and 
$A_3^{(vac)}(t)$ coincide (up to a normalisation 
of the spectral parameter $t$) with $A^{(vac)}_+(t)$ and
$A^{(vac)}_-(t)$ 
from ref. \cite{BLZ97a} \footnote{
This correspondence extends to certain excited states as well (see
also Section~8.2), but 
we will not elaborate this point here}.
The relevant parameters used therein are related to our notation as follows
\begin{equation}	
c^{sl_2}=-\frac{25}{2}, \qquad \beta^2=\frac{1}{4},\qquad p^{sl_2}={p\over 2},
\qquad  \Delta={\Delta_2\over2}=p^2-9/16.
\end{equation}	
Comparing now the asymptotic expansion \rf{Qas} with the corresponding result
of ref \cite{BLZ97a} (the eq.(4.19) therein)
 and using the expressions \rf{Ivac} and \rf{H1vacc} 
one can reproduce the values of first three local IM 
$I^{(sl_2)(vac)}_k(\Delta,c^{(sl_2)}=-25/2)$ given in eq.(40) of ref.
 \cite{BLZ96}
\begin{eqnarray}	
I^{(sl_2)(vac)}_1(\Delta,c^{sl_2})
&=&{1\over2}
I^{(vac)}_1(\Delta_2,0,c)=
\Delta+{25\over48}, \\
I^{(sl_2)(vac)}_3(\Delta,c^{sl_2})
&=&-{3\over32}H^{(1)(vac)}_1(1/g,p/g) 
=\Delta^2+{7\over8}
\Delta+{45\over256}, \\
I^{(sl_2)(vac)}_5(\Delta,c^{sl_2})&=&{1\over8}I^{(vac)}_5(\Delta_2,0,c)
=\Delta^3+{17\over16}\Delta^2-{175\over98304}\Delta+{15275\over774144}. 
\end{eqnarray}
\subsubsection{The case $g=2/3$, $p=1/6$}
Consider now the vacuum eigenvalue \rf{Tvacsp} of the operator $\Tb(t)$ when 
\begin{equation}
c=-2,\qquad g=2/3,\qquad p=1/6.\label{case2}
\end{equation}	
The value $g=2/3$ does lie in the domain \rf{quasiclas}
 therefore the results of the previous Sections 
do not directly apply. In particular the definitions
\rf{Tfund}-\rf{Jdef} require a
renormalization since the nonlocal IM \rf{Gdef} diverge logarithmically at
$g=2/3$. To obtain the asymptotic of $\Tb(t)$ at large $t$ in this case we set 
$g=2/3-\epsilon$ in \rf{Tvacsp}
 and let $\epsilon\to0$ (this corresponds to an analytic
regularization of divergent integrals). The first few terms of the 
formula \rf{Tvacsp} then read 
\begin{equation}
\log {\mathbb T}^{(vac)}(t)|_{g=2/3,p=1/6}=
-\sqrt{3}\,at\,\log t+{\mathbb
C}\,t+{\sqrt{3}\over4!}(at)^{-1}-{31\sqrt{3}\over8!}(at)^{-5}+\ldots
\label{tas23} 
\end{equation}	
where ${\mathbb C}$ is a renormalization constant and
\begin{equation}	
a=\Gamma(1-g)^3=\Gamma(1/3)^3.
\end{equation}	
On the other hand the eigenvalues of ${\mathbb T}(t)$ can be computed
independently.
Using the results of \cite{EB} one can show that in the case \rf{case2} the 
fusion equations \rf{fusrel} for the eigenvalues of ${\mathbb T}(t)$
simplify to a single  equation
\begin{equation}
{\mathbb T}(tq^{{1\over2}}){\mathbb T}(tq^{-{1\over2}})=2\cosh(a\pi t)\ 
{\mathbb T}(t),\qquad q=e^{2\pi i/3} \label{se}
\end{equation}	
which with an assumption that ${\mathbb T}(t)$  is an entire function of
$t$ and has  the leading asymptotics at large $t$ as in \rf{tas23}
can be completely solved 
\begin{equation}	
{\pmb {\mathbb T}}(t)={\mathbb T}^{(vac)}(t) \prod_{k=1}^{L} 
{(1+2n_k-2at e^{i\pi/6})(1+2n_k-2at e^{-i\pi/6})
\over
(1+2n_k+2at e^{i\pi/6})(1+2n_k+2at e^{-i\pi/6})},
\end{equation}
\begin{equation}
T^{(vac)}(t)={2\pi e^{{\mathbb C}\/'\,t}\over
\Gamma({1\over2}+ate^{i\pi/6}) 
\Gamma({1\over2}+ate^{-i\pi/6})}.
\end{equation}
where the integer $L\ge0$; $\{n_k\}$ is an arbitrary increasing
sequence of integers $0\le n_1<n_2<\ldots<n_L$ and ${\mathbb C}'$ is 
an arbitrary constant. The vacuum eigenvalue ${\mathbb T}^{(vac)}(t)$
is a unique solution of \rf{se} with no zeroes in the strip $|\mbox{arg}(t)
|<\pi/3$. One can easily check that its asymptotic expansion 
at large $t$ perfectly agrees with \rf{tas23}.

\subsubsection{The case $g=5/6$}

Below we will compare the asymptotic expansion \rf{Tvacsp} 
for $g=5/6$ against the 
numerical results of  \cite{EB} for 
two values of $p$
\begin{equation}	
p=1/6,\qquad p=1/3, \label{psp}
\end{equation}	 
which corresponds to
\begin{equation}
\Delta_2=0,\qquad \Delta_2=1/10, \label{deltasp}
\end{equation}	
respectively.
Again, for $g=5/6$ the formula \rf{Tvacsp}
requires a renormalization. Using the analytic regularization as in
the previous subsection one obtains
\begin{equation}
T^{(vac)}(t)|_{g=5/6}=-{2\over\sqrt{3}}(at)^2\log t+{\mathbb C}\,t-
{3\sqrt{3}\over 2}\,(at)^{-2}\, I^{(vac)}_1+
{\frac {32805\, \sqrt {3}}{43472}} (at)^{-10}\, I^{(vac)}_5+\ldots\label{tas56}
\end{equation}	
where 
\begin{equation}	
a=\Gamma(1-g)^3=\Gamma(1/6)^3,
\end{equation}	
and $I_1^{(vac)}$ and $I_5^{(vac)}$ are given in \rf{Ivac}.

\begin{table}[h]
\renewcommand{\baselinestretch}{1.0}
\normalsize
\centering
\small
\begin{tabular}{|lll|r|}
\hline\hline
$I^{(vac)}_k(c,\Delta_2,\Delta_3)$
&&
&Numerical values \cite{EB}\hfil \\
\hline\hline
$I^{(vac)}_1(6/5,0,0)$&$=-\frac{1}{20}$&$=-0.05$&$-0.499999999599\ 10^{-1}$\\ \hline
$I^{(vac)}_5(6/5,0,0)$&$=-\frac{121}{10500}$&$=-0.115238095238\ldots\   10^{-1}$
&$-0.115238095240\ 10^{-1}$\\ \hline\hline
$I^{(vac)}_1(6/5,1/10,0)$&$=\frac{1}{20}$&$=0.05$&$0.500000000002\ 10^{-1}$\\ \hline
$I^{(vac)}_5(6/5,1/10,0)$&$=\frac{209}{42000}$&$=0.497619047619\ldots
\ 10^{-2}$&$0.497619047634\  10^{-2}$\\
\hline\hline
\end{tabular}
\caption{Exact values of $I^{(vac)}(c, \Delta_2, \Delta_3)$ for
$g=5/6$, and $p$ given by \rf{psp}, compared with the corresponding
numerical results from \cite{EB}. \label{Table2}}
\end{table}
\renewcommand{\baselinestretch}{1.2}
\normalsize

As shown in \cite{EB} the eigenvalues of ${\mathbb T}(t)$ in this case 
satisfy the equation 
\begin{equation}	
\Tb(tq)\Tb(t\q)=e^{\frac{a^2\pi}{3}t^2}\Tb(-t)+e^{-\frac{a^2\pi}{3}t^2}\Tb(t).
\end{equation}
In \cite{EB} this equation was transformed to TBA-like integral equations,
due to Kl\"umper and Pearce \cite{KP91}, and studied numerically for 
a few vacuum eigenvalues of 
${\mathbb T}(t)$. In  particular, the numerical values of the
coefficients in \rf{tas56} were calculated. 
We have verified  that the values of
the local IM extracted from this numerical data  are in a good agreement 
with the corresponding exact values determined by \rf{Ivac} (see
Table~\ref{Table2}).

\setcounter{equation}{0}
\section{Applications}
\subsection{Boundary Affine Toda Theory} \label{BATFT}
The Hamiltonian corresponding to the (non-equilibrium) 
boundary affine Toda theory (BAT) with zero bulk mass reads
\begin{equation}	
{\bf H}_{{\rm BAT}} = {\bf H}_0+{\bf H}_1,\label{BAT}
\end{equation}
\begin{equation}	
{\bf H}_0={1\over {4\pi g}}\int_{-\infty}^{0}dx\, 
\big(\, 
{{\boldsymbol \Pi}}^2 + 
{{\boldsymbol \Phi}}_{x}^2\,  \big),\qquad
{\bf H}_1=
-{\kappa\over 2g}\sum_{i=1}^3\ u_i\ e^{i({\mbf e}_i,{\boldsymbol \Phi}_{B} +
{\bf V}t)} 
\ ,
\ed
where $g$, $\kappa$ and ${\mbf V}=(V_1,V_2)$ are parameters;
${\boldsymbol \Phi(x)}=(\Phi_1(x),\Phi_2(x))$,
${\boldsymbol{\Pi}}(x)=(\Pi_1(x),\Pi_2(x))$
are field operators obeying canonical 
commutation relations 
\bn
\big[\Pi_a(x)\, , \Phi_b(x')\big] = -2\pi i g\  \delta (x-x')\delta_{ab}
,\qquad a,b=1,2\label{bosecomm}
\ed
the variables $u_1,u_2,u_3$ 
\bn
u_1u_2=qu_2u_1,\qquad u_2u_3=qu_3u_2,\qquad u_3u_1=qu_1u_3,\qquad
u_1u_2u_3=q^{\frac{1}{2}},
\ed
describe boundary degree of freedom and $\Phi_{B} \equiv \Phi(0)$. 
The way $g$ enters the Hamiltonian \rf{BAT} allows one to interpret it
as a quantum parameter, because it always appears in the combination
$g\hbar$; in what follows we will set $\hbar=1$.
We will also assume that the parameter ${\mbf V}$ satisfies the requirement
\begin{equation}
({\bf e}_1{\mbf V})>0,\qquad ({\bf e}_2{\mbf V})>0. \label{7.5}
\end{equation}

At a nonzero ${\mbf V}$ and a temperature $T$ the system \rf{BAT} develops a
stationary non-equilibrium state 
which can be thought of as the result of an
infinite time evolution of the equilibrium state of the corresponding
``free'' system, with the interaction term
(the last term in \rf{BAT}) adiabatically switched on.
We will denote by 
$\langle\,{\bf A}\, \rangle_{\rm N}$  the
expectation value of an observable ${\bf A}$ over this 
non-equilibrium stationary state.

The density matrix ${\bf P}(t)$ of the system 
is determined by the Schr\"odinger equation
\begin{equation}	
\partial_t\, {\bf P}(t)=-i{[}{\bf H}(t),\,{\bf P}{]},
\end{equation}
with the initial  condition 
\begin{equation}	
{\bf P}(t)|_{t=-\infty}={\bf P}_0=Z_0^{-1}\, e^{-R {\bf
H}_0},\qquad R=g/T, \label{Pnot}
\end{equation}
where ${\bf P}_0$ 
is the equilibrium density matrix of the free system at the temperature $T$.
Using  the interaction representation corresponding to the Hamiltonian 
${\bf H}_0$ in \rf{BAT} one can write 
the density matrix ${\bf P}(t)$ as
\begin{equation}	
{\bf P}(t)=e^{-i {\bf H}_0  t}\ 
{\bf S}(t,-\infty)\
{\bf P}_0 \ {\bf  S}(-\infty,t)\
e^{i {\bf H}_0  t}\ ,\label{denmat}
\end{equation}
where
\begin{equation}\begin{array}{rcl}
{\bf S}(t,t_0)&=&\ds{\cal T} \exp\Big\{-i\int_{t_0}^{t} d\tau\
{\bf H}_1^{(int)}(\tau)\Big\}=\\
&=&\ds 1+\sum_{k=1}^\infty\,(-i\/)^k\,
\int_{t_0}^{t} {\cal D}_k(\{\tau\})\  
{\bf H}_1^{(int)}(\tau_1)\, {\bf H}_1^{(int)}(\tau_2) 
\cdots {\bf H}_1^{(int)}(\tau_n),
\end{array}\label{smat}\end{equation}
with 
\begin{equation}
{\bf H}_1^{(int)}(t)=e^{i{\bf H}_0  t}\ {\bf H}_1\
e^{-i{\bf H}_0  t}\ .
\label{Hint}
\end{equation}
In \rf{smat}\  and below the shorthand notation for the multiple ordered
integrals
\begin{equation}	
\int_{t_0}^{t}{\cal D}_k(\{\tau\})
=\int_{t_0}^{t}d\tau_1\int_{t_0}^{\tau_1}d\tau_2\cdots
\int_{t_0}^{\tau_{k-1}}d\tau_k.\label{Ddef}
\end{equation}
is used. 
The expectation value  of an
arbitrary operator ${\bf A}$ over the stationary  non-equilibrium state 
\rf{denmat} (which will be marked with the subscript ${\rm N}$) reads
\begin{equation}	
\vev{{\bf A}}_{{\rm N}}=
{\rm Tr}_{\cal H}\big[\, {{\bf  P}(t)\, {\bf A} }\label{expdef}
\big].
\end{equation}	
The trace is taken over the space ${\cal H}={\cal F}\otimes
{\cal U}$ where ${\cal F}$ is the Fock space representing the Bose
commutation relations \rf{bosecomm} and ${\cal U}$ is the space of
states of the boundary variables  $u_i$, $i=1,2,3$ (with the trace over
these states defined exactly as in \rf{faddtrace}).
Substituting
\rf{denmat} into \rf{expdef} one gets
\begin{equation}	
\vev{{\bf A}}_{{\rm N}}=
{\rm Tr}_{\cal H}\big[\, {{\bf P}_0\,  
{\bf S}(-\infty,t) \,{\bf A}^{(int)}(t)\, 
{\bf S}(t,-\infty)}\, \big]=
\vev{{\, {\bf S}(-\infty,t) \,{\bf A}^{(int)}(t)\,
{\bf S}(t,-\infty)\, }}_0\, ,\label{aver}
\end{equation}	
where $\vev{\, \ldots\, }_0$ denotes the expectation value over 
the equilibrium state 
\rf{Pnot} of the free system, and the superscript 
``$(int)$'' means that this operator is taken in the interaction
representation, i.e. ${\bf A}^{(int)}(t)=
e^{i {\bf H}_0  t}\, {\bf A}\, e^{-i{\bf  H}_0  t}$.
Equivalently, one may write the above expectation value as
\begin{equation} 
\langle\,  {\bf A}\, \rangle_{{\rm N}}= {\rm Tr}_{\cal H} \big[\,
{\bf P}\,  {\bf A}(t)\, \big]\ ,\label{heiex}
\end{equation}  
where ${\bf P}$ stands for
density matrix of the system at $t=0$, i.e.
${\bf P}={\bf P}(0)$, and ${\bf A}(t)$ is the
full Heisenberg operator
\begin{equation}	
{\bf A}(t)=
{\bf S}(0,t) \,{\bf A}^{(int)}(t)\,
{\bf S}(t,0)\ .\label{Ahei}
\end{equation}	

All above formulae are very well known (see e.g. \cite{Kel}); 
the notation used here is exactly the same as in \cite{BLZnon}. 
We are
interested in the expectation values  
$\vev{\, {\mathbb V}_j\, }_{{\rm N}}$ of the Heisenberg operators
\begin{equation}	
{\mathbb V}_j(t)={\bf S}(0,t)\,u_j\, {\bf V}^{(int)}_j(t)\, {\bf S}(t,0)\ .
\label{vhei}
\end{equation}
where 
\begin{equation}
{\bf V}^{(int)}_j(t)=e^{i({\bf e}_j,{\boldsymbol \Phi}^{(int)}_{B}(t) +
{\bf V}t)},\qquad j=1,2,3
\end{equation}	
and ${\boldsymbol\Phi}^{(int)}_B(t)$ is the  boundary field
${\boldsymbol\Phi}_B$ in the interaction
representation. It is convenient to introduce also auxiliary
operators
\begin{equation}	
{\mathbb V}_j (t,t_0)={\bf S}(t_0,t)\,u_j\,{\bf V}^{(int)}_j(t)\,
{\bf S}(t,t_0)\ ,\label{vtt}
\end{equation}	
where $t_0$ is a parameter. For $t_0 =0$ \rf{vtt}\ coincides with the
Heisenberg operators \rf{vhei}, and according to \rf{aver}\ the expectation
values of \rf{vhei} can be expressed through \rf{vtt} as follows
\begin{equation}	
\vev{\, {\mathbb V}_\sigma(t)\, }_{{\rm N}}=\lim_{t_{0}\to-\infty}\,\vev{\, 
{\mathbb V}_\sigma(t,t_{0})\, }_0 \ ,\label{vexp}
\end{equation}	
where $\sigma=1,2,3$.

Using the series expansion for the evolution operators \rf{smat}
and the commutation relations 
\begin{equation}	
{\bf V}^{(int)}_{i}(t_1){\bf V}^{(int)}_{j}(t_2)=
q^{2({\bf e}_i{\bf e}_j)}
{\bf V}^{(int)}_{j}(t_2){\bf V}^{(int)}_{i}(t_1),\qquad t_1>t_2
\end{equation}	
one can represent \rf{vtt} as a series of time-ordered integrals 
of products of ${\bf V}^{(int)}_j$ 
\begin{equation}\begin{array}{rl}
&{\mathbb V}_\sigma (t, t_0 )=
u_\s \,{\bf V}^{(int)}_\sigma  (t)\\
&\ds
+{\bf V}^{(int)}_\sigma  (t)
\sum_{k=1}^\infty 
\ \sum_{\s_1,\ldots,\s_k=1,2,3} \  {\mathbb C}_k(\s,\, \s_1,\ldots,\s_k)\, 
\int_{t_0}^t{\cal D}_k(\{t\})  \ 
{\bf V}^{(int)}_{\s_1}(t_1)\cdots {\bf V}^{(int)}_{\s_k}(t_k)\,
\ ,\end{array}
\label{vser}
\end{equation}	
where the sum is taken over all arrangements of the 
``charges'' $\s_1,\ldots,\s_k=1,2,3$. 
The 
coefficients  
$ C_k(\s|\, \s_1,\ldots,\s_k)$ can be easily calculated following the
arguments of ref. \cite{BLZnon}. From the definition \rf{vtt} it follows  
that  
\begin{equation}
i\frac{\partial}{\partial t_0}{\mathbb V}_\s(t,t_0)=[{\bf
H}_1^{(int)}(t_0),{\mathbb V}_\s(t,t_0)].
\end{equation}
Substituting the the series \rf{vser} into this differential equation 
one gets a recurrence relation for the coefficients $ {\mathbb C}_k(\s|\,
\s_1,\ldots,\s_k)$, which (with the initial condition 
${\mathbb C}_0(\s)=u_\s$) has a unique solution 	
\begin{equation}	
{\mathbb C}_k(\s,\, \s_1,\ldots,\s_k)=
u_1^{({\mbf w}_2 {\mbf X}_k)}u_2^{-({\mbf w}_1 {\mbf X}_k)}\ 
\Big(\frac{q^{\frac{1}{2}}\kappa}{g}\Big)^k\ q^{F({\mbf X}_k)}
\prod_{j=1}^k\sin( \pi g\,{\mbf w}_{\sigma'_j}{\mbf X}_{j-1})\ ,\label{Ck}
\end{equation}	
where
\begin{equation}	
\s'_j=\s_j-1 \pmod 3,\qquad {\mbf X}_j=\sum_{s=0
}^{j} {\mbf e}_{\s_s},\qquad \s_0\equiv\s \label{cumcharges}
\end{equation}	
and 
\begin{equation}
F({\mbf X})=-\frac{1}{2}\Big({\mbf w}_1{\mbf X}\Big)^2-\frac{1}{2}
\Big({\mbf w}_2{\mbf X}\Big)^2+\frac{1}{2}\ .
\end{equation}	

Using \rf{vser}, \rf{Ck} one obtains an infinite series representation for 
the expectation values \rf{vexp}\footnote{Representations  of this type 
for a similar problem related to the boundary sine-Gordon model
were first derived  in \cite{WES95} by combinatorial methods.}
\begin{equation}	
\vev{{\mathbb V}_\s}_{{\rm N}}=\frac{2\pi T}{\kappa\, \sin\pi g}
\sum_{n=1}^\infty (-it)^n\ \sum_{\s_1,\ldots,\s_{3n-1}}
\Big(\prod_{j=1}^{3n-1}{\sin(\pi g\,{\mbf w}_{\s'_j}{\mbf X}_{j-1})\over
\sin(\pi g)}\Big)\  J(\s,\s_1,\ldots,\s_{3n-1}|{\mbf p}),\label{expval}
\end{equation}	
\begin{equation}	
J(\s,\s_1,\ldots,\s_{3n-1}|{\mbf p})=
\int_{-\infty}^0{\cal D}_{3n-1}(\{\tau\})e^{-2{\mbf p}\sum_{j=1}^{3n-1}{\mbf
e}_{\s_j}\tau_j }\prod_{0\le j\le k\le
3n-1}\Big(2\sin\big({\tau_j-\tau_k\over 2}\big)\Big)^{g\,a_{jk}}.
\end{equation}
where $\s_i=1,2,3$ and the internal sum taken over all possible of values of
$\{\s_i\}$, \ $i=1,\ldots,3n-1$ such that
\bn
{\mbf e}_{\s_0}+{\mbf e}_{\s_1}+\ldots+{\mbf e}_{\s_{3n-1}}=0\ .\label{zc1}
\ed
It is convenient to represent the sequence of the ``charges''
$\s_0,\s_1,\ldots,\s_k$ for each term in \rf{vser} by  a path staring
from the origin ${\bf X}=0$ of the
triangular lattice 
\begin{equation}	
{\bf X}=n_1{\bf e}_1+n_2{\bf e}_2,\qquad  n_1,n_2\in {\mathbb Z}, 
\end{equation}	
and 
formed by the sequence of the ``step'' vector ${\bf
e}_{\s_0},{\bf e}_{\s_2},\ldots,{\bf e}_{\s_k}$. The vectors ${\bf
X}_j$ defined in \rf{cumcharges} then represent the end point of the
path after $j+1$ steps. It is important to 
note, that not every such path contributes to \rf{vser}. 
For example, for $\s_0=1$ only paths which are
\begin{description}
\item{i)}
always contained in the sector (see Fig~\ref{triangle})
\begin{equation}	
{\mbf w}_1{\mbf X}_m\ge 0,\qquad {\mbf w}_2{\mbf X}_m\ge 0,\qquad
m=0,1,\ldots k \label{sector1}
\end{equation}	
\item{ii)}
do not return to the origin (except for the final step), i.e.,
\begin{equation}	
{\mbf X}_m\not=0,\qquad m=0,1,\ldots k-1
\end{equation}	
\end{description}
correspond to nonzero coefficients $C_k(\s_0,\s_1,\ldots,\s_k)$, since 
otherwise one of the factors $\sin( \pi g\,{\mbf w}_{\sigma'_j}{\mbf
X}_{j-1})$ in \rf{Ck} vanishes. Obviously, the  paths \rf{zc1} contributing 
to \rf{expval} return to the origin at the last step.
As a result all integrals appearing in 
\rf{expval} converge at large $\tau$'s for 
\begin{equation}	
2({\bf e}_1{\mbf p})>-g,\qquad 2({\bf e}_2{\mbf p})<g.
\end{equation}	
\begin{figure}[ht]
\begin{center}
\epsfig{scale=.9, file=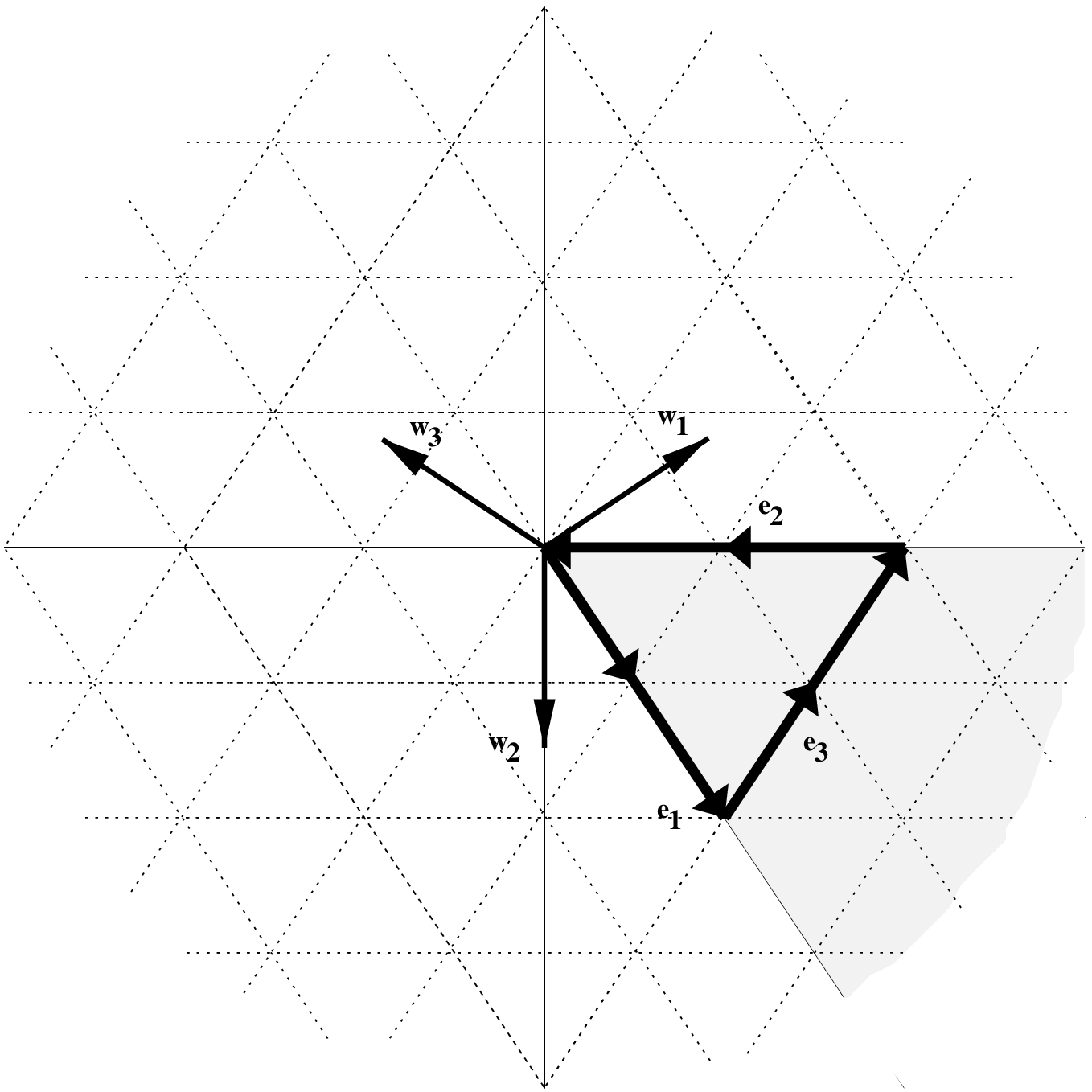}
\end{center}
\caption{The shaded area is the region defined by \rf{sector1}. We
also show an admissible path ${\mbf X}_5={\mbf e}_1+{\mbf e}_1+{\mbf
e}_3+{\mbf e}_3+{\mbf e}_2+{\mbf e}_2.$} 
\label{triangle}
\end{figure}

The expectation values \rf{expval} coincide with {\it equilibrium} expectation 
values of certain operators in the system which is similar to \rf{BAT} 
but contains different boundary degrees of freedom obeying the $q$-oscillator 
algebra \rf{qos}.
Consider the Hamiltonian
\begin{equation}	
\overline{\bf H}={1\over {4\pi g}}\int_{-\infty}^{0}dx\, 
\big(\, 
{{\boldsymbol \Pi}}^2 + 
{{\boldsymbol \Phi}}_{x}^2\,  \big)+\frac{1}{2}{\mbf h}{\mbf V}
-{\kappa\over 2g}
\sum_{i=1}^3\ y_i\ e^{i{\mbf e}_i{\boldsymbol \Phi}_{B}} 
\ , \label{Ham}
\ed
where ${\boldsymbol \Phi(x)},\, {\boldsymbol \Pi(x)}$ 
are again the Bose field operators
obeying the same commutation relations \rf{bosecomm} 
\begin{equation}	
{\mbf h}=\frac{2}{3}(h_1{\mbf e}_1+h_2{\mbf e}_2+h_3{\mbf e}_3),
\end{equation}	
and
\begin{equation}
{\begin{array}{rclrclrcl}	
y_1&=&(q-q^{-1})\ab_2^-,\quad&
y_2&=&(q-q^{-1})\ab_1^+ q^{\hbc_2},\quad&
y_3&=&q^{\frac{1}{2}}(q-q^{-1})^2q^{-\hbc_2}\ab_1^-\ab_2^+,
\end{array}\atop
\begin{array}{rclrclrcl}	
h_1&=&-\hbc_1-2\hbc_2,\quad&
h_2&=&2\hbc_1+\hbc_2,\quad&
h_1&=&-\hbc_1+\hbc_2,
\end{array}}
\end{equation}	
where the operators ${\bf a}_i^\pm$ and ${\cal H}_i$, $i=1,2$, commute 
with ${\boldsymbol\Phi}(x)$, ${\boldsymbol\Pi}(x)$ and satisfy 
the commutation relations \rf{qos}. 
Let $\rho$ be any representation of the algebra \rf{qos} such that 
the spectra of ${\mathcal H}_1$ and  ${\mathcal H}_2$ are
both real and bounded from above. The Hamiltonian \rf{BAT} acts in the 
space
\begin{equation}	
\overline{\cal H}={\cal F}\otimes\rho,
\end{equation}	
where  ${\cal F}$ is the Fock space representing the Bose
commutation relations \rf{bosecomm} and 
for the values of ${\mbf V}$
satisfying  \rf{7.5} this Hamiltonian is bounded from below. Then the
 system \rf{BAT} 
has a thermal equilibrium state described by the standard density matrix
\begin{equation}	
\overline{\bf P}={\overline Z}(\kappa,{\mbf V})^{-1}e^{-R{\overline{\bf H}}},
\qquad R=g/T
\end{equation}	
where 
\bn
{\overline Z}(\kappa,{\mbf V})={\rm Tr}_{\overline{\cal
H}}e^{-R{\overline{\bf H}}}, \label{pfunction}
\ed
is the partition function. Denote by
$\vev{{\bf A}}_{\rm E}$ the expectation value of an observable ${\bf
A}$ over this thermal equilibrium state
\begin{equation}	
\vev{{\bf A}}_{\rm E}={\rm Tr}_{\overline{\cal H}}\,({\bf A} \overline{{\bf
P}}).\label{eexp}
\end{equation}
We will be particularly interested in the expectation value
$\vev{\overline{\mathbb V}_1}_E$ of the operator
\begin{equation}
\overline{\mathbb V}_1=y_1 \, e^{i{\mbf e}_1{\boldsymbol \Phi}_{B}},
\end{equation}	
and will show below that 
\begin{equation}
\vev{\overline{\mathbb V}_1}_E=\vev{{\mathbb V}_1}_N, \label{7.39}
\end{equation}	
where $\vev{{\mathbb V}_1}_N$ is given by \rf{expval} above. 

The 
proof is closely parallel to that used in the proof of the analogous
result for the Boundary Sine-Gordon model, found in \cite{BLZnon}. We
will use the interaction representation with the  
Hamiltonian 
\begin{equation}	
\overline{\bf H}_0={1\over {4\pi g}}\int_{-\infty}^{0}dx\, 
\big(\, 
{{\boldsymbol \Pi}}^2 + 
{{\boldsymbol \Phi}}_{x}^2\,  \big)+\frac{1}{2}{\mbf h}{\mbf V},
\end{equation}	
corresponding to $\kappa=0$ in \rf{Ham}. 
Consider an auxiliary operator
\begin{equation}	
\overline{\mathbb V}_1(t,t_0)={\bf {\overline S}}(t_0,t)\ 
e^{i{\bf {\overline H}}_0t}\  {\overline {\mathbb V}}_1\,  e^{-i{\bf {\overline
H}}_0t}{\bf {\overline S}}(t,t_0),\label{vbtt}
\end{equation}	
where ${\bf {\overline S}}(t,t_0)$ is defined  as in \rf{smat} but with the
interaction Hamiltonian replaced by 
\begin{equation}	
{\bf {\overline
H}}^{(int)}_1(t) =e^{i{\bf {\overline H}}_0 t}\, 
{\bf  {\overline H}}_1\,  e^{-i{\bf {\overline H}}_0 t},\qquad
{\bf  {\overline H}}_1=-{\kappa\over 2g}
\sum_{i=1}^3\ y_i\ e^{i{\mbf e}_i{\boldsymbol \Phi}_{B}}. 
\end{equation}	
The operator \rf{vbtt} can be represented in a form similar to \rf{vser}
\begin{equation}\begin{array}{rl}
&\overline{{\mathbb V}}_1 (t, t_0 )=
{\bf a}_2^- \,{\bf V}^{(int)}_1  (t)\\
&\ds
+{\bf V}^{(int)}_\sigma  (t)
\sum_{k=1}^\infty 
\ \sum_{\s_1,\ldots,\s_k=1,2,3} \ \overline{\mathbb C}_k(\s,\, \s_1,\ldots,\s_k)\, 
\int_{t_0}^t{\cal D}_k(\{t\})  \ 
{\bf V}^{(int)}_{\s_1}(t_1)\cdots {\bf V}^{(int)}_{\s_k}(t_k)\,
\ ,\end{array}
\label{vbser}
\end{equation}	
where 
\begin{equation}	
\overline{{\mathbb C}}_k(\s,\, \s_1,\ldots,\s_k)=
({\bf a}_2^-)^{({\mbf w}_2 {\mbf X}_k)}({\bf a}_1^-q^{-{\cal H}_2})
^{-({\mbf w}_1 {\mbf X}_k)}\ 
\Big(\frac{q^{\frac{1}{2}}\kappa}{g}\Big)^k\ q^{F({\mbf X}_k)}
\prod_{j=1}^k\sin( \pi g\,{\mbf w}_{\sigma'_j}{\mbf X}_{j-1})\ ,\label{Ck2}
\end{equation}	
and the rest of notation is the same as in \rf{vser}. In particular,
the same remarks about the paths contributing to \rf{vser} apply to
\rf{vbser}. We can now use 
the representation \rf{vbser} in \rf{eexp}. Repeating the arguments
which used in derivation of eq.(3.15) in ref.\cite{BLZnon} one obtains
\bn
\vev{\overline{{\mathbb V}}_1}_{\rm E}=\lim_{t_0\to-\infty}
{\ds{\rm Tr}_{\overline{\cal H}}[e^{-R\overline{H}_0} \overline{{\mathbb
V}}_1(0,t_0)] \over\ds {\rm Tr}_{\overline{\cal H}}[e^{-R\overline{H}_0}]}
\label{Vexpval}\ed
Using now \rf{vbser} and the simple property of the trace
\bn
{\rm Tr}_{\overline{\cal H}}[e^{-R\overline{H}_0}({\bf a}_1^-)^n
({\bf a}_2^-)^m]=\delta_{n,0}\delta_{m,0}{\rm Tr}_{\overline{\cal
H}}[e^{-R\overline{H}_0}]
\ed
one arrives exactly to the series \rf{expval} for the expectation
value \rf{Vexpval}. This proves the relation \rf{7.39}.

Further, comparing the Matsubara representation for the partition
function  \rf{pfunction} with the vacuum eigenvalues of the operators ${\bf
A}_i(t)$ defined by \rf{Adef}  one can conclude that 
\begin{equation}	
\frac{{\overline Z}(\kappa,{\mbf V})}
{{\overline Z}(0,{\mbf V})}=A^{(vac)}_3(t,{\mbf p}),
\end{equation}	
where the parameters $t$ and ${\mbf p}$ related to $\kappa$, ${\mbf
V}$ and the temperature $T$ as 
\begin{equation}	
t=i\Big[\,\frac{\kappa \sin \pi g}{2\pi T}\Big(\frac{g}{2\pi
T}\Big)^{-g}\,\Big]^3,\qquad {\mbf p}=-i\frac{g}{4 \pi T}{\mbf V}.
\end{equation}
Expressing now the equilibrium expectation value
$\vev{\overline{\mathbb V}_1}_E$  through the
partition function \rf{pfunction}  
\begin{equation}	
<\overline{{\mathbb V_1}}>_{E}=\frac{2}{3}T\partial_\kappa\log 
{\overline Z}(\kappa,{\mbf V}),
\end{equation}	
and using \rf{7.39} we obtain
\begin{equation}	
\vev{{\mathbb V}_1}_N=\frac{2}{3}T\partial_\kappa\log 
A^{(vac)}_3(t,{\mbf p}). \label{expval1}
\end{equation}	
In a similar way one can show that 
\begin{equation}	
\vev{{\mathbb V}_2}_N=\frac{2}{3}T\partial_\kappa\log 
A^{(vac)}_1(t,{\mbf p}),\qquad 
\vev{{\mathbb V}_3}_N=\frac{2}{3}T\partial_\kappa\log 
A^{(vac)}_2(t,{\mbf p}). \label{expval23}
\end{equation}	

Finally note 
the $g\to1/g$ duality properties of the expectation
values \rf{expval1}, \rf{expval23}. Introduce the following combinations 
expectation values 
\begin{equation}	
J_\a={i\pi \kappa}\,\vev{\,{\mathbb V}_{\s-1}}_N-
{i\pi \kappa}\,\vev{{\mathbb V}_\s}_N,\qquad \a=1,2,3 \label{jalpha}
\end{equation}
then
\begin{equation}	
J_\a=2\pi i T \,t\partial_t \log \frac{A^{(vac)}_\b(t,{\mbf p})}
{A^{(vac)}_\g(t,{\mbf p})},
\end{equation}	
where 
$(\a,\b,\g)$ is a cyclic permutation of $(1,2,3)$.
  
According to \rf{Qas} the quantities \rf{jalpha} satisfy the following
``strong  
weak'' duality relations
\begin{equation}	
J_\a(g,{\mbf V},-\kappa)=-{\bf e}_\a{\mbf V}-g^{-1}
J_\a(g^{-1},g{\mbf V},-C(g)(-\kappa)^{-\frac{1}{g}}),\label{JJdual}
\end{equation}	
\begin{equation}
C(g)=\frac{\Gamma(g^{-1})}{\pi}	
\Big(\frac{\Gamma(g)}{\pi}\Big)^{\frac{1}{g}}.
\end{equation}

\subsection{Relation to third order differential equations}

Consider the differential equation
\begin{equation}	
\Big\{\partial_x^3 +\frac{\ell_1\ell_2+\ell_1\ell_3+\ell_2\ell_3-2}{x^2}
\partial_x
- \frac{\ell_1\ell_2\ell_3}{x^3} +x^{3\a}\Big\}\Psi(x)=E\Psi(x), \label{Shro1}
\end{equation}		
where
\begin{equation}	
\ell_1+\ell_2+\ell_3=3.
\end{equation}	
For       
\begin{equation}	
\Re e(\ell_1+3)>\Re e( \ell_2),\qquad \Re e(\ell_1+3)>\Re e( \ell_3),
\label{domain} 
\end{equation}	
the equation \rf{Shro1} has a unique solution satisfying the condition
\begin{equation}	
\psi(x,E,\ell_1,\ell_2,\ell_3)=x^{\ell_1}+O(x^{\ell_1+3}),
\end{equation}	
This solution can be analytically continued outside the domain
\rf{domain}. Obviously  
the functions $\psi(x,E,\ell_2,\ell_1,\ell_3)$ and
$\psi(x,E,\ell_3,\ell_1,\ell_2)$  defined in this way satisfy the same 
equation \rf{Shro1} and for generic $\ell_1,\ell_2,\ell_3$ the solutions 
\begin{equation} 
\psi_1(x)=\psi(x,E,\ell_1,\ell_2,\ell_3),\qquad
\psi_2(x)=\psi(x,E,\ell_2,\ell_1,\ell_3),\qquad
\psi_3(x)=\psi(x,E,\ell_3,\ell_1,\ell_2), \label{basisdiff}
\end{equation}	
are linearly independent since
\begin{equation}	
{\rm Wr} \left[\psi_1(x),\psi_2(x),\psi_3(x)\right]
=(\ell_1-\ell_2)(\ell_2-\ell_3)(\ell_3-\ell_1),
\end{equation}	
where ${\rm Wr}$ denotes the usual Wronskian.

Further, for all $E$ the equation \rf{Shro1} has a unique solution $\chi(x,E)$ 
which decays at $x\to+\infty$. We normalise this solution as
\begin{equation}
\chi(x)=x^{-\a}\exp\Big\{-\frac{x^{\a+1}}{\a+1}+O(x^{\a-2})\Big\}.
\end{equation}	
It can be expanded in the basis \rf{basisdiff}
\begin{equation}	
\chi(x,E)=R_1(E)\psi_1+R_2(E)\psi_2+R_3(E)\psi_3. \label{chi}
\end{equation}
It was conjectured in \cite{DT00a} that the connection coefficients $R_i(E)$ 
are simply related to the vacuum eigenvalues of the $\bf Q$-operators
\rf{Adef}.
The exact correspondence is as follows: 
\begin{equation}
{R_i(E)\over R_i(0)}=A_i^{(vac)}(t,{\mbf p}),\qquad i=1,2,3 \label{7.62}
\end{equation}	
where
\begin{equation}	 
\a={1\over g}-1
,\qquad E=i\,\rho\, t,\qquad \ell_i=-\frac{3({\mbf p}{\mbf w}_i)}{g}+1,
\end{equation}	
with the parameter $\rho$ and the vectors ${\mbf w}_i$ defined in \rf{edef}
and \rf{rho} respectively. The conjecture \rf{7.62} was supported in
\cite{DT00a} by extensive numerical work, but no analytical proof is
currently known (see Section~8.3 below for a discussion).

\begin{table}[h]
\renewcommand{\baselinestretch}{1.0}
\normalsize
\centering
\small
\begin{tabular}{| c|rr||rr| }
\hline
$n$& $E^{(1)}_n$ \cite{DT00a}&$E^{(1)}_n$ (AS)&$E^{(3)}_n$ \cite{DT00a}& $E^{(3)}_n$
(AS)\\
\hline 
1\,    & \,2.88682816& 2.8787210&  0.890944821   &   0.66892875 \\
2\,    & \,5.15225193& 5.1513633&  3.643901890   &   3.64527385 \\
3\,    & \,7.13037330& 7.1300266&  5.817136367   &   5.81718647 \\
4\,    & \,8.94036211& 8.9401849&  7.737709470   &   7.73772076 \\
5\,    & \,10.6356087&10.6355029&  9.508485338   &   9.50848916 \\ 
6\,    & \,12.2451641&12.2450946&  11.17453999   &  11.17454166 \\
7\,    & \,13.7870634&13.7870146&  12.76111150   &  12.76111235 \\
8\,    & \,15.2734900&15.2734540&  14.28420813   &  14.28420862 \\
9\,    & \,16.7131734&16.7131460&  15.75481968   &  15.75481998 \\
\hline
\end{tabular}
\caption{\small 
Numerical values for $E^{(1)}_n$ and $E^{(3)}_n$ from Table 2 
in \cite{DT00a}, truncated to 8 decimal places, compared with
$E^{(1)}_n$(AS) and $E^{(3)}_n$(AS), obtained from our
asymptotic formula, \rf{Easspecial}. \label{Diffeqn}}
\vspace{-0.1cm}
\end{table}
\renewcommand{\baselinestretch}{1.2}
\normalsize

Note that, due to the correspondence \rf{7.62}, the formulae \rf{Qas}
provide asymptotic expansions for the zeroes of the coefficients 
$R_i(E)$. 
As an example consider the following  particular case of \rf{Shro1},
\begin{equation}	
\partial_x^3\Psi(x) 
+x\ \Psi(x)=E\Psi(x),
\end{equation}		
which corresponds to 
\begin{equation}	
\a=1/3, \qquad \ell_1=0,\qquad \ell_2=1,\qquad \ell_3=2,
\end{equation}	
and 
\begin{equation}	
g=3/4, \qquad ({\mbf p}{\mbf w}_1)=1/4, \qquad
({\mbf p}{\mbf w}_2)=0,\qquad ({\mbf p}{\mbf
w}_3)=-1/4. \label{valuesofp}
\end{equation}	
The zeroes $R_1(E)$ and $R_3(E)$ in this case have been extensively
studied numerically both from the solutions of the differential
equation as well as from the non-linear integral equation
\cite{KBP91, DdV95, DT00a}
which is equivalent to the Bethe Ansatz equations \rf{BAE},
\rf{phases}
 for the vacuum 
eigenvalues $A_i^{(vac)}(t)$. 
We have  compared this numerical data against the asymptotic
expansion \rf{asseries}. For $g=3/4$ the expressions \rf{series1},
\rf{series2} allow us to derive only  
one non-trivial term in \rf{asseries1}, \rf{asseries2} (note that the first dual nonlocal IM also
contribute). For the values of ${\mbf p}$ given in \rf{valuesofp} the 
 corresponding formulae read 
\begin{equation}	
\begin{array}{rcl}
E^{(1)}_n&=&\Big({\ds\frac{8\pi}{3\sqrt{3}}}\Big)^{\frac{3}{4}}\ 
(n-\frac{1}{6})^{3/4}\ \Big[1+\frac{21}{256\pi^2}
(n-\frac{1}{6})^{-2}+O(n^{-4})\Big],\\
E^{(3)}_n&=&\Big({\ds\frac{8\pi}{3\sqrt{3}}}\Big)^{\frac{3}{4}}\ 
(n-\frac{5}{6})^{3/4}\ \Big[1-\frac{15}{256\pi^2}
(n-\frac{5}{6})^{-2}+O(n^{-4})\Big].
\end{array}	\label{Easspecial}
\end{equation}	
The numerical values are presented in Table~\ref{Diffeqn}.

As shown in \cite{BLZsdet} the correspondence  \rf{7.62} allows us to derive 
a certain duality relation for the vacuum eigenvalues $A_i^{(vac)}(t,
{\mbf p})$. 
It is convenient to make  a change of variables $\Psi(x)=e^y\tilde\Psi(y)$, 
$x=e^y$ which brings the equation \rf{Shro1} to the form
\begin{equation}	
\Big\{\partial_y^3 - (\nu_1\nu_2+\nu_1\nu_3+\nu_2\nu_3)\partial_y
+ i\nu_1\nu_2\nu_3 +e^{3(\alpha+1)y}-Ee^{3y}\Big\}\tilde\Psi(y)=0,
\label{Shro2} 
\end{equation}	
where 
\begin{equation}	
\nu_j=-i(\ell_j-1)=\frac{3i({\mbf p}{\mbf w}_j)}{g}.
\end{equation}	
It is easy to see that the change of variables 
\begin{equation}	
y=g y'+y_0,\qquad e^{-3y_0}=-g^3 E,
\end{equation}	
transform \rf{Shro2} to an equation of the same form but with the
parameters $g$, $\{\nu_i\}$ and replaced by
\begin{equation}	
g\to g^{-1}, \qquad \nu_i=g\nu_i,\qquad
E\to-g^3g^{-\frac{3}{g}}(-E)^{-\frac{1}{g}}. 
\end{equation}	
Using this fact in \rf{chi} one obtains the following duality 
relation for the connection coefficients $R_i$
\begin{equation}	
\frac{e^{iy_0\nu_\a}\,R_\a(g,\nu_1,\nu_2,\nu_3,E)}
{e^{iy_0\nu_\b}\,R_\b(g,\nu_1,\nu_2,\nu_3,E)}=
\frac{R_\a(g^{-1},g\nu_1,g\nu_2,g\nu_3,-c(g)\,(-E)^{-\frac{1}{g}})}
{R_\b(g^{-1},g\nu_1,g\nu_2,g\nu_3,-c(g)\,(-E)^{-\frac{1}{g}})},
\qquad c(g)=g^{3(1-g^{-1})},\qquad 
\end{equation}	
where $\a,\b=1,2,3$. Note that due to \rf{7.62}, this relation is in
agreement with \rf{Qas} and \rf{JJdual}.


\section{Discussion}

In conclusion we make a few remarks on 
 some open problems relevant to the topics of this paper.
\subsection{Truncation of the functional relations} It is well known that
the fusion equation for the transfer matrices in solvable lattice
models truncate when $q^2$ is a root unity. A similar
truncation, of course takes place for the CFT ${\bf T}$-operators.
For the theories  related to the quantum affine $U_q(\widehat{sl}(2))$
this truncation is well understood. For example, consider the CFT ${\bf
T}$-operators ${\bf T}_j(\lambda)$,
$j=0,\frac{1}{2},1,\frac{3}{2}\ldots$, of ref. \cite{BLZ96}, associated with
$2j+1$-dimensional representations of $U_q(\widehat{sl}(2))$. If
$q=e^{i\pi \beta^2}$
is a primitive root
of $1$ or $-1$, such that  $q^N=\pm1$ with some integer $N\ge2$ 
then one can easily show (see, e.g., \cite{BLZ99a}) that 
\begin{equation}
{\bf T}_{{N\over2}}(\lambda)=2\cos(2\pi Np)+{\bf T}_{{N\over2}-1}(\lambda)\ ,
\label{sl2reduc}
\end{equation}
where $p$ and $\beta$ are related to the Virasoro highest weight $\Delta$ and
the central charge $\hat{c}$ as
\begin{equation}
\Delta={p^2\over \beta^2}+{\hat{c}-1\over 24},
\qquad \hat{c}=1-6(\beta-\beta^{-1})^2,
\qquad q=e^{i\pi \beta^2} \label{chat}
\end{equation}
As is shown in \cite{BLZ97a} this allows one to bring the fusion relations
for ${\bf 
T}_j(\lambda)$  to the form
identical to the functional TBA equations  (the $Y$-system) of 
$D_N$ type \cite{ZamTBA, FLS95b}.

Further reduction occurs at particular values of the vacuum parameter $p$
\begin{equation}
{\bf T}_{{N\over2}-{1\over2} }(\lambda) =0, \qquad {\rm when}\qquad  
z^N=\pm1,\qquad z=e^{2\pi i p}
\end{equation}
which corresponds to the RSOS reduction in the six-vertex lattice
model \cite{BR89}. 
This latter RSOS-type reduction is also well known for
the case of $U_q(\widehat{sl}(3))$ \cite{BR90}. For example, it
readily follows from 
the determinant formulae \rf{aweyl}, \rf{bweyl}: when the vacuum
parameters $z_i$ defined 
in \rf{zdef} 
become roots of unity $z_i^N=1$, $i=1,2,3$ along with $q^N=1$ and
either $\mu_1+2=N$ or $\mu_2+1=N$ the determinants \rf{aweyl},
\rf{bweyl} vanish. However, no 
simple analogues of the reduction \rf{sl2reduc} valid for arbitrary
values of $z_i$ 
is known (except for the particular case of Sect~\ref{sect632} which
accidentally 
reduces to $U_q(\widehat{sl}(2))$). Similar remarks apply to related
problems such as the string structure of the Bethe Ansatz equations and the
TBA equations. In particular, no  $U_q(\widehat{sl}(3))$ analogues of the
$D_N$ type TBA equations are known.
\subsection{Connection with  $U_q(A_2^{(2)})$ related IQFT}
The set of classical IM \rf{locintintro} admits consistent reductions 
 $V(u)=0$ or
$V(u)=U'(u)/2$
causing all even IM $I_{k}^{(cl)}$, $k=2,4\pmod 6$ to vanish.
Remarkably, the non-vanishing  IM $I^{(cl)}_k$, $k=1,5 \pmod 6$
remain in involution \cite{KM89}
with respect to Virasoro Poisson bracket of the field
$U(u)$. 
The quantization of these conserved quantities
leads to new sets  (with respect to the quantum KdV case
\cite{SY88,EY89,Zam89b,BLZ96}) of 
local IM in $c<1$ CFT. Actually, there are 
two new non-equivalent sets of IM related to the 
different reductions mentioned above.
This structure of the local IM  naturally arises in the conformal limit
of the massive IQFT associated with the twisted quantum affine $U_q(A_2^{(2)})$
\cite{Zam89b, Smi91}. The universal CFT ${L}$-operator for this case reads 
\begin{equation}
{\cal L}=e^{i\pi Ph}{\cal P} \exp\Big\{\int_0^{2\pi}
\Big(:e^{-2\phi(u)}: 
y_0+:e^{\phi(u)}: y_1 \Big)  du\Big\},
\end{equation}
where $\phi(u)$ is the chiral Bose field normalised exactly as in
\cite{BLZ96} and $h$, $y_0$ and $y_1$ are the generators of the Borel
subalgebra of $U_q(A_2^{(2)})$ where $q$ is related to the Virasoro central
charge $\hat{c}$ as in \rf{chat}. In ref. \cite{FRS96} Fioravanti,
Ravanini and Stanishkov 
constructed an infinite set of finite-dimensional 
representations $\pi_s(t)$, $s=0,1,2,\ldots$, of this subalgebra 
(with the dimensions $\dim[\pi_s(t)]=(s+1)(s+2)/2$) and defined a
set of ${\bf T}$-operators 
\begin{equation}
{\bf T}_s(t)={\rm Tr}_{\pi_s(t)} [e^{i\pi Ph}{\cal L}], \qquad
s=0,1,2,\ldots \label{TFRS} 
\end{equation}
Here we use the spectral parameter $t$ which is related to $\lambda$
of \cite{FRS96} as $t=\lambda^3$. These operators act in the Virasoro
highest weight module 
${\cal V}_\Delta$ with $\Delta$ defined in \rf{chat}; 
they are entire functions of $t$ and 
their leading asymptotics at large $t$ can be easily estimated 
\begin{equation}   
\log {\bf T}_s(t)\sim m_s \ t^{2/(6-3\beta^2)}, \qquad |t|\to \infty
\end{equation}
where $m_s$ are some numerical constants.
Further, at small $t$ one has
\begin{equation}
{\bf T}_1(t)=1+2\cos(2\pi p)+{\bf G}_1\  t+ O(t^2), \label{TFRSas}
\end{equation}
where the vacuum eigenvalue ${\bf G}_1$ in ${\cal V}_\Delta$ can be 
explicitly calculated
$$
G_1^{(vac)}\vert_{A^{(2)}_2}=2\pi\ 
e^{-2\pi i p}\ 
\int_0^{2\pi}\,du_1\int_0^{u_1}\,du_2{\displaystyle
e^{ip(u_1+u_2)}\Big(2\sin\big(\frac{u_1-u_2}{2}\big)\Big)^{\frac{\b^2}{2}}
\over\displaystyle
\Big(2\sin\frac{u_1}{2}\Big)^{\b^2}\Big(2\sin\frac{u_2}{2}\Big)^{\b^2}} 
$$
\bn
={4\pi^3\Gamma(1+\b^2/2)\Gamma(1-3\b^2/4)\Gamma(1-\b^2)\over
\Gamma(1+\beta^2/4)\Gamma(1+p-\b^2/2)\Gamma(1-p-\b^2/2)\Gamma(1+p-\b^2/4)
\Gamma(1-p-\b^2/4)}.\label{G1A22int}
\ed
In \cite{FRS96} it was shown\footnote{Although the calculations in
\cite{FRS96}  
have been performed in the
lowest non-trivial order in $t$ the relation \rf{a22FE} is expected to hold
in all orders in $t$.} that these operators satisfy the functional 
relation \cite{KS95} 
\begin{equation}
{\bf T}_s(tq^{1/2}){\bf T}_s(tq^{-1/2})={\bf T}_s(-t)+
{\bf T}_{s-1}(t){\bf T}_{s+1}(t),\qquad q=e^{i\pi \beta^2} \label{a22FE}
\end{equation}

Now turn back to the ${\cal W}_3$ theory. Consider the ${\cal W}_3$
module ${\cal V}_{\Delta_2,0}$, with the weight $\Delta_3=0$.
Let ${\cal V}_{sym}$ be a subspace ${\cal V}_{sym}\in{\cal V}_{\Delta_2,0}$
of this module
invariant with respect to the action of the automorphism $\sigma_{13}$
defined in Sect.~\ref{symmetry}
Then from the definitions \rf{Tquant}, \rf{Tquant2}   and
the relations \rf{eval2}, \rf{Laction}  one concludes that the operators
${\mathbb T}_m(t)$ and $\overline{\mathbb T}_m(t)$ from  \rf{T}, \rf{Tbar} 
obey the relation 
\begin{equation}
{\mathbb T}_m(t)=\overline{\mathbb T}_m(-t)\label{symrel}
\end{equation}
when acting in the subspace ${\cal V}_{sym}$. It follows then from \rf{Tas2}
and the discussion there that all even local IM ${\bf I}_k$,
$k=2,4 \pmod 6$, vanish in this subspace. Denote by ${\mathbb
T}_m(t|g,p))$ the eigenvalues ${\mathbb T}_m(t)$ in ${\cal
V}_{sym}\in{\cal V}_{\Delta_2,0}$ 
where $g$ is related to the central charge $c$ of the ${\cal W}_3$
algebra \rf{gdef}  
and $p$ parameterises the weight $\Delta_2=p^2/g+(c-2)/24$.
Taking into account \rf{symrel}
the functional equations \rf{fusrel} become equivalent to \rf{a22FE} provided 
one makes a replacement $q\to q^{\frac{1}{2}}$. Thus we have good reasons 
to expect that with a suitable 
redefinition of parameters the above eigenvalues ${\mathbb
T}_m(t|g,p))$ of  ${\mathbb T}_m(t)$  are identical to the eigenvalues 
$T_s(t|\beta^2,p)$ of the operators \rf{TFRS} acting in the Virasoro module
${\cal V}_\Delta(\hat{c})$ with $p$ and $\hat{c}$ already defined in \rf{chat}.
The exact correspondence is
\begin{equation} 
T_s(t|\beta^2,p)={\mathbb T}_s(\rho_1 t|{\beta^2/2},p), \qquad s=0,1,2,\ldots
\end{equation}
where the factor 
\begin{equation} 
\rho_1={\Gamma(1-\frac{\b^2}{4})\ \Gamma(1+\frac{\b^2}{2})\ \Gamma(1-{\b^2})
\over\Gamma(1+\frac{\b^2}{4})\ \Gamma^2(1-\frac{\b^2}{2})},  
\label{rho1} 
\end{equation}
is calculated from the comparison of \rf{G1A22int} with
\rf{G1vacspec}. Further, the   
subspace ${\cal V}_{sym}$ should support the action of the Virasoro
algebra with the central charge $\hat{c}$. It would be interesting to 
obtain an explicit realization of this Virasoro algebra in terms of
the generators of the ${\cal W}_3$ algebra. It would also be
interesting to obtain an explicit construction of the ${\bf Q}$ operators 
for the $U_q(A_2^{(2)})$-related ${\bf T}$ operators \rf{TFRS}.

\subsection{Concluding remarks}

The study of the integrable structure of CFT in
\cite{BLZ96,BLZ97a,BLZ99a}
and in this paper once again demonstrates the fundamental role of Baxter's
${\bf Q}$-operators in the theory of integrable quantum systems. 
Some further applications of ${\bf Q}$-operators in CFT can be found in 
\cite{Smi00}.

Here we adopted the approach of \cite{BLZ97a,BLZ99a} to construct 
the ${\bf Q}$-operators as traces of certain quantum monodromy
matrices associated with infinite-dimensional representations of 
the $q$-oscillator algebra. As remarked in \cite{BLZ99a} this construction 
is rather general and can be applied to the lattice models as well.
In particular, as the structure of the functional relations considered in 
Section~5 is completely
determined by the decomposition properties of the products of 
representation of the
quantum affine algebra $U_q(\widehat{sl}(3))$, all these functional 
relations remain valid  for the associated lattice models,
 with minor modifications
related to the normalisation conventions of the lattice
transfer matrices. Note that for certain lattice models 
there exists an alternative approach to ${\bf Q}$-operators \cite{BS90}
(based on Baxter's original work \cite{Bax73a}) 
which allows the explicit calculation of their matrix elements in a simple 
product form (see also  \cite{BBP90,BKMS,FK95,Pro99,Smidual,Kas,Hik} 
for further examples). 
It would be interesting to
understand a relationship between the above two approaches to the 
${\bf Q}$-operators.

The conjecture \rf{7.62} of \cite{DT00a} is a generalisation of an earlier 
conjecture of the same authors \cite{DT99b} which established an exact
relation between the spectral determinants of the (second order)
Schr\"odinger operator with homogeneous potentials \cite{Vor92,Vor94}
and the vacuum eigenvalues of the ${\bf Q}$-operators in the  
$c<1$ CFT \cite{BLZ97a}. This earlier conjecture of \cite{DT99b} 
has been extended
for the more general Schr\"odinger operator (which includes the
``angular momentum'' term)  
\bn
 \partial_x^2 \Psi(x)+ \Big\{\, 
E-x^{2\alpha}-{l (l+1)\over x^2}\, \Big\}
\Psi(x)=0\ \label{difsl2}
\ed
and proved in \cite{BLZsdet}. Here the parameters $l$ and $\alpha$ are
related to the  Virasoro vacuum parameter $p$ and central charge $\hat{c}$ 
defined in \rf{chat}  as $l=2p/\beta^2-1/2$ and $\alpha=1/\beta^2-1$.
The most essential 
ingredients of this proof can be summarized as follows: 
\begin{enumerate}[(i)]
\item {\it The analytic properties (both in the spectral parameter $E$ and the 
angular momentum $l$).} For the spectral determinants of \rf{difsl2}
these properties are deduced from the standard  
theory of the Schr\"odinger equations (see, e.g., \cite{newton}). 
On the other hand, in \cite{BLZ99a} and \cite{BLZnon}
the same analytic properties for the vacuum eigenvalues of the ${\bf
Q}$-operators  were establish in $c<1$ 
CFT. Note that the analysis of these analytic properties 
in the variable $p$ in \cite{BLZnon} (corresponding to  $l$ in \rf{difsl2}) 
was based on the ``real time perturbation theory'' results
(analogous to the representations \rf{expval} of this paper).
\item {\it The quasiclassical approximation.} The same 
real time perturbation theory results were used in \cite{BLZnon} 
to obtain the leading 
large $p$ asymptotics of the coefficient of the expansion of the 
eigenvalues of the ${\bf Q}$-operators in the spectral parameter. 
These asymptotics exactly matched \cite{BLZsdet} 
 the corresponding asymptotics for the 
spectral determinants of \rf{difsl2} calculated from the standard 
quasiclassical approximation to   \rf{difsl2} for large values of $l$.
\item {\it Quantum Wronskian relation and the Riemann-Hilbert problem}.
The final step of the proof of \cite{BLZsdet} was based on the key
observation of \cite{DT99b} that 
the functional relation for spectral determinants of the Schr\"odinger
equation obtained in \cite{Vor92,Vor94} (see also \cite{BLZsdet}
for the case of general $l$) exactly coincide with the quantum 
Wronskian relation for the eigenvalues of the ${\bf Q}$-operators  
in $c<1$ CFT obtained in \cite{BLZ97a}. In fact, it was shown 
\cite{BLZnon} that the above properties (i) and (ii) transform this 
functional relation to a simple Riemann-Hilbert problem which has a
unique solution.
\end{enumerate}
It should be noted that no additional assumptions (like the 
phase assignment in the Bethe Ansatz equations for vacuum eigenvalues, 
similar to \rf{phases}) were used in \cite{BLZsdet}.

It would be interesting to see whether is it possible to modify
the above arguments to prove the conjecture \rf{7.62} related to 
the third order differential equation \rf{Shro1}. It seems that there no
difficulties in generalising the above properties (i) and (ii).
In particular, we have verified that the asymptotics of the connection
coefficients $R_i(E)$ in \rf{chi} obtained from the quasiclassical
approximation to the differential equation \rf{Shro1} for large values
of $\ell_i$ exactly match the ${\mbf p}\to\infty$ asymptotics 
of the eigenvalues of the ${\bf Q}$-operators obtained from  
\rf{Aint} and \rf{psiq}. It is also easy to prove that the connection
coefficients $R_i(E)/R_i(0)$ satisfy the same quantum Wronskian
relation \rf{qwron}. However, it is  currently not clear how transform
this relation to a tractable 
Riemann-Hilbert problem which now should involve two complex ``angular
momentum'' variables.  

\section*{Acknowledgements}

VB thanks S.L.Lukyanov and A.B.Zamolodchikov for useful discussions.
The work of S.Khoroshkin was supported by
RFBR grant  01-01-00539, grant for the support of
scientific schools  00-15-96557, and INTAS 99-1705.

\def\qi{q^{h_i}}
\def\qqi{q^{-h_i}}
\app{Universal $R$-matrix}
Here we give a short proof of Proposition 1, which allows us to
 derive the basic properties of the universal $L$-operator \rf{Loper}.

Due  to \rf {3.6} and \rf{1} the basic property
\rf{comult}  of the universal $R$-matrix
implies the following relations:
\bn
(1\otimes y_i+y_i\otimes q^{-h_i}){\overline{\mathcal R}}=
{\overline{\mathcal R}}(1\otimes y_i+y_i\otimes q^{h_i}),
\label{app11}
\ed
that is
\bn
[1\otimes y_i, {\overline{\mathcal R}}]=
{\overline{\mathcal R}}(y_i\otimes q^{h_i})-(y_i\otimes
q^{-h_i}) {\overline{\mathcal R}}.
\label{app12}
\ed
Sum up \rf{app12} over $i$  and put $\overline{I}=\sum_i y_i$. We have
\bn
[\overline{I},{\overline{\mathcal R}}]=
{\overline{\mathcal R}}(\sum_{i=1}^3y_i\otimes q^{h_i})
-(\sum_{i=1}^3y_i\otimes q^{-h_i}){\overline{\mathcal R}}.
\label{app13}
\ed
Let us imagine now that we check equality (\ref{app13}) directly.
For this we compute the commutator on the left hand side using the relations
\bn
[y_i,x_j]=\delta_{i,j}\frac{q^{h_i}-q^{-h_i}}{q-q^{-1}}.
\label{app14}
\ed
After application of the Leibnitz rule we have certain series over
 $y_i\otimes 1$ and $1\otimes x_i$ with $q^{h_i}$ and $\qqi$ inserted somewhere
 in the right tensor space.

We move all $\qqi$ to the left and all $\qi$ to the right using the relations
\bn
\qi x_j=q^{-a_{ij}}x_j\qi,\qquad
\qqi x_j=q^{a_{ij}}x_j\qqi.
\label{app15}
\ed
The result should coincide with right hand side of (\ref{app13}).

Suppose now that we have some $I$, $A_i$ and $B_i$,
satisfying the conditions \rf{cond1}, \rf{cond2} of Proposition 1, 
and we try to compute $[I\otimes 1,{\overline{\mathcal R}}]$.
Again, we first use \rf{cond1}, to get
 monomials over $y_i\otimes 1$ and $1\otimes x_i$ with $A_i$ and $B_i$
inserted in the second tensor component, then move $A_i$ to the left and $B_i$
 to the right by means of \rf{cond2}. Clearly, we should get the same result as
 in the first calculation with $\qi\otimes 1$  replaced by
$-B_i\otimes 1$ and $\qqi\otimes 1$  replaced by
$-A_i\otimes 1$, since
 we used identical commutation relations. This proves \rf{Rprop}.

\app{Functional relations} \label{Functional relations}
\def\kk{{\overline{k}}}
\def\nn{{\overline{n}}}
\def\mm{{\overline{m}}}
\def\eps{{\overline{\varepsilon}}}
\def\ppi{{\tilde{\pi}}}
\def\ZZ{{\mathbb Z}}
\def\DDelta{\overline{\Delta}}
\def\ggamma{(q-q^{-1})}
In this appendix we derive the determinant formulas
\rf{aweyl}, \rf{bweyl} for the operators
 ${\bf T}_\mu(t)$, and
$\To_\mu(t)$ and the functional relations \rf{QQ}, \rf{QQa}.
The main ingredient of the proof is similar to that of \cite{BLZ99a}
and consists of decomposition of the tensor product of certain degenerated
Verma modules into a sum of shifted evaluation Verma modules.

Let $\pi^+_{\mu}(t)$ be the evaluation Verma module of the algebra
 ${\mathcal B}_+$ with highest weight $\mu=(\mu_1,\mu_2,\mu_3)$
with respect to evaluation map \rf{eval1}. We fix the basis
 $v_{\nn}$, where $\nn\in\ZZ_{\geq0}^3$, that is
${\nn}=(n_1,n_2,n_3)$ and $n_i\geq 0$, $n_i\in\ZZ$,
 which is given
by the action of the generators $F_\a$,$F_\b$ and $F_{\a+\b}$ of $U_q(gl(3))$:
\begin{equation}
v_{\nn}=t^{n_2+\frac{n_1+n_3}{2}}
q^{-\frac{n_2}{2}}F_\a^{n_1}F_{\a+\b}^{n_2} F_\b^{n_3}v_0,
\end{equation}
 where $v_0$ is the vacuum vector. Then the action of the  generators of the
algebra ${\mathcal B}_+$ is given by the relations:

\begin{eqnarray}
\begin{array}{lll}
y_1v_\nn&=&t^{\hf}[n_3]_q[\mb-n_3+1]_qv_{\nn-\eps_3}+
t^{\hf}q^{\mb-2n_3}[n_2]_q
v_{\nn+\eps_1-\eps_2},\\
y_2v_\nn&=& q^{n_3+\mu_1+\mu_3+\frac{1}{2}}v_{\nn+\eps_2},\\
y_3v_{\nn}&=& t^{\hf}[n_1]_q[\ma-n_1-n_2+n_3+1]_qv_{\nn-\eps_1}-
t^{\hf}q^{n_2-n_3-2-\ma}[n_2]_qv_{\nn-\eps_2+\eps_3},
\label{yaction}\\
h_1v_\nn&=&(\mu_\b +n_1-n_2-2n_3)v_\nn,\qquad\\
h_2v_\nn&=&(-\mu_{\a+\b}+n_1+2n_2+n_3)v_\nn,\qquad\\
h_3v_\nn&=&(\mu_\a-2n_1-n_2+n_3)v_\nn,\qquad
\end{array}
\end{eqnarray}
where 
\begin{equation}
\begin{array}{lllllllll}
\mu_\a&=&\mu_1-\mu_2,\qquad &\mu_\b&=&\mu_2-\mu_3,\qquad
 &\mu_{\a+\b}&=&\mu_1-\mu_3,\\
\eps_1&=&(1,0,0),\qquad & \eps_2&=&(0,1,0),\qquad& \eps_3&=&(0,0,1).
\end{array}
\end{equation}

Any weight $\mu$ defines a shift automorphism $p_\mu$ of the algebra
${\mathcal B}_+$: 
\begin{equation}
\begin{array}{llllll}
p_\mu(y_i)&=&y_i,\qquad &p_\mu(h_1)&=&h_1-\mu_2+\mu_3,\\
p_\mu(h_2)&=&h_2+\mu_1-\mu_3,\qquad &p_\mu(h_3)&=&h_2-\mu_1+\mu_2.
\end{array}
\end{equation}
 For a representation $\pi$ of ${\mathcal B}_+$  denote by $\pi[\mu]$
the shifted representation $\pi\cdot p_\mu$. In particular, denote by
 $\ppi_\mu^+(t)$ the shifted evaluation Verma module
$\pi_\mu^+(t)[\mu]$. In this module the action of generators
$y_i$ is given by \rf{yaction} while the  vacuum vector is of
 zero weight and
\begin{equation}
\begin{array}{c}
h_1v_\nn=(n_1-n_2-2n_3)v_\nn,\;\qquad
h_2v_\nn=(n_1+2n_2+n_3)v_\nn,\;\\
h_3v_\nn=(n_3-2n_1-n_2)v_\nn,
\label{C8}
\end{array}
\end{equation}
so  the traces are related by
\begin{eqnarray}
{\rm Tr}_{\pi_\mu^+(t)}\, e^{i\pi{\bf P\,h}} {\mathcal L}
=z_1^{\mu_1}z_2^{\mu_2}z_3^{\mu_3}
{\rm Tr}_{\ppi_\mu^+(t)}\, e^{i\pi{\bf P\,h}} {\mathcal L}.
\label{trcon}
\end{eqnarray}
Let us  define three degenerations $\ppi^i(t)$, $i=1,2,3$ of
 shifted evaluation Verma module $\ppi_{\mu}^+(t)$. They correspond to
the limit $q\to +\infty$ in three
different
 asymptotic zones
 \begin{equation}
\ma\gg 0,\, \mb\gg 0;\qquad \,\ma\ll 0,\,\mb\gg 0\,\qquad{\rm and}
 \qquad
\ma\ll 0,\,\mb\ll 0\,
\end{equation}
 under assumptions that all the numbers are real.

 All the degenerated Verma modules  $\ppi^i(t)$ are equipped with bases
$v_\nn$,  $\nn_i\in\ZZ_{\geq0}^3$, on which the Cartan generators act as in
 \rf{C8} and
\begin{equation}
\begin{array}{lll}
y_1v_\nn&=&\ggamma^{-1}q^{x-n_3+1}[n_3]_q
v_{\nn-\eps_3}+q^{x-2n_3}[n_2]_qv_{\nn+\eps_1-\eps_2},\\
y_2v_\nn&=&q^{n_3+\frac{1}{2}}v_{\nn+\eps_2},\\
\label{M0}
y_3v_\nn&=&
\ggamma^{-1}q^{x-n_1-n_2+n_3+1}[n_1]_qv_{\nn-\eps_1},
\end{array}
\end{equation}
 for the module  $\ppi^1(t)$, where we use the notation $t=q^{2x}\,$;
\begin{equation}
\begin{array}{lll}
y_1v_\nn&=&\ggamma^{-1}q^{x-n_3+1}[n_3]_q
v_{\nn-\eps_3}+q^{x-2n_3}[n_2]_qv_{\nn+\eps_1-\eps_2},\\
y_2v_\nn&=&q^{n_3+\frac{1}{2}}v_{\nn+\eps_2},\\
\label{M1}
y_3v_\nn&=&-\ggamma^{-1}
q^{x+n_1+n_2-n_3-1}[n_1]_qv_{\nn-\eps_1}-
q^{x+n_2-n_3-2}[n_2]_qv_{\nn-\eps_2+\eps_3},
\end{array}
\end{equation}
 for the module  $\ppi^2(t)$ and
\begin{equation}
\begin{array}{lll}
y_1v_\nn&=&-\ggamma^{-1}q^{x+n_3-1}[n_3]_qv_{\nn-\eps_3},\\
y_2v_\nn&=&q^{n_3+\frac{1}{2}}v_{\nn+\eps_2},\\
\label{M2}
y_3v_\nn&=&
-\ggamma^{-1}q^{x+n_1+n_2-n_3-1}[n_1]_qv_{\nn-\eps_1}-
q^{x+n_2-n_3-2}[n_2]_qv_{\nn-\eps_2+\eps_3},
\end{array}
\end{equation}
 for the module  $\ppi^3(t)$.

The determinant formula \rf{aweyl}   is based on the decomposition of the
 tensor product
\begin{eqnarray}
\ppi^1(tq^{2\mu_1})\otimes\ppi^2(tq^{2\mu_2})
\otimes\ppi^2(tq^{2\mu_3}).
\label{tprod0}
\end{eqnarray}
The claim is
 that this tensor product admits a filtration,  labelled
 by six arbitrary nonnegative integers $k_1,k_2$, $l_1,l_2$, $m_1,m_2$
with factormodules, isomorphic to
$$\ppi^+_\mu(t)[(k_1+k_2){ \alpha}+(l_1+l_2){\b}+
(m_1+m_2){ (\a+\b)}],$$
where
$\alpha=(1,-1,0),$  $\b=(0,1,-1)$ are simple roots for the Lie algebra $sl(3)$.

On the level of characters this means an equality of the form
\begin{eqnarray}
\begin{array}{c}
\ppi^1(tq^{2\mu_1})\otimes\ppi^2(tq^{2\mu_2})
\otimes\ppi^2(tq^{2\mu_3})=\\
\sum_{{k_i,l_i,m_i\geq 0}}
\ppi^+_\mu(t)[(k_1+k_2+m_1+m_2){\alpha}+(l_1+l_2+m_1+m_2){\b}],
\label{MMM}
\end{array}
\end{eqnarray}
or, if we put $ \widetilde{\A}_i(t)=
{\rm Tr}_{\ppi^i(t)}\, e^{i\pi{\bf P\,h}} {\mathcal L}$,
which converges in a region \rf{Pdom},
 \begin{eqnarray}
 \widetilde{\A}_1(tq^{2\mu_1})
 \widetilde{\A}_2(tq^{2\mu_2}) \widetilde{\A}_3(tq^{2\mu_3})=
\DDelta^{-2}
{\rm Tr}_{\ppi_\mu^+(t)}\, e^{i\pi{\bf P\,h}} {\mathcal L},
\label{tr1}
\end{eqnarray}
where
\begin{eqnarray}
\DDelta=(1-{z_2}/{z_1})(1-{z_3}/{z_1})(1-{z_3}/{z_2}).
\label{Ddelta}
\end{eqnarray}
We construct the decomposition of tensor product \rf{tprod0}
in two steps. First, consider the second order tensor product
$\ppi^1(t_1)\otimes\ppi^2(t_2)$, where we choose
 the basis $v_\nn^\mm$, $\nn,\mm\in\ZZ_{\geq0}^3$
 \begin{eqnarray}
 v_\nn^\mm=\sum_{k_1=0}^{n_1}\sum_{k_2=0}^{n_2}
 {n_1\choose k_1}_q{n_2\choose k_2}_q
 q^{\Omega(\nn,\mm,\kk)}
 \left(t_1/t_2\right)^{k_1/2}
 v_{\mm+\kk}\otimes v_{\nn-\kk},
 \label{vec1}
 \end{eqnarray}
where  $\kk=(k_1,k_2,0)$,
\begin{equation}
(n)_q=\frac{q^{2n}-1}{q^2-1},\qquad 
 {n\choose k}_q=\frac{(n)_q!}{(k)_q!(n-q)_q!},
\end{equation}
 and the phase $\Omega(\nn,\mm,\kk)$ is
 \begin{equation}
\Omega(\nn,\mm,\kk)=(n_1-k_1)k_2+(k_2-k_1)m_3+k_2m_2.
\end{equation}
  
Define the following partial order  in $\ZZ_{\geq0}^3$:
$\mm'\leq\mm$ if
 \begin{equation}
{m'}_3\leq {m}_3, \;{m'}_2+{m'}_3\leq {m}_2+{m}_3\quad {\rm and}
\quad
{m'}_1+{m'}_2+{m'}_3\leq {m}_1+{m}_2+{m}_3.
\end{equation}
 Let $V^\mm$ be 
${\mathcal B}_+$ submodules of
$\ppi^1(t_1)\otimes\ppi^2(t_2)$,
 generated by the free action of the $y_i$, $i=1,2,3,$
\begin{eqnarray}
V^\mm=\sum_{\mm'\leq\mm}{\mathcal B}_+(v_{\mm'}\otimes v_0).
\end{eqnarray}
Then the vectors $v_\nn^\mm$, $\nn\in\ZZ_{\geq 0}^3$ form a basis of
 the factormodule
\begin{equation}
\widetilde{V}^\mm=V^\mm/\sum_{\mm'\leq\mm}^{\mm'\not=\mm}V^{\mm'}
\end{equation}
and the action of the generators of ${\mathcal B}_+$ in $\widetilde{V}^\mm$
is given by
\begin{equation}
\begin{array}{lll}
y_1v_\nn^\mm&=&\ggamma^{-1}q^{x_2-n_3+1}[n_3]_q
v_{\nn-\eps_3}^\mm+q^{x_2-2n_3}[n_2]_qv_{\nn+\eps_1-\eps_2}^\mm,\\
y_2v_\nn^\mm&=&q^{n_3+\frac{1}{2}}v_{\nn+\eps_2}^\mm,\\
y_3v_\nn^\mm&=&
q^{x_1}[x_1-x_2-n_1-n_2+n_3+1]_q
[n_1]_qv_{\nn-\eps_1}^\mm -\\
&&-q^{x_2+n_2-n_3-2}[n_2]_qv_{\nn-\eps_2+\eps_3}^\mm,\\
h_1v_\nn^\mm&=&(n_1-n_2-2n_3+m_1-m_2-2m_3)v_\nn^\mm,\qquad\\
h_2v_\nn^\mm&=&(n_1+2n_2+n_3+m_1+2m_2+m_3)v_\nn^\mm,\qquad\\
h_3v_\nn^\mm&=&(-2n_1-n_2+n_3-2m_1-m_2+m_3)v_\nn^\mm.\qquad
\end{array}
\label{V12}
\end{equation}
where $t_1=q^{2x_1}$, $\, t_2=q^{2x_2}$.

Let $\ppi^{12}(t_1,t_2)$ be the following representation of  ${\mathcal B}_+$
with a basis $v_\nn$, $\nn\in\ZZ_{\geq 0}^3$:
\begin{equation}
\begin{array}{lll}
y_1v_\nn&=&\ggamma^{-1}q^{x_2-n_3+1}[n_3]_q
v_{\nn-\eps_3}+q^{x_2-2n_3}[n_2]_qv_{\nn+\eps_1-\eps_2},\\
y_2v_\nn&=&q^{n_3+\frac{1}{2}}v_{\nn+\eps_2},\\
y_3v_\nn&=&
q^{x_1}[x_1-x_2+1-n_1-n_2+n_3]_q
[n_1]_qv_{\nn-\eps_1} -q^{x_2+n_2-n_3-2}[n_2]_qv_{\nn-\eps_2+\eps_3},
\end{array}
\label{M12}
\end{equation}
(the Cartan generators act as in \rf{C8}).
 Then, the relations \rf{V12} mean that the tensor product
$\ppi^1(t_1)\otimes\ppi^2(t_2)$ admits a filtration with factormodules
 isomorphic to
 $$\ppi^{12}(t_1,t_2)[(m_1+m_2)\a+(m_2+m_3)\b],\qquad m_i\geq0$$
and
\begin{eqnarray}
 \widetilde{\A}_1(t_1)
 \widetilde{\A}_2(t_2) =
\DDelta^{-1}
{\rm Tr}_{\ppi^{12}(t_1,t_2)}\, e^{i\pi{\bf P\,h}} {\mathcal L}.
\label{A1A2}
\end{eqnarray}
Next, we decompose the tensor product
$\ppi^{1,2}(t_1,t_2)\otimes\ppi^3(t_3)$. Here we use the basis
 $v_\nn^\mm$, $\nn,\mm\in\ZZ_{\geq0}^3$,
\begin{equation}
\begin{array}{lll}
v_\nn^\mm&=&\ds\sum_{j=0}^{{\rm min }(n_1,n_3)}\sum_{k_2=0}^{n_2}
\sum_{k_3=0}^{n_3-j}
(-1)^j (j)_q!{n_2\choose k_2}_q{n_3-j\choose k_3}_q{n_3\choose j}_q
{n_1\choose j}_q\times\\
&&\\\ds
&& \ds \times q^{\Omega_j(\nn,\mm,\kk)}
\left(t_3/t_2\right)^{k_3/2}v_{\nn-\kk-j(\eps_1+\eps_3)}\otimes
v_{\mm+\kk+j\eps_2},
\end{array}
\label{basis}
\end{equation}
where 
 $\kk=(0,k_2,k_3)$ and  the phase $\Omega_j(\nn,\mm,\kk)$ is
\begin{eqnarray}
\begin{array}{lll}
\Omega_j(\nn,\mm,\kk)&=&
j(n_2-k_2-n_3+k_3-m_1-2m_2-m_3)+k_2(k_3-n_3-m_1-2m_2+m_3)\\
&&+n_1(2m_1+m_2-m_3)+n_2(m_1+2m_2)+n_3(m_2-m_1+2m_3)+\frac{3j(j+1)}{2}.
\end{array}  \nonumber
\end{eqnarray}
In this case we
define the another partial order  in $\ZZ_{\geq0}^3$:
$\mm'\leq\mm$ if
 $${m'}_1\leq {m}_1,\quad {m'}_1+{m'}_2\leq {m}_1+{m}_2\quad{\rm and}
\quad{m'}_1+{m'}_2+{m'}_3\leq {m}_1+{m}_2+{m}_3.$$
 Let $V^\mm$ be the following
${\mathcal B}_+$ submodules of
$\ppi^{12}(t_1,t_2)\otimes\ppi^3(t_3)$:
\begin{eqnarray}
V^\mm=\sum_{\mm'\leq\mm}{\mathcal B}_+(v_{0}\otimes v_{\mm'}).
\end{eqnarray}
 The vectors $v_\nn^\mm$, $\nn\in\ZZ_{\geq 0}^3$ form a basis of the
 factormodule
\begin{equation}
\widetilde{V}^\mm=V^\mm/\sum_{\mm'\leq\mm}^{\mm'\not=\mm}V^{\mm'}
\end{equation}
in which
\begin{equation}
\begin{array}{lll}
y_1v_\nn^\mm&=&q^{x_3}[x_2-x_3-n_3+1]_q[n_3]_q
v_{\nn-\eps_3}^\mm+q^{x_2-2n_3}[n_2]_qv_{\nn+\eps_1-\eps_2}^\mm,\\
y_2v_\nn^\mm&=&q^{n_3+\frac{1}{2}}v_{\nn+\eps_2}^\mm,\\
y_3v_\nn^\mm&=&
q^{x_1}[x_1-x_2-n_1-n_2+n_3+1]_q
[n_1]_qv_{\nn-\eps_1}^\mm -\\
&&-q^{x_2+n_2-n_3-2}[n_2]_qv_{\nn-\eps_2+\eps_3}^\mm,\\
\end{array}
\label{V123}
\end{equation}
with  $t_i=q^{2x_i}$ 
and the Cartan generators acting as in \rf{V12}. We see that the factormodules
are shifts of the same ${\mathcal B}_+$-module which is isomorphic to
$\ppi_\mu^+(t)$, if we put $t_i=tq^{2\mu_i}$. This in combination with
\rf{A1A2}, proves the main statements \rf{MMM} and \rf{tr1}.

In order to deduce \rf{aweyl} from \rf{tr1} we notice that the degenerated
 evaluation Verma modules $\ppi^i(t_i)$ are not irreducible; each of them
  admits
a filtration indexed by a single nonnegative integer such that all the factors
 are the shifts of oscillator representations:
the representation $\ppi^1(t)$ admits an increasing filtration by submodules
$V^n$, where $V^n$ is generated over ${\mathcal B}_+$ by the vectors
$v_{(0,0,k)}$, $k\leq n$ with factormodules isomorphic to representation
$\rho_1(q^2t)[n\b]$ ,$n\geq0$;
 the representation $\ppi^3(t)$ admits an increasing filtration by submodules
$V^n$, which are generated over ${\mathcal B}_+$ by the vectors
$v_{(k,0,0)}$, $k\leq n$ with factormodules isomorphic to representation
$\rho_3(q^{-2}t)[n\a]$ ,$n\geq0$ and
the representation $\ppi^2(t)$ admits a decreasing filtration by submodules
$V^n$, which are generated over ${\mathcal B}_+$ by the vector
\begin{equation}
\sum_{k=0}^n\frac{[n]_q!}{[k]_q![n-k]_q!}
q^{\frac{-k(k-1)}{2}}(q^{-1}-q)^kv_{(k,n-k,k)}.
\end{equation}
The factormodules are isomorphic to
$\rho_3(t)[n(\a+\b)]$, $n\geq0$.

This means that
\begin{eqnarray}
\begin{array}{ccccc}
\widetilde{\A}_1(t)&=&(1-z_3/z_2)^{-1}
{\rm Tr}_{\rho_1(q^2t)}\, e^{i\pi{\bf P\,h}} {\mathcal L}&=&
\DDelta^{-1}\A_1(q^2t),\\
\widetilde{\A}_2(t)&=&(1-z_3/z_1)^{-1}
{\rm Tr}_{\rho_2(t)}\, e^{i\pi{\bf P\,h}} {\mathcal L}&=&
\DDelta^{-1}\A_2(t),\\
\widetilde{\A}_3(t)&=&(1-z_2/z_1)^{-1}
{\rm Tr}_{\rho_3(q^{-2}t)}\, e^{i\pi{\bf P\,h}} {\mathcal L}&=&
\DDelta^{-1}\A_3(q^{-2}t).
\end{array}
\label{parab}
\end{eqnarray}
The relation \rf{tr1} combined with \rf{parab} gives
\begin{eqnarray}
{\rm Tr}_{\ppi_\mu^+(t)}\, e^{i\pi{\bf P\,h}} {\mathcal L}=
\DDelta^{-1} {\A}_1(tq^{2\mu_1+2})
 {\A}_2(tq^{2\mu_2}) {\A}_3(tq^{2\mu_3-2}).
\label{tr2}
\end{eqnarray}
The determinant formula \rf{aweyl} is a direct corollary of \rf{tr2} and
of the Bernstein-Gel'fand-Gel'fand resolution of finite-dimensional module
$\pi_\mu$, which implies the relation
\begin{eqnarray}
\T_\mu(t)=\sum_{\sigma\in S_3}(-1)^{l(\sigma)}\T^+_{\sigma(\mu+\rho)-\rho}(t),
\label{BGG}
\end{eqnarray}
where $\rho=(1,0,-1)$.

The determinant formula \rf{bweyl} can be deduced in an analogous manner
with the use of the second evaluation map or can be deduced from \rf{aweyl} in
the following way.
 Due to the definition \rf{eval2}, we
conjugate the action of the generators of ${\mathcal B}_+$ in the shifted Verma module
$\ppi_\mu^+(t)$ by $\sigma_{13}$.
Taking into account the fact that the $U_q(gl(3))$-Verma module $\pi_{\mu_1,\mu_2,\mu_3}\cdot{\sigma_{13}}$
is isomorphic to the Verma module $\pi_{-\mu_3,-\mu_2,-\mu_1}$,
 we see that
$$
\overline{\mathbf T}_{\mu}(t)
={\rm Tr}_{\pi_{-\sigma_{13}(\mu)}(-t)
\cdot(1\ot\sigma_{13})}\, e^{i\pi{\bf P\,h}} {\mathcal L}
\qquad {\rm and}\qquad
{\mathbf T}_\mu^+(t)={\rm Tr}_{\pi^+_{-\sigma_{13}(\mu)}(-t)
\cdot(1\ot\sigma_{13})}\,
 e^{i\pi{\bf P\,h}} {\mathcal L},
$$
which means that we can use the statement of \rf{aweyl} for the highest weight
$-\sigma_{13}(\mu)$ combined with automorphism $\sigma _{13}$ of the
algebra ${\mathcal B}_+$. Due to the definition \rf{rhoi} we get the
statement of \rf{bweyl}.

Let us turn to the relations \rf{QQ}, \rf{QQa}. Their proof consists of an
interpretation of both sides of the equalities as of ${\mathcal B}_+$-modules.
 Consider, for instance, the  tensor product
$$\ppi^2(t_2)\otimes \ppi^3(t_3),$$ of degenerated evaluation Verma modules.
Analogously to \rf{M12}, \rf{A1A2} it is isomorphic, on the level of characters, to a
sum 
$$\sum_{m_i\geq 0}\ppi^{23}(t_2,t_3)[(m_1+m_2)\alpha+(m_2+m_3)\beta)],$$
where the module $\ppi^{23}(t_2,t_3)$
is the following representation of  ${\mathcal B}_+$
with a basis $v_\nn$, $\nn\in\ZZ_{\geq 0}^3$:
\begin{eqnarray}
\begin{array}{lll}
y_1v_\nn&=&q^{x_3}[x_2-x_3-n_3+1]_q[n_3]_q
v_{\nn-\eps_3}+q^{x_2-2n_3}[n_2]_qv_{\nn+\eps_1-\eps_2},\\
y_2v_\nn&=&q^{n_3+\frac{1}{2}}v_{\nn+\eps_2},\\
y_3v_\nn&=&
-\ggamma^{-1}q^{x_2+n_1+n_2-n_3-1}
[n_1]_qv_{\nn-\eps_1} -q^{x_2+n_2-n_3-2}[n_2]_qv_{\nn-\eps_2+\eps_3},
\end{array}
\label{M23}
\end{eqnarray}
where $h_i$ act as in \rf{C8}, $t_i=q^{2x_i}$. 
When $t_2=t_3=t$, $\ppi^{23}(t_2,t_3)$ has a submodule
 generated by the vector $v_{\eps_3}$ with a basis $\widetilde{v}_\nn=v_{\nn+\eps_3}$,
 $\nn\in\ZZ_{\geq 0}^3$. One can see from \rf{P0b}, \rf{C8} that   this submodule is isomorphic to
 $\ppi^{23}(q^{-2}t,q^2t)[\beta]$. The factormodule 
$$\ppi^{23}(t,t)/\ppi^{23}(q^{-2}t, q^2 t)[\b]$$
has  a basis $v_{n,m}$, $n,m\geq 0$ with
  the action of ${\mathcal B}_+$ given by
\begin{equation}
\begin{array}{lll}
$$y_1v_{n,m}&=&q^{x-2p}[m]_q
v_{n+1,m-1},\\
y_2v_{n,m}&=&q^{\frac{1}{2}}v_{n,m+1},\\
\label{P0b}
y_3v_{n,m}&=& -\ggamma^{-1}q^{x+n+m-1}[n]_qv_{n-1,m},
\end{array}
\end{equation}
$$h_1v_{n,m}=(n-m)v_{n,m},\qquad
h_2v_{n,m}=(n+2m)v_{n,m},\qquad
h_3v_{n,m}=(-2n-m)v_{n,m}.$$
{}From \rf{P0b} we can check directly that this
factormodule is isomorphic to the representation $\rrho_1^+(q^{-1}t)$, so
$$\DDelta\left(\widetilde{\A}_2(t)\widetilde{\A}_3(t)-
{z_3}/{z_2}\widetilde{\A}_2(q^{-2}t)\widetilde{\A}_3(q^2t)\right)=
(1-z_3/z_1)^{-1}(1-z_2/z_1)^{-1}\overline{{\mathbf A}}_1(q^{-1}t),$$
or
\begin{eqnarray}
{\A}_2(qt){\A}_3(q^{-1}t)-
{z_3}/{z_2}{\A}_2(q^{-1}t){\A}_3(qt)=
(1-z_3/z_2)\overline{{\mathbf A}}_1(t),
\label{QQb}
\end{eqnarray}
which is equivalent to one of the relations \rf{QQ}. The others follow from
\rf{QQb} after application of the symmetries \rf{tauaut} and \rf{sigmaaut}.

\app{The ${\mbf p}\to \infty$ asymptotics of ${\bf \Psi}(\nu)$.} \label{DDV}
In this Appendix we derive the leading asymptotics of the function ${\bf
\Psi}(\nu)$ at large values of ${\mbf p}=(p_1,p_2)$. We assume $p_1$ and
$p_2$ tend to infinity 
while their ratio $p_2/p_1$ is kept fixed
\begin{equation}	
p_1,p_2\to \infty,\qquad p_2/p_1={\rm finite}.
\end{equation}	
First consider the classical case. Writing the formula \rf{T-one} in full
one gets 
\begin{equation}
{\bf T}^{(cl)}(\l)={\bf \Lambda}^{(cl)}_1(\l)+{\bf \Lambda}^{(cl)}_2(\l)+
{\bf \Lambda}^{(cl)}_3(\l),\qquad {\bf \Lambda}^{(cl)}_1(\l){\bf
\Lambda}^{(cl)}_2(\l){\bf \Lambda}^{(cl)}_3(\l)=1,
\end{equation}	
where ${\bf \Lambda}^{(cl)}_j(\l)$, $j=1,2,3$ are the eigenvalues 
of the classical monodromy matrix ${\bf M}(\l)$.
The leading ${\mbf p}\to \infty$ 
contribution to these eigenvalues can be easily calculated directly from
the definition  \rf{Mdef}. According to \rf{cmiura} the potentials 
$U(u)$ and $V(u)$ in \rf{Lthird} in this limit become independent of $u$  
\begin{equation}
U(u)|_{{\mbf p}\to \infty}\sim x_1 x_2+x_1 x_3 + x_2 x_3,\qquad 
V(u)|_{{\mbf p}\to \infty}\sim i x_1 x_2 x_3,
\end{equation}	
where 
\begin{equation}	
x_j=({\mbf w}_j {\mbf p}), \qquad j=1,2,3 \label{xjdef}
\end{equation}	
with ${\mbf w}_j$ defined in \rf{edef}. Therefore the eigenvalues of ${\bf
M}(\l)$
\begin{equation}	
{\bf \Lambda}^{(cl)}_j(\l)=e^{2\pi i x_j(\l)}, 
\end{equation}	
are expressed though the roots $x_j(\l)$,  $j=1,2,3$, of the cubic 
equation 
\begin{equation}
x(\l)^3+g_2\,x(\l)+g_3-i\l^3=0, \label{cubic}
\end{equation}	
where we have introduced the notation
\begin{equation}
g_2=x_1x_2+x_1x_3+x_2x_3,\qquad g_3=-x_1x_2x_3.\qquad
\end{equation}	
The roots of \rf{cubic} can be written as 
\begin{equation}	
x_j(\l)=2\Big[\Big(-{g_2\over3}\Big)^{\frac{1}{2}}
{}_2F_1(-\frac{1}{6},\frac{1}{6},\frac{1}{2};
1+\frac{27\,(g_3-i\l^3)^2}{4\,g_2^2})\Big]_{j-th}, \label{hyperxj}
\end{equation}	
where ${}_2F_1$ stands for the Gauss hypergeometric function. The
expression in the square brackets above is a multivalued function and the
symbol $[...]_{j-th}$ denotes an appropriate  branch of this function
fixed by the requirement $x_j(0)=x_j$ where $x_j$, $j=1,2,3$ defined
in \rf{xjdef}.
Expanding this expression in a series in $\l$ and using the
Sommerfeld-Watson transformation one can rewrite \rf{hyperxj} as 
\begin{equation}		
\log{\bf \Lambda}^{(cl)}_j(\l)=2\pi i r_j+
\int_{C_\nu}{d\nu\over \nu}\,
{\Gamma(1-{i\nu\over3})
\Gamma(-{1\over3}+{i\nu\over3})\Gamma({1\over3}+{i\nu\over3})
}(-i\l)^{{i\nu\over3}} \ {\bf \Psi}^{(cl)}_j(\nu), \label{lamclas}
\end{equation}	
where
\begin{eqnarray}	
{\bf \Psi}^{(cl)}_j(\nu)\Big|_{{\mbf p}\to \infty}&=&
{2^{-{2i\nu\over3}}\over \sqrt{3\pi}\Gamma(1/2+i\nu/3)}
\times\nonumber\\
&&\Big[\Big(-{g_2\over3}\Big)^{(1-i\nu)/2}\ {}_2F_1(-\frac{1}{6}+
\frac{i\nu}{6},\frac{1}{6}+
\frac{i\nu}{6},\frac{1}{2}+
\frac{i\nu}{3},1+\frac{27 g_3^2}{4 g_2^3})\Big]_{j{\rm -th}}, \label{psi}
\end{eqnarray}	
where the integration contour goes along the line $\Im m\, 
\nu=-1-\epsilon$ and the branches of the expression in the square 
brackets are chosen in accordance with \rf{hyperxj}
\begin{equation}	
{\bf \Psi}^{(cl)}_j(0)={({\mbf w}_j{\mbf p})\over 2\pi\sqrt{3}}.
\end{equation}	

Now turn to the quantum case. As before we restrict our considerations
to the case $0<g<{2\over3}$. 
Examining expressions \rf{Aint} it is easy to
conclude that the leading ${\mbf p}\to \infty$ asymptotics of 
${\bf \Psi}(\nu)$ is the same for all its eigenvalues, i.e. 
${\bf \Psi}(\nu)|_{{\mbf p}\to \infty}$ is proportional to the unit
operator ${\bf I}$. Further we make an assumption (which again
can be justified by studying the ${\mbf p}\to \infty$ of the Coulomb
integrals \rf{g1vac}) that the ${\mbf p}$ and $g$ dependence of 
${\bf \Psi}(\nu)|_{{\mbf p}\to \infty}$ factorizes in the form
\begin{equation}	
{\bf \Psi}(\l)|_{{\mbf p}\to \infty}\sim \psi_1(\nu,g)\psi_2(\nu,{\mbf
p})\ {\bf I}.
\end{equation}
	
Comparing \rf{lamclas} with the $g\to 0$ limit of \rf{Aint} one 
gets 
\begin{equation}
{\bf \Psi}_j(\nu)|_{{\mbf p}\to \infty}={\psi_1(\nu,g)\over\psi_1(\nu,0)}
{\bf \Psi}^{(cl)}_j(\nu)|_{{\mbf p}\to \infty}. \label{clasquant}
\end{equation}	
To calculate the $g$-dependent multiplier in the last formula we
will use the DDV equation. Since the asymptotics \rf{Qas} are the same for
all eigenvalues of ${\bf \Psi}(\nu)$ it is enough to consider the
vacuum eigenvalue. Moreover, since the multiplier we want to calculate
does not depend on ${\mbf p}$ we will restrict ourselves to the 
case $p_2=0$ and denote
\begin{equation}
p_1=p, \qquad x_1=p, \qquad x_2=0,  \qquad x_3=-p. \label{pddv}
\end{equation}

In this case the vacuum eigenvalues $A_1^{(vac)}(t)$ and
$\overline{A}_3^{(vac)}(t)$ of the operator ${\bf A}_1(t)$ and 
$\overline{{\bf A}}_3(t)$ satisfy the relation 
\begin{equation}	
\overline{A}_3^{(vac)}(t)=A_1^{(vac)}(-t).
\end{equation}
Define the function 
\bn
a(t)=e^{2\pi
ip}\frac{A^{(vac)}_{1}(itq^{2})}{A^{(vac)}_{1}(itq^{-2})}
\frac{A^{(vac)}_{1}(-itq^{-1})}{A^{(vac)}_{1}(-itq)}.
\label{adefn} 
\ed
Recalling the assumption about the structure of the zeros of $A_1(t)$
in the beginning of Section~\ref{Asymptotic Expansions} (the zeroes of
$A_1^{(vac)}(t)$ lie on the positive imaginary axis of $t$) 
and using \rf{Aprod}
we can show for real $p$  
\bn
a(t)^*=a(t^*)^{-1}. \label{inverse}
\ed
Due to \rf{Aas} $a(t)$ has the asymptotics 
\bn
\log a(t)\sim -2iM\sin\left(\frac{\pi(1+\xi)}{3}\right)\
t^{{1+\xi}\over3}, \qquad 
t\rightarrow\infty,\qquad |\arg(t)|<\min[2\pi g,\pi(1-g)] \label{asymptotic} 
\ed
while for small $t$, we obviously have
\bn
a(t)=2\pi ip +O(t), \qquad t\rightarrow 0.\label{smallt}
\ed

Again using \rf{Aprod} we have
\bn
\log a(t)=2\pi ip +\sum_{k}F(itt_k^{-1}), \label{sum}
\ed
where
\bn
F(t)=\log \left( \frac{1-tq^2}{1-tq^{-2}}\frac{1+tq^{-1}}{1+tq}\right)
\label{F}. 
\ed
Using  the Bethe Ansatz equations \rf{BAE}, \rf{phases} we can write the sum in
\rf{sum} as a contour integral 
\bn
\log a(t)=2\pi ip +\int_C\frac{\,dv}{2\pi
i}F(tv^{-1})\partial_v\log(1+a(v)), \label{contour} 
\ed
where the contour $C$ goes from $+\infty$ to zero above the
real axis, winds about zero, and returns to $+\infty$ below
the real axis. Integrating by parts
(boundary terms vanish) one obtains 
$$
\log a(t)-2\pi ip = \frac{1}{2\pi
i}\int_{0}^{\infty}\frac{dv}{v}\ t\partial_tF(tv^{-1}) \left\{\log
(1+a(v-i0))-\log(1+a(v+i0))\right\},  
\nonumber
$$
$$
=\int_{0}^{\infty}\frac{dv}{\pi v}t\partial_tF(tv^{-1})\Im
m\log(1+a(v-i0))-\int_{0}^{\infty}\frac{dv}{2\pi
iv}t\partial_tF(tv^{-1})\log a(v), \nonumber 
$$
where we have used \rf{inverse} to obtain the second line.

We now introduce new variables
\bn
t=e^{\frac{3\t}{1+\xi}}, \quad v=e^{\frac{3\t'}{1+\xi}}, 
\ed
and recalling that $q=\exp(i\pi g)$, we obtain
\begin{eqnarray}
\log a(\t)=2\pi ip &-& 2i\int_{-\infty}^{\infty}R(\t-\t')\Im
m\log(1+a(\t'-i0))\,d\t' \nonumber \\ 
&+&\int_{-\infty}^{\infty}R(\t-\t')\log a(\t')\,d\t', \label{newvariables}
\end{eqnarray}
where
\bn
R(\t)=\frac{3i}{2\pi(1+\xi) }t\partial_{t}F(t).
\ed

With the standard notation for the convolution integral the
Eq. \rf{newvariables} becomes 
\bn
K*\log a(\t)=2\pi ip-2i R*\Im m\log(1+a(\t-i0)), \label{convolution}
\ed
with $K(\t)=\delta(\t)-R(\t)$.
We now apply the inverse of the integral operator $K(\t)$ to both
sides of \rf{convolution}. Since $K(\t)$ is a singular operator
one has  to add an appropriate zero mode of $K$ to the RHS of the
resulting equation. The required 
zero mode is completely fixed by the asymptotic conditions
\rf{asymptotic} and \rf{smallt}. In this way 
one  obtains 
\bn
i\log a(\t)=-\frac{2\pi p}{g}-2 
G*\Im m\log(1+a(\t-i0))+2M\sin\left(\frac{\pi(1+\xi)}{3}\right)e^{\t},
\label{zero modes} 
\ed
where $G(\t)=-(K^{-1}\ast R)(\t)$. 

Now define
\bn
B(p)=\frac{(1+\xi)}{3}\log (-it_1), \label{Bp}
\ed
where $t_1$ is the smallest (in absolute value) zero of $A^{vac}_1(t)$. 
The convolution term in \rf{zero modes} can be written as
\begin{eqnarray}
G*\Im m\log(1+a^{vac}(\t-i0))&=&\int_{-\infty}^{B(p)}G(\t-\t')\Im
m\log(1+a^{vac}(\t'-i0))\,d\t'\nonumber \\ 
&+&\int_{B(p)}^{\infty}G(\t-\t')\Im
m\log(1+a^{vac}(\t'-i0))\,d\t'. \label{intcutoff} 
\end{eqnarray}
Let us make the technical assumption that 
\bn
B(p)\sim \mbox{const}\log p, \quad p\rightarrow \infty. \label{Bp assump}
\ed
Then one can show the second term in \rf{intcutoff} vanishes when
$p\to+\infty$ and therefore can be  
dropped in the leading large $p$ approximation.  The Eq. 
\rf{zero modes} then becomes
\bn
-\frac{\pi p}{g}+M\sin\left(\frac{\pi(1+\xi)}{3}\right)e^{\t}=
\int_{-\infty}^{B(p)}\frac{d\t'}{2\pi i}\partial_\t
\log S(\t-\t')\Im m\log\left(1+a(\t')\right)\,, \label{Wiener-Hopf}
\ed
where we have used the notation
\begin{equation}	
G(\t)=\delta(\t)+{1\over 2 \pi i}\partial_\t \log S(\t),
\end{equation}	
\begin{equation}	
\log S(\t)=-i\int_{-\infty}^{\infty}{d\nu \over \nu}\sin(\nu \t)
\frac{\sinh\left(\frac{\pi
\nu(1+\xi)}{3}\right)\cosh\left(\frac{\pi
\nu}{6}\right)}{\sinh\left(\frac{\pi
\nu\xi}{3}\right)\cosh\left(\frac{\pi \nu}{2}\right)},
\label{Smatr} 
\ed
and the relation 
\bn
\log a^{vac}(\t)=2i\log\Im m\log(1+a^{vac}(\t)), \quad \t<B(p) 
\ed
which holds for $\t<B(p)$. The function $S(\t)$ is Smirnov's kink-kink $S$-matrix for the algebra $A^{(2)}_2$, \cite{Smi91}.

The Fourier transform of the integral kernel in \rf{Wiener-Hopf} factorizes as 
\begin{equation}	
-{1\over 2\pi i}\int_{-\infty}^{\infty}
e^{-i\nu \t}\partial_\t\log S(\t)d\t={1\over K_+(\nu)K_-(\nu)},
\end{equation}	
where 
\bn
K_+(\nu)=\frac{1}{\sqrt{g}}\frac{\Gamma\left(-\frac{i\nu(1+\xi)}{3}\right)
\Gamma\left(\frac{1}{2}-\frac{i\nu}{6}
\right)e^{i\nu\delta}}{\Gamma\left(-\frac{i\nu\xi}{3}\right)
\Gamma\left(\frac{1}{2}-\frac{i\nu}{2}\right)},\qquad K_-(\nu)=K_+(-\nu),
\label{K}
\ed
\bn
3\delta=(1+\xi)\log(1+\xi)-{\xi}\log\xi+\log{2\over3\sqrt{3}}.
\label{delta} 
\ed
The function $K_+(\nu)$  is analytic and nonzero 
in the upper half-plane $\Im m\  \nu>-1$ where it has the asymptotics
\begin{equation}	
K_+(\nu)\sim1+O\left(\frac{1}{\nu}\right), \qquad\nu\to \infty.
\end{equation}	
The Eq.\rf{Wiener-Hopf}  is an equation of Wiener-Hopf type, 
which can be solved by the standard technique.
Defining the Fourier transform
\bn
f_+(\nu)=\int_{-\infty}^{B(p)}\Im m\log(1+a^{vac}(\t))e^{-i\t \nu}\,d\t, 
\label{f+}
\ed
one obtains
\bn
\frac{f_+(\nu)}{K_+(\nu)}e^{i\nu B(p)}
=\frac{1}{2\pi i}\int_{-\infty+i\epsilon}^{\infty+i\epsilon}\frac{1}{\nu'-\nu}K_-(\nu')
G_+(\nu')\,d\nu',
\ed
where
\begin{eqnarray}
G_+(\nu)&=&\int_{-\infty}^{0}\left(\frac{\pi p}{g}-M\sin\left(\frac{\pi(1+\xi)}{3}
\right)e^{\t+B(p)}\right)e^{-i\t\nu}\,d\nu, \nonumber \\
&=&\frac{i\pi p}{g\nu}-\frac{iM}{\nu+i}\sin\left(\frac{\pi(1+\xi)}{3}\right)e^{B(p)},
\end{eqnarray}
and $\epsilon$ is a small positive constant.
Completing the contour in the lower half-plane, we have
\bn
\frac{f_+(\nu)}{K_+(\nu)}e^{i\nu B(p)}
=\frac{i\pi p}{\sqrt{g}\nu}
-\frac{iM\pi e^{B(p)-\delta}}{\sqrt{g}(\nu+i)}
\frac{\Gamma\left(\frac{2}{3}\right)}
{\Gamma\left(\frac{\xi}{3}\right)\Gamma\left(\frac{2-\xi}{3}\right)}.
\label{WH1} 
\ed
As follows from the definition \rf{f+} the leading $1/\nu$ term in the $\nu\rightarrow\infty$ asymptotic expansion of the RHS of \rf{WH1} is independent of $p$. This gives
\bn
(-it_1)^{\frac{1+\xi}{3}}=
p\frac{\Gamma\left(\frac{\xi}{3}\right)\Gamma\left(\frac{2-\xi}{3}
\right)}{M\Gamma\left(\frac{2}{3}\right)}e^\delta
\label{e-b}, 
\ed
and thus we have
\bn
f_+(\nu)=\frac{i\pi p}{2\nu}\frac{\Gamma\left(1-\frac{i\nu(1+\xi)}{3}\right)
\Gamma\left(\frac{1}{2}-\frac{i\nu}{6}\right)}
{\Gamma\left(1-\frac{i\xi\nu}{3}\right)
\Gamma\left(\frac{3}{2}-\frac{i\nu}{2}\right)}e^{i\nu\delta}(-it_1)^
{\frac{-i\nu(1+\xi)}{3}}. \label{WHfinal}
\ed

Given the function $f_+(\nu)$, \rf{WHfinal}, we can recover $\log A_1(t)$ from
\bn
\log A_1(t)=-\frac{i}{2\pi}\int_{C_\nu}
\frac{\cosh\left(\frac{\pi \nu}{6}\right)}
{\cosh\left(\frac{\pi \nu}{2}\right)
\sinh\left(\frac{\pi \xi\nu}{3}\right)}f_+(\nu)
(it)^{\frac{i\nu(1+\xi)}{3}}\,d\nu,
\ed
where the contour $C_\nu$ is along the line $\Im m \ \nu=-1-\epsilon$, with 
$\epsilon$ is a small positive constant. Thus we have
\bn
\log A^{vac}_1(t)=-\frac{3p}{4\pi \xi}\int_{C_\nu}\frac{e^{i\nu\delta}}{\nu^2}
\frac{\Gamma\left(1-\frac{i\nu(1+\xi)}{3}\right)
\Gamma\left(-\frac{1}{2}+\frac{i\nu}{2}\right)
\Gamma\left(1+\frac{i\xi \nu}{3}\right)}
{\Gamma\left(\frac{1}{2}+\frac{i\nu}{6}\right)}
\left(-\frac{t}{t_1}\right)^{\frac{i\nu(1+\xi)}{3}}\,d\nu. \label{Aintegral}
\ed
This expression can be written in the form of  \rf{Aint}
\begin{eqnarray}
\log \A_i(t)&=&{1\over 2\pi i}\int_{C_\nu}{d\nu\over \nu}\,
\Gamma(i\nu\xi/3)\Gamma(1-i\nu(\xi+1)/3)
\Gamma((-1+i\nu)/3)\times\nonumber\\
&&\phantom{{1\over 2\pi i}\int_{C_\nu}}
\Gamma((1+i\nu)/3)\Big(\Gamma(1-g)\Big)^{i\nu(1+\xi)}
(it)^{i\nu(1+\xi)/3} \ {\bf \Psi}(\nu)|_{p\rightarrow\infty}, 
\end{eqnarray}
where 
\bn
{\bf \Psi}(\nu)|_{p\rightarrow\infty}
={2^{-{2i\nu\over3}}\over
\sqrt{3}}\Big({p^2\over3}\Big)^{(1-i\nu)/2}{1\over
\Gamma(1/3+i\nu/6)\Gamma(2/3+i\nu/6)}.  
\ed
Comparing this with the specialization of \rf{psi} for the case \rf{pddv},     
one concludes  that the factor in \rf{clasquant}
\bn
{\psi_1(\nu,g)\over\psi_1(\nu,0)}\equiv 1.
\ed
Taking into account \rf{clasquant}, we obtain  the expression
\rf{psiq} given in the main text.

To get a small $t$ expansion of $A^{vac}_1(t)$, we complete the contour in the
 lower-half plane, and evaluate the residues at the points $\nu=3i(g-1)n, \ \ 
n=1,2,3\ldots $, and we obtain the large $p$ asymptotic equivalent of 
\rf{Hdef}.
Thus, we have from \rf{Aintegral}
\bn
H^{(1)(vac)}_{1}|_{p\rightarrow\infty}=-\frac{p}{2}
\frac{e^{3(1-g)\delta}}{\Gamma^3(1-g)}\frac{\Gamma\left(1-\frac{3}{2}g \right)
\Gamma(g)}{\Gamma\left(1-\frac{1}{2}g\right)}g^3(-it_1)^{-1}. \label{Hn1}
\ed
However, we know $H^{(1)(vac)}_1$ exactly \rf{A22H1}. 
Taking the large $p$ asymptotic of \rf{A22H1}, comparing with \rf{Hn1}, and
 using \rf{e-b} we see that we must take 
\bn
M=\frac{\Gamma\left(\frac{\xi}{3}\right)
\Gamma\left(\frac{2-\xi}{3}\right)}
{\Gamma\left(\frac{2}{3}\right)}\left(\Gamma(1-g)\right)^{\frac{1}{1-g}}, 
\label{M}
\ed
and then 
\bn
H^{(1)(vac)}_n|_{p\rightarrow\infty}=\frac{(-1)^n}{2n!}
\frac{\Gamma\left(\frac{3(1-g)n-1}{2}\right)\Gamma(gn)}
{\Gamma\left(\frac{(1-g)n+1}{2}\right)}
g^{3n}p^{1-3(1-g)n}. \label{Hnfinal}
\ed

We also generate an asymptotic expansion for large $t$ by completing the 
contour in the upper half-plane. Here we have poles at the following points:
\bn
\nu=i(2n-1),\quad n=0,1,2,\ldots, \quad \nu=3i(g^{-1}-1)n, \quad n=0,1,2,3
\ldots, 
\ed 
and we obtain
\begin{eqnarray}
A^{vac}_1(t)&=&C^{vac}(g,p)(s)^{-\frac{p}{g}} 
\exp\left\{\sum_{n=0}^{\infty}B_{2n-1}I^{vac}_{2n-1}|_{p\rightarrow\infty}
(s)^{-\frac{2n-1}{3(1-g)}}\right\} \nonumber \\
&\times &\exp\left\{-\sum_{n=1}^{\infty}
\tilde{H}^{(1)(vac)}_{n}|_{p\rightarrow\infty}
(s)^{-\frac{n}{g}}\right\}, \label{larget}
\end{eqnarray}
where
\bn
B_{2n-1}=\frac{(-1)^{n+1}}{3(1-g)(n)!}
\frac{\Gamma\left(\frac{2n-1}{3(1-g)}\right)
\Gamma\left(\frac{2n-1}{3(1-g^{-1})}\right)}
{\Gamma\left(\frac{2-n}{3}\right)}g^{\frac{n(1+g)-1}{g-1}}, \label{Bn}
\ed
\bn
C^{(vac)}(g,p)=\left(\frac{2^{\frac{2}{3}}p}{e}\right)^{\frac{3p}{g}-3p}(g)^{-\frac{3p}{g}}, \qquad \tilde{H}^{(1)(vac)}_{n}(g, p)=H^{(1)(vac)}_n(g^{-1}, g^{-1}p). \label{Htilde}
\ed
and $I^{vac}_{2n-1}|_{p\rightarrow\infty}$, $n=0,1,2,\ldots$ are the remaining
 Integrals of Motion after the $A^{(2)}_2$ specialization has been taken into
 account ($W(u)\equiv 0$),
\bn
I^{vac}_{-1}\equiv 1, \qquad I^{vac}_{2n-1}|_{p\rightarrow\infty}=
\frac{p^{2n}}{g^{n}}, \quad n=1,2,3,\ldots
\ed
Note also that $C^{(vac)}(g,p)$ is the vacuum asymptotic value of the
operator ${\bf C}_j\{g,g{\boldsymbol \phi}(u)\}$ defined in \rf{Qas}.   

\renewcommand{\baselinestretch}{1.0}
\normalsize

\app{Representation Matrices for the Fundamental Representations}
\label{repmatrix}  

\def\poa{\pi_{\omega_1}}
\def\pob{\pi_{\omega_2}}
For the first fundamental representation $\poa$ we have
\bn
\poa(H_{\a})=\left( \begin{array}{ccc}
		1 & 0 & 0 \\
		0 & -1 & 0 \\
		0 & 0 & 0 
			\end{array} \right) \quad
\poa(E_{\a})=\left( \begin{array}{ccc}
		0 & 1 & 0 \\
		0 & 0 & 0 \\
		0 & 0 & 0
			\end{array} \right) \quad
\poa(F_{\a})=\left( \begin{array}{ccc}
		0 & 0 & 0 \\
		1 & 0 & 0 \\
		0 & 0 & 0
			\end{array} \right) \quad
\ed
\bn
\poa(H_{\b})=\left( \begin{array}{ccc}
		0 & 0 & 0 \\
		0 & 1 & 0 \\
		0 & 0 & -1 
			\end{array} \right) \quad
\poa(E_{\b})=\left( \begin{array}{ccc}
		0 & 0 & 0 \\
		0 & 0 & 1 \\
		0 & 0 & 0 
			\end{array} \right) \quad
\poa(F_{\b})=\left( \begin{array}{ccc}
		0 & 0 & 0 \\
		0 & 0 & 0 \\
		0 & 1 & 0 
			\end{array} \right)
\ed

while for $\pi_{\omega_{2}}$ we have
\bn
\pob(H_{\a})=\left( \begin{array}{ccc}
		0 & 0 & 0 \\
		0 & 1 & 0 \\
		0 & 0 & -1
			\end{array} \right) \quad
\pob(E_{\a})=\left( \begin{array}{ccc}
		0 & 0 & 0 \\
		0 & 0 & 1 \\
		0 & 0 & 0
			\end{array} \right) \quad
\pob(F_{\a})=\left( \begin{array}{ccc}
		0 & 0 & 0 \\
		0 & 0 & 0 \\
		0 & 1 & 0 
			\end{array} \right) \ed
\bn
\pob(H_{\b})=\left( \begin{array}{ccc}
		1 & 0 & 0 \\
		0 & -1 & 0 \\
		0 & 0 & 0 
			\end{array} \right) \quad
\pob(E_{\b})=\left( \begin{array}{ccc}
		0 & 1 & 0 \\
		0 & 0 & 0 \\
		0 & 0 & 0
			\end{array} \right) \quad
\pob(F_{\b})=\left( \begin{array}{ccc}
		0 & 0 & 0 \\
		1 & 0 & 0 \\
		0 & 0 & 0 
		\end{array} \right) \ed

\renewcommand{\baselinestretch}{1.2}
\normalsize

\def\cprime{$'$}

\end{document}